\documentclass[12pt,twoside]{mitthesis}
\RequirePackage{algorithm}[0.1]
\usepackage{lgrind}
\usepackage{cmap}
\usepackage[T1]{fontenc}
\usepackage{graphicx,caption}
\usepackage{chngpage}
\graphicspath{ {./Figures/} }
\def\all{all}
\ifx\files\all  \else 

\begin{document}

\title{Numerical Characterization of Fragmentation \\ in Ionic Liquid Clusters}

\author{Madeleine Schroeder}
\department{Department of Aerospace and Astronautical Engineering}

\degree{Master of Science in Aerospace and Astronautical Engineering}

\degreemonth{May}
\degreeyear{2021}
\thesisdate{May 18, 2021}


\supervisor{Paulo Lozano}{M. Alem\'{a}n$-$Velasco Professor of Aeronautics and Astronautics, MIT}

\chairman{Zoltan Spakovszky}{Chair, Graduate Program Committee}

\maketitle



\cleardoublepage
\setcounter{savepage}{\thepage}
\begin{abstractpage}
%
%
%

Ionic liquid ion sources are a promising technology that can be used for many applications from space propulsion to focused ion beam microetching. The variety of ionic liquids that can be synthesized enables the selection of desired beam properties for optimizing propulsion and focused ion beam performance. Ionic liquid ion sources produce ion beams by extracting single ions and metastable solvated ion clusters from the surface of the ionic liquid and accelerating them using an electric field generated by applying a voltage between a sharp tip and a plate with an aperture. The solvated ion clusters often fragment in the electric field region, reducing the specific impulse and efficiency for propulsion applications and increasing the beam spot size for focused ion beam applications by broadening the energy distribution of the beam. 

Fragmentation behavior has previously been characterized in the region with no electric field. However, fragmentation under the effect of an electric field has not been investigated as experimental results are difficult to interpret for regions with electric fields. The goal of this work is to use various types of numerical methods to characterize fragmentation under the effect of an electric field. Molecular dynamics simulations are performed of various ionic liquid clusters under different conditions to determine the rate of fragmentation. These simulation results are also used to determine the different fragmentation pathways taken by each type of cluster, and the size of the different clusters as a result of energy content and electric field strength. Various physics-based models are compared to the molecular dynamics results with the goal of deriving a new model that accounts for the effect of the electric field on fragmentation. Approximate Bayesian computational methods are employed to infer the temperature of different ionic liquid cluster types and the percentage of the beam composed of each species by comparing simulated retarding potential analysis curves to experimental ones. Finally, the results of multi-scale N-body simulations are post-processed and compared to experimental data. Results show remarkable agreement between N-body simulations using the fragmentation rates determined by molecular dynamics and experimental data. 

\end{abstractpage}


\cleardoublepage

\section*{Acknowledgments}

I would first like to thank Professor Paulo Lozano for giving me the opportunity to work in SPL. Starting research in SPL as an undergrad is the reason that I stayed at MIT, and I am all the better for it. Thank you for the unwavering support and encouragement.

Thank you to my family for supporting me in everything I do. Thank you Mom and Dad for always listening to my ramblings. Thank you Grandma and Grandpa for the unconditional love and support, I wish that you could see my graduations in person. Thank you aunts and uncles and cousins for giving me perspective, and reminding me that my job is actually pretty cool.

Thank you to my friends for telling me to keep going. Reign and Levi, thank you for keeping me grounded and reminding me that the world outside is still spinning. Humby, MIT wouldn't have been the same without you. You were always there for me, and for that I am eternally grateful. 

I would also like to thank the people in the Space Propulsion Lab. Thank you to Charity, Matt, and Colin for helping me make it through quarantine with an ounce of sanity. Kyle and David, thank you for putting so much effort into your UROP work. Much of the data presented in this thesis was made possible by your efforts. Gustav, thank you for listening, even when you really needed to get work done. None of my experiments would have worked without your advice. Mia, thank you for always being in the lab. Breaking things is only fun when you are there. And last but not least, Elaine, thank you for introducing me to SPL. Your support during my undergraduate research made me want to stay at MIT, and I cannot thank you enough for that.


\pagestyle{plain}
\tableofcontents
\newpage
\listoffigures
\newpage
\listoftables

\chapter{Introduction}

\section{Motivation}

Small satellites such as Cubesats, provide a low complexity, low cost way to test new technologies in space. Cubesats have applications for many missions including Earth observation, science and technology demonstration, and even deep space exploration. The relative low complexity enables rapid prototyping and testing of new space technologies. The relative low cost offers opportunities for more people to be involved in space exploration, including university teams and small companies. The low cost also facilitates rapid technology development due to the lower economic risk. 

Most of the electronics for satellites including communication, power, and attitude control subsystems have been successfully miniaturized, facilitating a recent growth in small satellite development. However, propulsion systems are often not included on small satellites \cite{Lemmer2017PropulsionCubeSats}. The development of electric propulsion (EP) systems that fit the small satellite form factor further expands the opportunities available to such satellites by facilitating precise attitude control for imaging and larger delta-V missions for deep space exploration \cite{Krejci2015,Freeman2019DesignThrusters,Mier-Hicks2017ElectrospraySatellites,Jia-Richards2020AnalyticalSystems}. However, most of the EP systems used for larger satellites including hall effect thrusters and ion engines can not be scaled down to the size required by Cubesats and other small satellites while preserving high performance. One of the emerging electric propulsion technologies to address this problem is ionic liquid ion source (ILIS) electrospray propulsion. 

\begin{figure}
    \centering
    \includegraphics[width=4.5in]{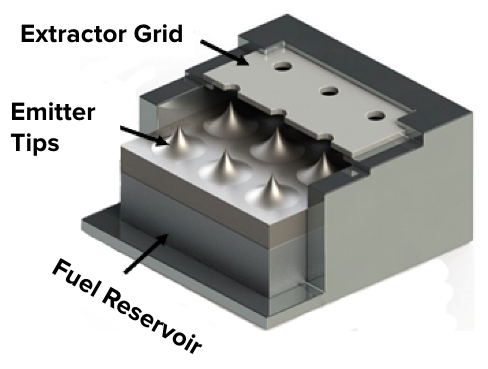}
    \caption{Diagram of electrospray emitter geometry. Image created by Catherine Miller.}
    \label{fig:EsprayExample}
\end{figure}

ILIS electrospray produces thrust by accelerating ions from an ionic liquid propellant using an electric field \cite{Krejci2017}. Ionic liquids are salts that are liquid at room temperature and have negligible vapor pressure, making them particularly well suited to space applications as they require no complex containment system \cite{Huddleston2001CharacterizationCation,Chiu2007IonicPropulsion,Prince2012IonicSystems}. A voltage is applied between a sharp tip covered in propellant and an extractor grid with a hole in it. The resulting electric field extracts pure ions and ion clusters from the surface of the ionic liquid and accelerates them, producing thrust. Typical electrospray emitter systems for use on small satellites consist of arrays of hundreds of sharp tips \cite{Krejci2015,Krejci2017,Freeman2019}. These arrays are made out of a porous material which is wetted with the propellant. Figure \ref{fig:EsprayExample} shows a diagram of one such electrospray emitter system. Electrospray technology is also applicable to other fields including use in chemical processes, superconductors, and focused ion beam (FIB) applications such as micromachining, imaging, and material deposition \cite{Zorzos2008TheApplications,VakilAsadollahei2019InvestigationApplications,Perez-Martinez2012VisualizationApplications,Fedkiw2009DevelopmentApplications,Perez-Martinez2010DevelopmentMicrofabrication}.

While electrospray propulsion systems have flown successfully on technology demonstration missions, further characterization of emission behavior is necessary to understand lifetime limiting mechanisms for the propulsion system as well as interactions of the ion beam with the satellite. Characterization of electrospray ion beams is also necessary to improve beam behavior for FIB applications. In particular, the process of fragmentation has significant impacts on propulsion thrust and specific impulse, hardware degradation, and FIB spot size \cite{Miller2020MeasurementSources,Perez-Martinez2010DevelopmentMicrofabrication,Perez-Martinez2012VisualizationApplications, Zorzos2008TheApplications}. During emission, both single ions and clusters composed of a single ion plus a neutral ion pair are extracted from the liquid surface and accelerated by the electric field. Fragmentation is a process in which the ion clusters break up, usually into a single ion and a number of neutral ion clusters. If the fragmentation process occurs during the acceleration of the cluster the resulting ion exits the extractor grid at a different velocity than the clusters that did not fragment. Emission of ions and clusters with different velocities reduces the efficiency and specific impulse of the propulsion system \cite{Coles2012InvestigatingBeams,Coles2013InvestigatingBeams,Lozano2005EfficiencyThrusters}. Neutral ion cluster impingement on the extractor grid is believed to lead to liquid accumulation and eventual electrical shorts. Energy spread in the beam makes it more difficult to focus the beam for FIB applications \cite{Perez-Martinez2010DevelopmentMicrofabrication,Perez-Martinez2012VisualizationApplications,Fedkiw2009DevelopmentApplications,Zorzos2008TheApplications}. Characterization of the fragmentation process is needed to understand its effects on hardware lifetime as well as propulsion and FIB system performance.

Previous experimental work to determine fragmentation behavior for electrospray emission has successfully determined the fragmentation rates of different ionic liquids when there is no electric field present \cite{Miller2019CharacterizationSources,Miller2020MeasurementSources}. This was done using retarding potential analysis (RPA) curves, which show the energy distribution of the ion beam and thus the fragmentation behavior. However, the fragmentation behavior in an electric field is difficult to characterize with RPA curves as the data from fragmentation of multiple cluster sizes appear in the same region of the graph and are difficult to interpret. Additionally, current models for cluster fragmentation in regions with electric fields have not been successful in predicting experimental behavior.

\section{Research Objectives}

The goal of this work is to characterize the behavior of ionic liquid clusters in regions with applied electric fields. There are three main areas in which this effort will be focused

\begin{enumerate}
    \item Determination of cluster fragmentation rates for clusters with temperatures between 300 K and 2500 K under the influence of electric fields between $5\times10^5$ V/m and $1\times10^{10}$ V/m.
    \item Derivation of a physics-based model for fragmentation rates under different conditions and comparison of this model to simulation data. 
    \item Determination of internal energy and fragmentation rates of clusters from experimental data.
\end{enumerate}

These goals are achieved with a combination of molecular dynamics (MD) simulations and approximate Bayesian computation (ABC). MD simulations are used to determine fragmentation rates of clusters of different ionic liquids and different sizes under different energy and electric field conditions. These results are also analyzed for trends in fragmentation behavior including geometry during fragmentation. The MD results from fragmentation are used to derive a physics-based approximation of the behavior of the clusters during fragmentation which is then used to calculate fragmentation rates for different conditions. ABC is used with the MD results and previously collected experimental data to determine the internal energy distribution and beam mass composition of electrospray emitters for various emission conditions.

\chapter{Background and Literature Review}

\section{Electrospray Emission}

\subsection{Ionic Liquid Properties}

Ionic liquids are salts that are liquid at room temperature. They are composed of a wide variety of organic or inorganic cations (A+) and anions (B-). Poor coordination of the anions and cations in the salt lead to low melting temperatures usually ranging between $0^{\circ}$C and $100^{\circ}$C \cite{Valderrama2016MeltingPrediction}. The forces between the ions in the liquid are Coulombic, which results in negligible vapor pressure under vacuum conditions \cite{Huddleston2001CharacterizationCation}. This makes ionic liquids good candidates for use in space as they do not require complex, heavy containment systems or dangerous pressurized conditions \cite{Chiu2007IonicPropulsion,Prince2012IonicSystems}. Many ionic liquids also have high conductivity and low viscosity, which makes them suitable for electrospray applications \cite{Huddleston2001CharacterizationCation,Chiu2007IonicPropulsion,Prince2012IonicSystems}. Ionic liquids are also currently being studied for applications in energy storage, chemical processes, and superconductors \cite{Qi2020High-VoltageBatteries,Hallett2011Room-temperature2,Ue2003ApplicationCapacitors}.

\begin{figure}
    \centering
    \includegraphics[width=6in]{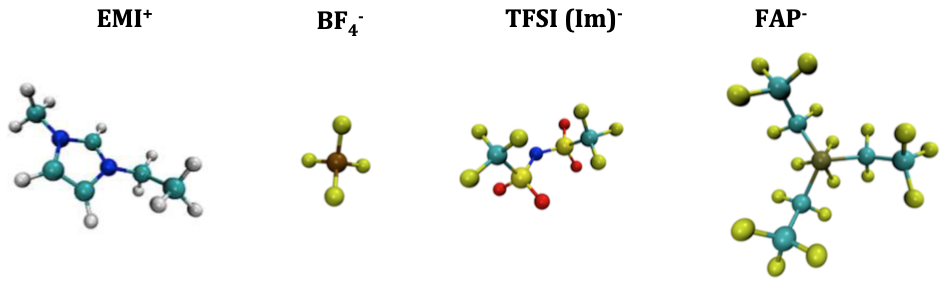}
    \caption{Molecular structure of ionic liquid ions.}
    \label{fig:ILvmd}
\end{figure}

This work focuses on three ionic liquids, EMI-BF$_4$, EMI-Im, and EMI-FAP. Figure \ref{fig:ILvmd} shows the molecular structure of the cation EMI and the various anions, BF4, Im, and FAP included in this work. Table \ref{tab:ILproperties1} and table \ref{tab:ILproperties2} show the properties of the ionic liquids that are used in this work \cite{Miller2019CharacterizationSources}. The dielectric constant for the ionic liquids listed here is likely approximately 12 F/m \cite{Huang2011Staticc}.

\begin{table}[ht]
\caption{Common ionic liquids \cite{Miller2019CharacterizationSources}}
\label{tab:ILproperties1}
\begin{center}
\begin{tabular}{l|l|l|l|l|l|l|}
\cline{2-7}
                                 & \multicolumn{3}{l|}{Cation}    & \multicolumn{3}{l|}{Anion}       \\ \hline
\multicolumn{1}{|l|}{Name}       & Formula        & Mass  & Atoms & Formula          & Mass  & Atoms \\ \hline
\multicolumn{1}{|l|}{$EMI-BF_4$} & $C_6N_2H_{11}$ & 111.2 & 19    & $BF_4$           & 86.8  & 5     \\ \hline
\multicolumn{1}{|l|}{$EMI-Im$}   & $C_6N_2H_{11}$ & 111.2 & 19    & $(C_2NS_2)_4F_6$ & 280   & 15    \\ \hline
\multicolumn{1}{|l|}{$EMI-FAP$}  & $C_6N_2H_{11}$ & 111.2 & 19    & $(C_2F_5)_3PF_3$ & 445   & 25    \\ \hline
\multicolumn{1}{|l|}{$BMI-I$}    & $C_8N_2H_{15}$ & 139.2 & 25    & I                & 126.9 & 1     \\ \hline
\end{tabular}
\end{center}
\end{table}

\begin{table}[ht]
\caption{Liquid properties of common ionic liquids \cite{Miller2019CharacterizationSources}}
\label{tab:ILproperties2}
\begin{center}
\begin{tabular}{|p{2.6cm}|p{2.6cm}|p{2.4cm}|p{2.8cm}|}
\hline
Name       & Conductivity (Si/m) & Viscosity (mPa s) & Surface Tension (mN/m) \\ \hline
$EMI-BF_4$ & 1.3                                                 & 27.5                                             & 52.0                                              \\ \hline
$EMI-Im$   & 0.88                                                & 34                                               & 34.9                                            \\ \hline
$EMI-FAP$  & 0.57                                                & 74.5                                             & 35.3                                                 \\ \hline
$BMI-I$    & $0.25 (50^{\circ}C$)                                 & $500 (50^{\circ}C$)                               & 54.7                                                  \\ \hline
\end{tabular}
\end{center}
\end{table}

\subsection{Ion Emission Principles}

\begin{figure}
    \centering
    \includegraphics[width=5in]{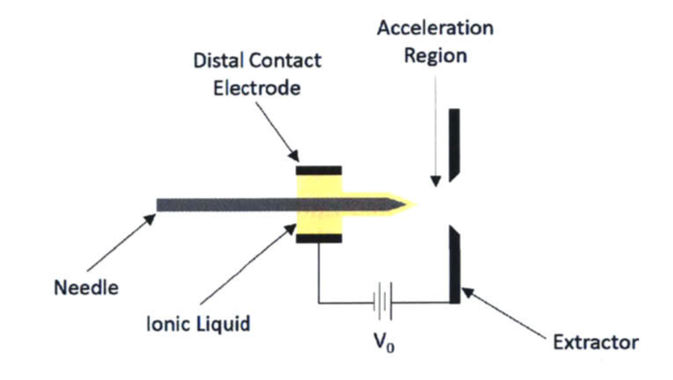}
    \caption{Geometry of a tungsten single electrospray emitter. Image created by Catherine Miller.}
    \label{fig:singleEmitterGeometry}
\end{figure}

\begin{figure}
    \centering
    \includegraphics[width=3.5in]{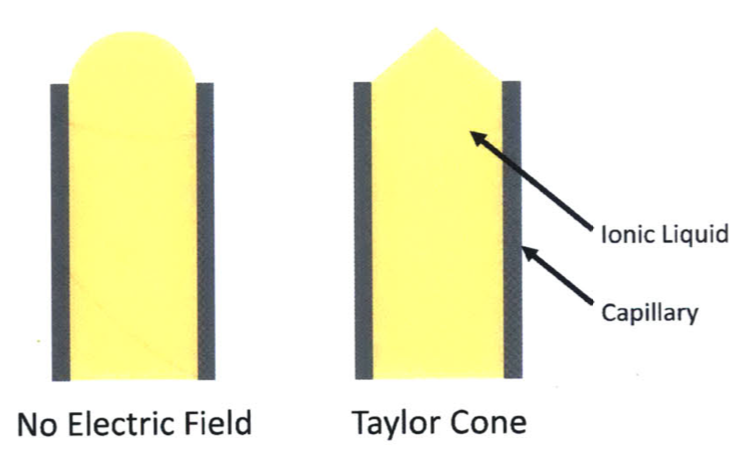}
    \caption{Taylor cone for a capillary electrospray source. Image created by Catherine Miller.}
    \label{fig:taylorCone}
\end{figure}

Electrospray emission occurs when an electric field is applied to an ionic liquid, evaporating single ions and ion clusters from the surface. Figure \ref{fig:singleEmitterGeometry} shows the geometry of a single electrospray emitter. This single emitter consists of a tungsten needle coated with ionic liquid. A voltage is applied between the distal electrode holding the ionic liquid and a plate with an aperture called the extractor grid. The radius of curvature of the tip is usually between 5 and 10 $\mu$m, which increases the electric field at the tip. When electric fields are applied to conductive liquids such as ionic liquids a Taylor cone forms \cite{Taylor1964DisintegrationField}. Figure \ref{fig:taylorCone} shows the formation of a Taylor cone for a capillary emitter geometry. 

The Taylor cone shape occurs when the traction from the applied electric field balances the force of the surface tension. This balance is given in Equation \ref{taylorBalance}

\begin{equation}\label{taylorBalance}
    \frac{1}{2}\epsilon_0 E^2 = \frac{\gamma}{R_c}
\end{equation}

where $\epsilon_0$ is the permittivity of free space, E is the magnitude of the electric field, $\gamma$ is the surface tension of the ionic liquid, and $R_c$ is the radius of curvature of the ionic liquid surface. When this balance holds the electric field is given by Equation \ref{TaylorEField}

\begin{equation}\label{TaylorEField}
    E = \sqrt{\frac{2 \gamma cot(\theta_T)}{\epsilon_r} }
\end{equation}

where $\theta_T$ is the half angle of the cone and r is the distance from the tip of the cone. Solving Laplace's equation gives the cone half angle as 49.3$^\circ$ \cite{Taylor1964DisintegrationField}. The electric field is highest at the tip of the cone. Theoretically there is a singularity at the tip of the cone where the radius of curvature is 0 and the electric field is infinite. In practice, this singularity cannot exist because ions take time to relax to the surface of a finite conductivity liquid. When the conductivity of a liquid is finite as it is in the case of ionic liquids, electrospray emission can occur in three different modes, the cone-jet mode, the mixed ion-droplet mode, and the pure ionic mode, also known as the pure ionic regime (PIR) \cite{Martinez-Sanchez2015SessionRegime,Martinez-Sanchez2015SessionThrusters}. See Reference \cite{Miller2019CharacterizationSources} for a literature review on the development of each mode. In the cone-jet mode back pressure is applied to a capillary filled with ionic liquid or electrolytic solvent   \cite{Martinez-Sanchez2015SessionRegime}. At the tip of the Taylor cone a jet forms, which breaks into droplets. The mixed ion-droplet regime occurs when the conductivity and surface tension of the liquid are high and the flow rate is low \cite{Martinez-Sanchez2015SessionRegime}. This results in the emission of single ions and ion cluster in addition to droplets \cite{Martinez-Sanchez2015SessionRegime}. The pure ionic mode occurs when the electric field strength is larger, resulting in a force great enough to evaporate single ions and ion clusters directly from the surface of the liquid \cite{Martinez-Sanchez2015SessionThrusters}. The PIR is achievable from capillary emitters as well as externally wetted single emitters and more recently, porous emitters with some of the common ionic liquids discussed previously \cite{Krejci2017,Martinez-Sanchez2015SessionThrusters,Legge2011ElectrosprayMetals,Courtney2012EmissionSources}. See References \cite{Miller2019CharacterizationSources, Castro2007EffectSources} for a review of some of the common ionic liquids that successfully emit in the pure ionic mode. 

In the pure ionic mode there is a characteristic distance from the tip of the emitter, $r^{\ast}$, where the charges are not relaxed \cite{Martinez-Sanchez2015SessionThrusters}. The force balance at the surface is then given by

\begin{equation}\label{taylorBalanceFull}
    \frac{1}{2}\epsilon_0 E_{\ast}^2 - \frac{1}{2}\epsilon_r\epsilon_0 E_{in}^2 = 2 \frac{\gamma}{r^{\ast}}
\end{equation}

where $E_{\ast}$ is the electric field outside the liquid surface, $\epsilon_r$ is the relative permittivity of the ionic liquid, and $E_{in}$ is the electric field inside the ionic liquid. We can then solve for the critical distance from the tip at which the charges are no longer relaxed as 

\begin{equation}\label{Rstar}
    r^{\ast} = \frac{4 \gamma}{\epsilon_0 E_{\ast}^2}(\frac{\epsilon_r}{\epsilon_r - 1})
\end{equation}

This gives the approximate area of the meniscus where emission occurs. Emission from this region is modeled as an activated process similar to other modes of evaporation \cite{Martinez-Sanchez2015SessionThrusters}. The current density for purely thermal emission is given by

\begin{equation}\label{thermalCurrent}
    j = \frac{\sigma k T}{h} exp(-\frac{1}{kT} \Delta G)
\end{equation}

where $j$ is the current density, $\sigma$ is the surface charge density of the liquid, $k$ is Boltzmann's constant, T is the temperature, h is Planck's constant, and $\Delta G$ is the solvation energy of the ion or ion cluster in the liquid. The effect of the electric field on the rate of emission is taken into account via the Schottky model, also known as the image point model \cite{Miller2020MeasurementSources,Petro2019DevelopmentModeling}. The Schottky model will be described further in section \ref{sec:Schottky}. Inclusion of the Schottky model for field enhanced ion emission yields Equation \ref{thermalCurrentFieldEnhanced} for the current density

\begin{equation}\label{thermalCurrentFieldEnhanced}
    j = \frac{\sigma k T}{h} exp(-\frac{1}{kT} (\Delta G - \sqrt{\frac{e^3 E_{\ast}}{4 \pi \epsilon_0}}))
\end{equation}

The critical electric field at which emission begins can be found by setting Equation \ref{CriticalEField} equal to zero and is given by 

\begin{equation}\label{CriticalEField}
    E_{\ast} = \frac{4 \pi \epsilon_0}{e^3} \Delta G^2
\end{equation}

Assuming the liquid at the tip is a hemisphere the total emitted current can be estimated from the characteristic emission area and the conductivity of the liquid. This is given in Equation \ref{FullCurrent}

\begin{equation}\label{FullCurrent}
    I = \frac{32 \pi K \gamma^2}{\epsilon_0^2 E_{\ast}^3} \frac{\epsilon_r}{\epsilon_r -1}
\end{equation}

The voltage required for emission can be calculating by solving Laplace's equation using prolate spheroidal coordinates \cite{Martinez-Sanchez2015SessionPropulsion}. This is given by Equation \ref{startupVoltage}

\begin{equation}\label{startupVoltage}
    V_{start} = \sqrt{\frac{\gamma R_C}{\epsilon_0}} ln (\frac{4 d}{R_C})
\end{equation}

where $R_c$ is the radius of curvature of the emitter and d is the distance between the tip and the extractor grid.



\subsection{Emitted Species}

\begin{figure}
    \centering
    \includegraphics[width=4.5in]{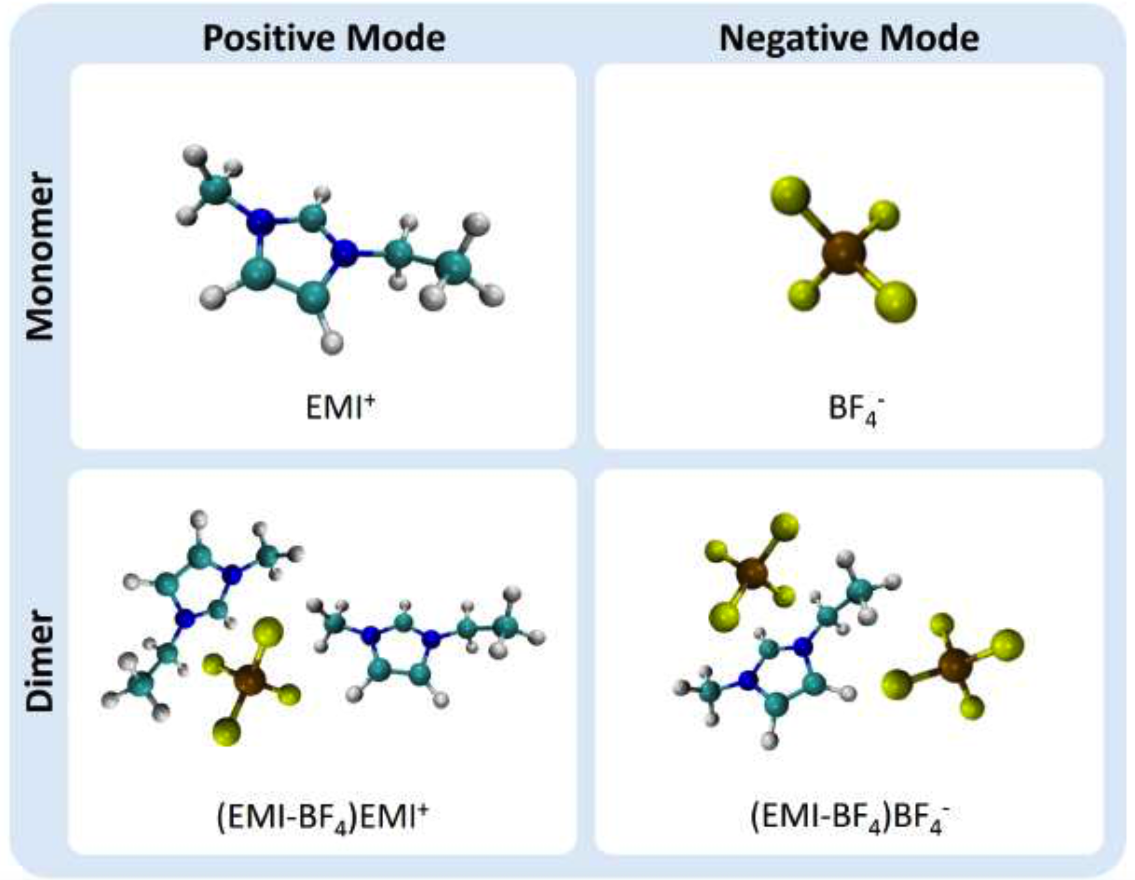}
    \caption{Monomers and dimers for EMI-BF$_4$. Image created by Catherine Miller.}
    \label{fig:clusterExample}
\end{figure}

Electrospray sources operating in the pure ionic regime produce primarily singly charged ions and singly charged ion clusters \cite{Martinez-Sanchez2015SessionThrusters,Castro2007EffectSources, Petro2020CharacterizationEmitters}. Single ions such as $A^+$ and $B^-$ are referred to as monomers. Clusters consist of a single ion with additional neutral ion pairs. The general cluster composition is given by $(A^+B^-)_nA^+$ or $(A^+B^-)_nB^-$ where $n$ denotes the number of neutral ion pairs in the cluster. Dimers are ion clusters with one neutral pair such as (A+B-)A+ or (A+B-)B-. Larger clusters such as trimers, quadmers, and pentamers follow the same convention with n values of 2, 3, and 4 respectively. Figure \ref{fig:clusterExample} shows an example of the monomers and dimers for the positive and negative mode for EMI-BF$_4$.

Electrospray sources operating in the PIR emit mostly monomers and dimers with beam trimer fractions less than 10\% \cite{Petro2020CharacterizationEmitters}. This work focuses on the behavior of dimers and trimers of various ionic liquids. The proportion of the ion beam composed of each cluster size depends on the firing conditions. Previous work has shown that emitter geometry, firing voltage, emitted current, and ionic liquid temperature all have effects on the beam mass composition \cite{Miller2019CharacterizationSources,Miller2015OnSources}. Emission of species of different masses affects the performance of electrospray propulsion systems as well as electrospray focused ion beam technology \cite{VakilAsadollahei2019InvestigationApplications,Fedkiw2009DevelopmentApplications,Coles2012InvestigatingBeams,Miller2019CharacterizationSources}. 


\subsection{Ion Acceleration Dynamics}\label{sec:Dynamics}

This section details the dynamics of emitted ions and ion clusters and how they are affected by fragmentation. Equations \ref{iontotE} through \ref{ionEmonFF} are taken from Reference \cite{Miller2019CharacterizationSources}.

Emitted ions have an initial velocity $v_0$. After emission the ions are accelerated in the electric field between the emission site and the extractor grid. This acceleration increases the velocity and thus the kinetic energy of the ions. The total energy of the emitted ion is then given by the sum of the potential energy from the potential field and the kinetic energy from the velocity of the ion

\begin{equation}\label{iontotE}
    E = q \phi_0 + \frac{1}{2} m v_0^2
\end{equation}

where $q$ is the total charge of the ion, $\phi_0$ is the potential of the ion source, $v_0$ is the initial velocity of the ion, and $m$ is the mass of the ion. After emission at some point x downstream from the emission site, the potential is given by $\phi(x)$ and the energy of the ion is given by

\begin{equation}\label{iontotEx}
    E(x) = q \phi(x) + \frac{1}{2} m v(x)^2
\end{equation}
 
where $v(x)$ is the velocity of the ion at the point x. The total energy of the ion is conserved as the force from the electric field is conservative. Thus we can equate Equation \ref{ionKE} and Equation \ref{iontotEx} to get

\begin{equation}\label{ionKEequal}
    q \phi_0 + \frac{1}{2} m v_0^2 = q \phi(x) + \frac{1}{2} m v(x)^2
\end{equation}

Solving for the velocity of the ion at point x yields

\begin{equation}\label{ionV}
    v(x) = \sqrt{v_0^2 + \frac{2 q (\phi_0 - \phi(x))}{m}}
\end{equation}

The velocity of the ion when it reaches the end of the potential field is given by evaluating Equation \ref{ionV} at the potential $\phi(l) = 0$ where $l$ is the distance to reach the end of the potential field

\begin{equation}\label{ionVexit}
    v_f = \sqrt{v_0^2 + \frac{2 q \phi_0}{m}}
\end{equation}

The kinetic energy of the ion is given by

\begin{equation}\label{ionKE}
    K(x) = \frac{1}{2} m (v(x))^2 = \frac{1}{2} m v_0^2 +  q(\phi_0 - \phi(x))
\end{equation}

Evaluating Equation \ref{ionKE} at $\phi(l) = 0$ gives the kinetic energy of the ion when it reaches the end of the potential field

\begin{equation}\label{ionKEexit}
    K_f = \frac{1}{2} m (v(x))^2 = \frac{1}{2} m v_0^2 +  q \phi_0
\end{equation}

The equations above hold for any emitted ions that do not fragment. These are called the monoenergetic ions as they all attain the same final kinetic energy which is equal to the source potential energy in the case where the initial velocity is negligible.

\subsubsection{Dimer Fragmentation}

Now consider a dimer that fragments into a monomer during emission. The mass of the dimer is given by $m_{di}$ and the mass of the monomer is given by $m_{mon}$. Suppose the dimer fragments at some point x downstream from the emission site. This means that when it fragments the dimer has been accelerated over a potential drop given by $\phi_0 - \phi_{di}$ where $\phi_{di}$ is the potential at which the dimer fragmented. Thus the velocity of the dimer when it fragments is given by

\begin{equation}\label{ionVdi}
    v_{di} = \sqrt{v_0^2 + \frac{2 q (\phi_0 - \phi_{di})}{m_{di}}}
\end{equation}

When fragmentation occurs the solvation energy of the cluster is assumed to be much smaller than the kinetic energy of the ion which means the products of fragmentation continue in the same direction with the same velocity as the dimer before fragmentation. The energy of the monomer is then given by the sum of the potential and kinetic energies as

\begin{equation}\label{ionEmon}
    E(x) = q \phi_{di} + \frac{1}{2} m_{mon} v_{di}^2
\end{equation}

which can be simplified to

\begin{equation}\label{ionEmon2}
    E(x) = q \phi_{di} + \frac{1}{2} m_{mon} v_0^2 + q \frac{m_{mon}}{m_{di}} (\phi_0 - \phi_{di})
\end{equation}

The energy of the monomer at any given point on its path is given by

\begin{equation}\label{ionEmon3}
    E(x) = q \phi(x) + \frac{1}{2} m_{mon} v_{mon}^2
\end{equation}

This quantity is conserved. Setting Equation \ref{ionEmon2} and Equation \ref{ionEmon3} equal to each other we can see that

\begin{equation}\label{ionEmon4}
    q \phi_{di} + \frac{1}{2} m_{mon} v_0^2 + q \frac{m_{mon}}{m_{di}} (\phi_0 - \phi_{di}) = q \phi(x) + \frac{1}{2} m_{mon} v_{mon}^2
\end{equation}

This yields the velocity of the monomer as a function of the distance from the emission site

\begin{equation}\label{ionVmon}
    v_{mon} = \sqrt{v_0^2 + \frac{2 q }{m_{mon}}(\phi_{di}-\phi(x)) + \frac{2 q}{m_{di}} (\phi_0 - \phi_{di})}
\end{equation}

Evaluating this result at the exit potential $\phi(x) = 0$ gives the velocity of the ion when it reaches the end of the potential field.

\begin{equation}\label{ionVmonf}
    v_{mon,f} = \sqrt{v_0^2 + \frac{2 q }{m_{mon}}\phi_{di} + \frac{2 q}{m_{di}} (\phi_0 - \phi_{di})}
\end{equation}

The kinetic energy of the monomer is given by

\begin{equation}\label{ionKEmon4}
    KE_{mon} = \frac{1}{2} m_{mon} v_0^2 + q(\phi_{di} - \phi(x)) + q\frac{m_{mon}}{m_{di}}(\phi_0 - \phi_{di})
\end{equation}

Evaluating this at the exit potential $\phi(x) = 0$ gives the kinetic energy of the ion when it reaches the end of the potential field.

\begin{equation}\label{ionEmonf}
    KE_{mon,f} = \frac{1}{2} m_{mon} v_0^2 + q\phi_{di} + q\frac{m_{mon}}{m_{di}}(\phi_0 - \phi_{di})
\end{equation}

In the limiting case where the dimer fragments in the field free region the fragmentation potential $\phi_{di} = 0$ which yields the following kinetic energy when the monomer reaches the current collector

\begin{equation}\label{ionEmonFF}
    KE_{mon,FF} = \frac{1}{2} m_{mon} v_0^2 + q\frac{m_{mon}}{m_{di}}\phi_0
\end{equation}

\subsubsection{Trimer Fragmentation}

Now consider a trimer that fragments into a dimer and then again into a monomer after emission. The mass of the trimer is given by $m_{tri}$, the mass of the dimer is given by $m_{di}$, and the mass of the monomer is given by $m_{mon}$. Suppose the trimer first fragments into a dimer at a potential $\phi_{tri}$. After this fragmentation the dimer has been accelerated by a potential $\phi_0 - \phi_{tri}$. The velocity of the dimer at breakup is the same as the velocity of the trimer after being accelerated over this potential, which is given by

\begin{equation}\label{ionVdifromDi}
    v_{di} = \sqrt{v_0^2 + \frac{2 q (\phi_0 - \phi_{tri})}{m_{tri}}}
\end{equation}

The energy of the dimer is then given by the sum of the potential and kinetic energies as

\begin{equation}\label{XionEdi}
    E(x)_{di} = q \phi_{tri} + \frac{1}{2} m_{di} v_{di}^2
\end{equation}

which can be simplified to

\begin{equation}\label{XionEdi2}
    E(x)_{di} = q \phi_{tri} + \frac{1}{2} m_{di} v_0^2 + q \frac{m_{di}}{m_{tri}} (\phi_0 - \phi_{tri})
\end{equation}

The energy of the dimer at any given point on its path is given by

\begin{equation}\label{XionEdi3}
    E(x)_{di} = q \phi(x) + \frac{1}{2} m_{di} v_{di}(x)^2
\end{equation}

The total energy is conserved. Setting Equation \ref{XionEmon2} and Equation \ref{XionEmon3} equal to each other we can see that

\begin{equation}\label{XionEdi4}
    q \phi_{tri} + \frac{1}{2} m_{di} v_0^2 + q \frac{m_{di}}{m_{tri}} (\phi_0 - \phi_{tri}) = q \phi(x) + \frac{1}{2} m_{di} v_{di}(x)^2
\end{equation}

This yields the velocity of the dimer as a function of the distance from the emission site

\begin{equation}\label{XionVdi}
    v_{di} = \sqrt{v_0^2 + \frac{2 q }{m_{di}}(\phi_{tri}-\phi(x)) + \frac{2 q}{m_{tri}} (\phi_0 - \phi_{tri})}
\end{equation}

Assuming that the dimer reaches the detector without fragmenting we can evaluate this result at the exit potential $\phi(x) = 0$ giving the velocity of the dimer when it reaches the end of the potential field.

\begin{equation}\label{XionVdif}
    v_{di,f,unfrag} = \sqrt{v_0^2 + \frac{2 q }{m_{di}}\phi_{tri} + \frac{2 q}{m_{tri}} (\phi_0 - \phi_{tri})}
\end{equation}

The kinetic energy of the unfragmented dimer is then given by

\begin{equation}\label{XionKEdi}
    KE_{di,unfrag}(x) = \frac{1}{2} m_{di} v_0^2 + q (\phi_{tri}-\phi(x)) + \frac{q m_{di}}{m_{tri}} (\phi_0 - \phi_{tri})
\end{equation}

Evaluating this at the exit potential $\phi(x) = 0$ gives the kinetic energy of the unfragmented dimer when it reaches the end of the potential field.

\begin{equation}\label{XionKEdif}
    KE_{di,f,unfrag}(x) = \frac{1}{2} m_{di} v_0^2 + q \phi_{tri} + \frac{q m_{di}}{m_{tri}} (\phi_0 - \phi_{tri})
\end{equation}

Now suppose that the resulting dimer fragments into a monomer at a potential $\phi_{di}$. The resulting monomer has been accelerated across an additional potential given by $\phi_{tri} - \phi_{di}$ resulting in a velocity given by 

\begin{equation}\label{XionVmon}
    v_{mon} = \sqrt{v_0^2 + \frac{2 q }{m_{di}}(\phi_{tri}-\phi_{di}) + \frac{2 q}{m_{tri}} (\phi_0 - \phi_{tri})}
\end{equation}

The energy of the monomer is then given by the sum of the potential and kinetic energies as 

\begin{equation}\label{XionEmon}
    E(x)_{mon} = q \phi_{di} + \frac{1}{2} m_{mon} v_{mon}^2
\end{equation}

which can be simplified to

\begin{equation}\label{XionEmon2}
    E(x)_{mon} = q \phi_{di} + \frac{1}{2} m_{mon} v_0^2 + q \frac{m_{mon}}{m_{di}} (\phi_{tri} - \phi_{di}) + \frac{m_{mon}}{m_{tri}} (\phi_{0} - \phi_{tri})
\end{equation}

The energy of the monomer at any given point on its path is given by

\begin{equation}\label{XionEmon3}
    E(x)_{mon} = q \phi(x) + \frac{1}{2} m_{mon} v_{mon}(x)^2
\end{equation}

Total energy is conserved. Setting Equation \ref{XionEmon2} and Equation \ref{XionEmon3} equal to each other we can see that

\begin{equation}\label{XionEmon4}
    q \phi_{di} + \frac{1}{2} m_{mon} v_0^2 + q \frac{m_{mon}}{m_{di}} (\phi_{tri} - \phi_{di}) + \frac{m_{mon}}{m_{tri}} (\phi_{0} - \phi_{tri}) = q \phi(x) + \frac{1}{2} m_{mon} v_{mon}(x)^2
\end{equation}

This yields the velocity of the monomer as a function of the distance from the emission site

\begin{equation}\label{XionVmonfromtri}
    v_{mon} = \sqrt{v_0^2 + \frac{2 q }{m_{di}}(\phi_{tri}-\phi_{di}) + \frac{2 q}{m_{tri}} (\phi_0 - \phi_{tri}) + \frac{2q}{m_{mon}}(\phi_{di} - \phi(x))}
\end{equation}

Evaluating this result at the exit potential $\phi(x) = 0$ gives the velocity of the ion when it reaches the end of the potential field.

\begin{equation}\label{XionVmonf}
    v_{mon,f} = \sqrt{v_0^2 + \frac{2 q }{m_{di}}(\phi_{tri}-\phi_{di}) + \frac{2 q}{m_{tri}} (\phi_0 - \phi_{tri}) + \frac{2q}{m_{mon}}\phi_{di}}
\end{equation}

The kinetic energy of the monomer is given by

\begin{equation}\label{XionKEmon}
    KE_{mon}(x) = \frac{1}{2} m_{mon} v_0^2 + q\frac{m_{mon}}{m_{di}} (\phi_{tri} - \phi_{di}) + q\frac{m_{mon}}{m_{tri}} (\phi_0 - \phi_{tri}) + q (\phi_{di} - \phi(x))
\end{equation}

Evaluating this at the exit potential $\phi(x) = 0$ gives the kinetic energy of the ion when it reaches the end of the potential field.

\begin{equation}\label{XionKEmonf}
    KE_{mon,f} = \frac{1}{2} m_{mon} v_0^2 + q\frac{m_{mon}}{m_{di}} (\phi_{tri} - \phi_{di}) + q\frac{m_{mon}}{m_{tri}} (\phi_0 - \phi_{tri}) + q \phi_{di}
\end{equation}

The limiting cases of the final kinetic energies for different fragmentation potentials will be investigated in section \ref{sec:RPA}.

\section{Ion Cluster Fragmentation}

This section describes the current models for ionic liquid cluster fragmentation including physics-based rate models as well as some experimental evidence for these models. For a full review of fragmentation experimental methods and previous work see Reference \cite{Miller2019CharacterizationSources}.

\subsection{Arrhenius Rate Model}

Ionic liquid cluster fragmentation in a region with no electric field is considered an activated process, which can be modelled using an Arrhenius rate law given by

\begin{equation}\label{Arrhenius}
    K = A exp(-\frac{E_a}{k T}) = \frac {1}{\tau}
\end{equation}

where A is a constant rate coefficient, $E_a$ is the activation energy, k is Boltzmann's constant, T is the internal temperature of the cluster, and $\tau$ is the mean lifetime of the cluster \cite{Miller2020MeasurementSources}. This rate equation relates the rate at which clusters fragment to the energy required for the process to occur, $E_a$, and the characteristic energy of the cluster, $kT$. The temperature of the cluster takes into account all energy in the cluster including rotational and vibrational degrees of freedom as well as the potential energy from the Coulombic interactions of the molecules. The rate coefficients and activation energies for some ionic liquids including EMI-Im and EMI-FAP have recently been determined experimentally using differential mobility analysis \cite{Miller2019CharacterizationSources,Hogan2009TandemNanodrops}.

Given that cluster fragmentation is assumed to be a constant rate process, the probability of a cluster fragmenting in a given time period is given by

\begin{equation}\label{probFrag}
    p = (1-e^{-\frac{\delta t}{\tau}})
\end{equation}

where $\delta t$ is the time period and $\tau$ is the mean lifetime of the cluster \cite{Petro2019DevelopmentModeling}.

\subsection{Schottky Electric Field Model}\label{sec:Schottky}

In the case where fragmentation is occurring in a region with an electric field the force of the electric field on the molecules in the cluster will affect the fragmentation rate. Previous work has used the Schottky model to modify the Arrhenius rate law to account the for the effect of the electric field \cite{Miller2020MeasurementSources,Petro2019DevelopmentModeling}. This Schottky model is the same model that is used to determine the rate of field assisted evaporation of ions from the liquid meniscus to find the current emitted from electrospray under different conditions. The effect of the electric field on fragmentation is to reduce the effective energy barrier of the process by some energy $E_e$. This is shown in Equation \ref{Schottky1}

\begin{equation}\label{Schottky1}
    K = A exp(-\frac{E_a - E_e}{k T}) = \frac {1}{\tau}
\end{equation}

\begin{figure}
    \centering
    \includegraphics[width=4.5in]{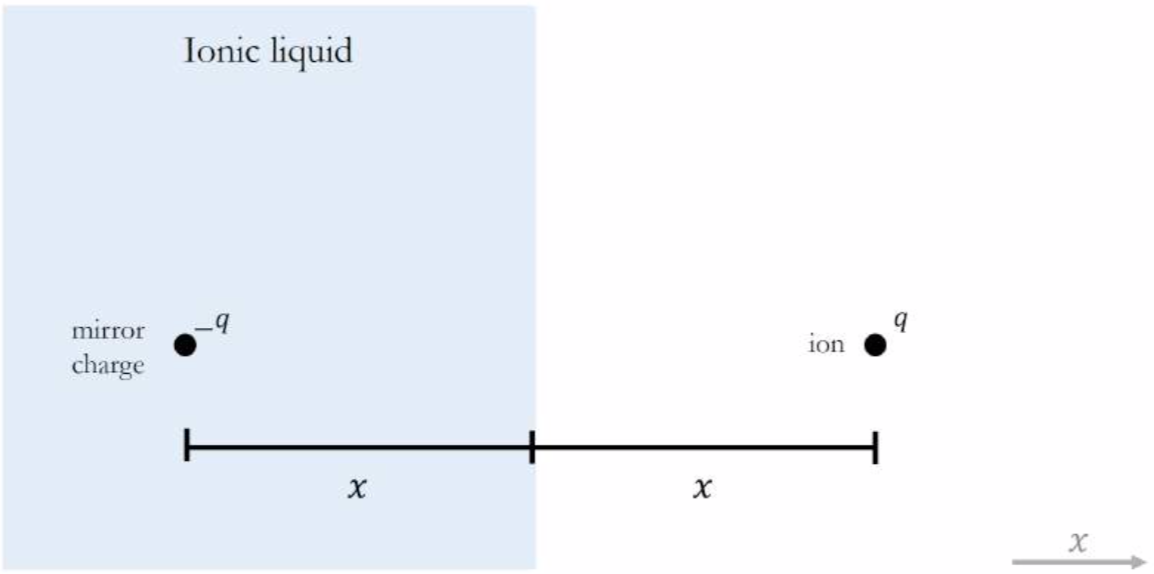}
    \caption{Geometry assumed by the image point model. Image created by Catherine Miller.}
    \label{fig:IPM}
\end{figure}

The reduction in the fragmentation energy barrier, $E_e$, is given by the image point model \cite{Miller2020MeasurementSources,Iribarne1976OnDroplets}. This model assumes that the ion escaping from the cluster is a single ion being removed from a flat, perfectly conducting liquid surface \cite{Martinez-Sanchez2015SessionRegime}. The geometry of this model is shown in Figure \ref{fig:IPM}. There are two forces on the escaping ion, the force of the mirror charge left in the liquid surface, and the force of the electric field. These forces are given by Equations \ref{mirrorChargeF} and \ref{eFieldF} respectively.

\begin{equation}\label{mirrorChargeF}
    F_q = \frac{-q^2}{4 \pi \epsilon_0 (2x)^2}
\end{equation}

\begin{equation}\label{eFieldF}
    F_E = q E
\end{equation}

where q is the charge of the ion and its image, x is the distance of the ion from the liquid surface, and E is the strength of the electric field. The total force on the ion is the sum of these two forces. The work required to remove the ion from the surface is then given by the integration of these forces between x and $\infty$ where the electric field and the force of the mirror charge are assumed to be 0

\begin{equation}\label{WSchottky}
    W = - \int_{\infty}^x \frac{-q^2}{4 \pi \epsilon_0 (2x')^2} + qE dx' = qEx + \frac{q^2}{16 \pi \epsilon_0 x}
\end{equation}

At some point $x_{W_{min}}$ the force of the electric field will balance the force of the image charge. The total work to move the charge to this point is the minimum of the work function, which is given by

\begin{equation}\label{WminSchottky}
    W_{min} = \sqrt{\frac{q^3E}{4 \pi \epsilon_0}}
\end{equation}

\begin{figure}
    \centering
    \includegraphics[width=4in]{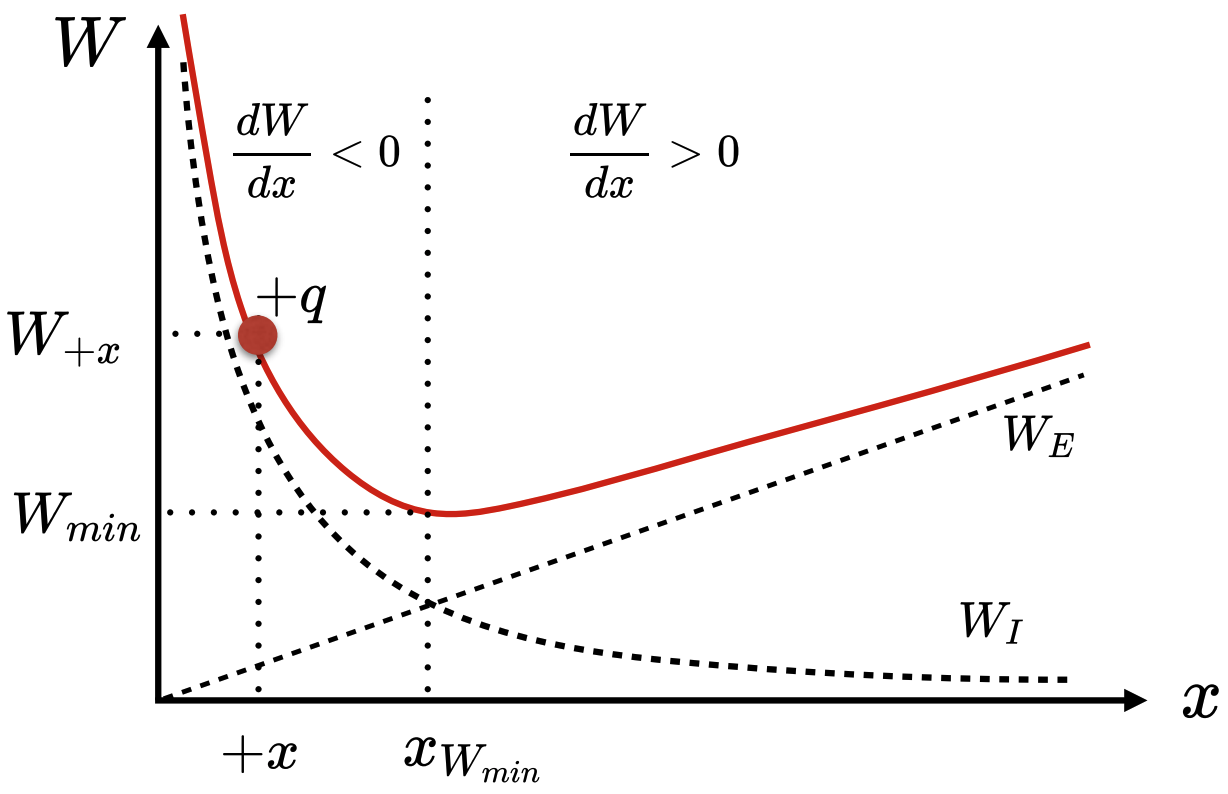}
    \caption{Minimum work reduction of activation energy \cite{Martinez-Sanchez2015SessionRegime}}
    \label{fig:Wmin}
\end{figure}

To escape the flat surface of liquid the ion need only be moved to the x location. After reaching x the force of the electric field is greater than the force of the image charge and the escaping ion will be accelerated away from the liquid surface. The minimum amount of work to remove the ion to this point x is the reduction of the activation energy needed for fragmentation, $E_e$. This is shown in Figure \ref{fig:Wmin}. This can be incorporated into the Arrhenius rate model as given in Equation \ref{Schottky1}

\begin{equation}\label{Schottky2}
    K = A exp(-\frac{1}{k T}(E_a - \sqrt{\frac{q^3E}{4 \pi \epsilon_0}})) = \frac {1}{\tau}
\end{equation}

Previous work has shown that electrohydrodynamic models that use the Schottky model to predict emitted current from electrospray conditions such as voltage, geometry, surface tension, and conductivity yield results on the same order of magnitude as experimental results from similar conditions \cite{Cidoncha2019InformingEvaporation}. Additionally, it has been used to predict fragmentation rates in the acceleration region where the electric field is relatively low and nearly constant \cite{Miller2019CharacterizationSources, Miller2020MeasurementSources}. Miller reported the results of applying the Schottky model with a computational model of the Laplacian electric field to determine the rates of fragmentation in the acceleration region for various ionic liquids at temperatures estimated from experimental results. The Schottky model and the electric field model predicted near constant rates of dissociation in the majority of the acceleration region, which agrees with the constant slope seen in regions of experimental energy distribution curves \cite{Miller2019CharacterizationSources}. Additionally, the predicted rates of fragmentation resulted in total fragmentation percentage results that were similar to those seen in the experimental data. 

More recent efforts to apply the Schottky model to cluster fragmentation, however, have not been successful. Work done by Petro et. al. with a full N-body simulation of an electrospray ion beam based on the previously mentioned electrohydrodynamic (EHD) model determined that fragmentation rates predicted by the Schottky model resulted in nearly all clusters fragmenting immediately after emission \cite{Petro2019DevelopmentModeling}. In contrast, experimental data shows that fragmentation occurs throughout the acceleration region as well as in the field free region. It is suspected that the reason the results of applying the Schottky model in this case differ from the work of Miller is the accuracy of the electric field model. The electric field model from EHD computation in the work of Petro et. al. takes into account the small radius of curvature of the liquid meniscus, which is not accounted for in Miller's model. Thus, it results in a higher electric field at the meniscus surface. The electric field strength at the meniscus surface from the EHD model is approximately $2\times10^9 V/m$ as compared to the $2\times10^7 V/m$ used by Miller \cite{Miller2019CharacterizationSources}. This higher electric field combined with the Schottky model results in higher fragmentation rates near the emission site in the N-Body simulation. This indicates that the Schottky model is not fully appropriate for determining mean lifetimes of ionic liquid clusters at the high electric field strengths seen near the emission site of electrospray emitters. 

\subsection{Curvature and Surface Energy Effects}

The Schottky model assumes that an ion is being extracted from a flat, perfectly conducting liquid surface. This geometry may be adequate for electric field enhanced ion emission from the liquid meniscus, however, the geometry of cluster fragmentation is much different. Previous work by de la Mora et. al. produced an estimation of the solvation energy of removing an ion from a charged spherical droplet \cite{Labowsky2000AEnergy}. This model accounts for the electrostatic interaction of the ion and the droplet as well as the surface energy changes during the evaporation process. This model is given in Equation \ref{CurvatureEffects}. 

\begin{equation}\label{CurvatureEffects}
    \Delta G = (\frac{e^2}{8 \pi \epsilon_0 R_i} + 4 \pi \gamma R_i^2) - (\frac{e^2[2 F(z) + 1]}{8 \pi \epsilon_0 R'} + 4 \pi \gamma \frac{2 R_i^3}{3 R'})
\end{equation}

$\Delta G$ is the total change in energy of the ion and the droplet during the evaporation process, $R_i$ is the radius of the escaping ion, $\gamma$ is the surface tension of the liquid, $F(z)$ is a constant that accounts for the net charge in the drop, and $R'$ is the radius of the droplet left over after the ion escapes. The first two terms on the right-hand side represent the solvation energy for removing an ion from a charged surface as derived previously by Born \cite{Atkins1982TheSolvation}. Of these, the first term is the change in energy due to the electrostatic interaction with the droplet and the second term is the change in energy due to the surface energy of the droplet. The third term is a correction accounting for the net charge in the original droplet, and the fourth term is a correction accounting for the surface energy of the escaping ion after leaving the droplet.

\subsection{Experimental Data}

This section describes some of the previous experimental work done to determine the fragmentation behavior of ionic liquid clusters under different conditions. There are three common experimental ways of characterizing fragmentation, retarding potential analysis, differential mobility analysis, and collision induced dissociation. Details of how each of these experimental methods are used are given in the following sections.

\subsubsection{Retarding Potential Analysis}\label{sec:RPAexp}

Previous work with ionic liquid electrospray RPA data has shown that for EMI-BF$_4$, approximately half of the beam is monoenergetic \cite{Petro2020CharacterizationEmitters,Lozano2006EnergySource}. The components of the beam with lower energies correspond to products of fragmentation. Similar results have been obtained for BMII electrospray sources \cite{Fedkiw2009DevelopmentApplications}. Previous work by Miller has shown that the percentage of the beam composed of products of fragmentation depends heavily on the temperature of the ionic liquid and the voltage applied to the source \cite{Miller2015OnSources}. However, the exact dependence is not well characterized due to the limitations of partial beam characterization methods \cite{Miller2015OnSources}. Miller also determined fragmentation rates of EMI-BF$_4$, EMI-Im, and EMI-FAP dimers in the field free region of electrospray emitters \cite{Miller2019CharacterizationSources}. Additionally, evidence suggests that the complexity of the ionic liquid molecules affects fragmentation rates \cite{Miller2019CharacterizationSources,Miller2020MeasurementSources,Coles2012InvestigatingBeams}. Ionic liquids with more complex anions such as EMI-FAP and EMI-Im show lower fragmentation rates than ionic liquids with less complex anions such as EMI-BF$_4$ \cite{Miller2020MeasurementSources}. While this is suggested by the experimental work of Miller, the fragmentation rates were not compared between clusters having the same energy content as it is difficult to determine the exact energy content of electrosprayed ions \cite{Miller2019CharacterizationSources}.

\subsubsection{Differential Mobility Analysis}\label{sec:DMAexp}

Work by Hogan and de la Mora has been done using differential mobility analysis (DMA) to characterize fragmentation of room temperature ionic liquid clusters \cite{Hogan2009TandemNanodrops,Hogan2010Ion-pairClusters}. In DMA experiments ion clusters are emitted from capillary electrospray sources using ionic liquids diluted in solvents such as acetonitrile. Once droplets are emitted from the source the solvent evaporates, leaving behind an ion cluster. DMA uses the different velocities of clusters of different size to selectively analyze cluster behavior by mass. The results of these studies have resulted in calculations of the activation energies and rate coefficients of fragmentation for dimers of some ionic liquids \cite{Miller2019CharacterizationSources,Miller2020MeasurementSources}. While useful for determining fragmentation rates of clusters in the field free region, DMA does not provide fragmentation behavior for ion clusters under the effect of an electric field. Additionally, the clusters resulting from the evaporation of the solvent are assumed to be at the same temperature as the liquid mixture used in the electrospray source, usually somewhere between 298 K and 320 K. Ion clusters emitted from electrospray sources are believed to be at a much higher temperature between 500 K and 1500 K due to the addition of energy to the clusters during evaporation \cite{Miller2020MeasurementSources, Lozano2006EnergySource}. 

\subsubsection{Collision Induced Dissociation}\label{sec:CIDexp}

Collision induced dissociation (CID) in an experimental technique that determines fragmentation rates as a function of cluster energy \cite{Armentrout2007StatisticalThresholds}. In CID experiments ion clusters are obtained by electrospraying ionic liquids mixed with solvents from capillary sources. The solvent evaporates and the ion clusters are accelerated into a gas bath that is used to equilibrate the temperature of the ions \cite{Carpenter2017HowIons}. The ions dissociate during collisions with the gas. The mass of the products of these collisions is measured to determine the rate of fragmentation and how it is related to the energy of the clusters \cite{Carpenter2017HowIons}. Prince et al. performed CID experiments in 2017 and 2019, which were compared with molecular dynamics simulations results \cite{Prince2017ADissociation,Prince2019SolvatedPropellants}. Improvements to the experiment were made to better control the temperature of the ions and determine the fragmentation rates of the clusters. Temperatures between 422 K and 431 K were examined. More recently, Roy et al. used CID to measure dissociation rates of ionic liquid clusters such as EMI-BF$_4$, BMI-BF$_4$, HMI-BF$_4$, and OMI-BF$_4$ \cite{Roy2020Gas-PhaseClusters}. These results agreed with quantum mechanical calculations. However, similar to DMA experiments, CID experiments can not be used to characterize fragmentation rates of ionic liquid clusters under the effect of an electric field and have not been used to characterize fragmentation in clusters at the temperatures expected with electrospray sources.

\subsection{Computational Modelling}

Molecular dynamics (MD) modelling has been performed to characterize electrospray emission. These results can provide an understanding of the conditions of the emitted clusters in ion beams, which in turn can provide information on the conditions under which clusters will fragment. In 2008 Takahashi and Lozano developed simulations of small droplets of EMI-BF$_4$ under the effect of an electric field to model electrospray emission \cite{Takahashi2008ComputationalThrusters}. Results demonstrated the effect of increasing voltage on the current emitted. In 2009 Takahashi and Lozano produced results from smaller, 27 molecule clusters on a tungsten surface \cite{Takahashi2009AtomisticThrusters}. The emission from the tungsten surface required a larger electric field due to the charge on the tungsten \cite{Takahashi2009AtomisticThrusters}. Both of these works used the AMBER force field with parameters from Andrade et al. in LAMMPS \cite{DeAndrade2002ComputationalValidation,Plimpton1995FastDynamics}. In 2012 Borner et al. developed a coarse-grained model of EMI-BF$_4$ based on the AMBER force field, which was demonstrated to match classical full field MD simulations for colloid emission from a capillary \cite{Borner2012ModelingModel}. Emitted species matched those of previous simulations including the direct emission of some neutral species and a decrease of the number of solvated species with increasing applied electric field. Geometry parameters of some trimers were determined under the effect of an electric field, revealing the change in direction of the molecules in the cluster as a result of the electric field force.

In 2012 and 2013 Coles and Lozano used LAMMPS to simulate emission from clusters of 4913 ion pairs \cite{Coles2012InvestigatingBeams, Coles2013InvestigatingBeams}. The work in 2012 used the Liu force field while the work from 2013 used the force field of Canogia, Padua, and Lopes \cite{Liu2004ALiquids,CanongiaLopes2004ModelingField}. The larger cluster size was enabled by the use of NAMD \cite{Phillips2005ScalableNAMD}. The width of the simulated droplet was larger than the longest significant distance for Coulombic interactions, which made it more accurate to electrospray dynamics \cite{Coles2012InvestigatingBeams}. Simulation procedures were similar to those used by Takahashi \cite{Takahashi2008ComputationalThrusters,Takahashi2009AtomisticThrusters}. Results showed that the energy of clusters may be significantly increased due to stretching of the molecules during emission. Energy distributions of emitted clusters were generally Maxwellian with temperatures between 300 and 2000 K, however, methods of calculating temperature for individual ions are not entirely trustworthy. These simulations also showed some doubly charge ions and neutrals emitted directly from the ionic liquid droplet. 

In 2015 Prince et al. performed simulations of 125 ion-pair droplets of EMI-Im in LAMMPS \cite{Prince2015MolecularNanodrops}. Parameters for the anions were taken from Canogia, Padua, and Lopes (CLP) \cite{CanongiaLopes2004ModelingField}. Dihedrals were modelled using the CHARMM potential. The goal of the droplet simulation was to model cone jet emission from a 4 nm diameter cone jet. Data on the angular and species distribution of emission were recording with results similar to those of previous experimental work \cite{Perez-Martinez2012VisualizationApplications,Perez-Martinez2015IonMicrotips}. Mostly monomers and dimers were emitted with significant fragmentation occurring near the emission site on the droplet. The energy evolution of clusters was examined, and the results did not match those from the work of Coles \cite{Coles2012InvestigatingBeams}. It is unclear if the difference was due to the different ionic liquids being studied or the methods employed in simulating them.

In 2017 Mehta and Levin developed course-grained models similar to Borner et al. to simulate electrospray emission in the droplet mode from a capillary \cite{Mehta2018MolecularEMIM-BF4}. They compared the emission characteristics of EAN and EMI-BF$_4$, noting that the tighter bonds between EAN molecules resulted in emission beginning at higher electric fields. Their results for the beam composition were similar to those found in previous simulations and experimental work \cite{Borner2012ModelingModel,Lozano2003StudiesThrusters}

MD has also been used to characterize fragmentation rates of ionic liquid clusters. In 2012 Coles and Lozano performed simulations of EMI-BF$_4$ dimers using LAMMPS with the Liu force field. They noted that significant distances can build between molecules in a cluster without fragmentation, increasing the difficulty of defining the distance at which fragmentation occurs. Additionally, they reported the dependence of the fragmentation on electric field and cluster temperature, with both serving to increase the rate of fragmentation. This work was also the first to use the atom position and velocity coordinates from emission simulations to initiate fragmentation simulations. In 2013 Coles performed simulations of EMI-BF$_4$, EMI-FSI, EMI-Im, EMI-FAP, EMI-Br, and EMI-Cl using the CLP force field \cite{Coles2013InvestigatingBeams}. They compared fragmentation rates for each ionic liquid with the same total amount of excess energy after emission. They found that the clusters with anions with more degrees of freedom such as EMI-Im and EMI-FAP fragmented slower than the clusters with anions with fewer degrees of freedom. They hypothesized that this result was due to the larger number of vibrational degrees of freedom within each molecule to distribute the energy of the cluster. 

In 2017 and 2019 Prince et al. performed simulations of EMI-Im clusters of various sizes using the same force field as used for the emission simulations in the same work \cite{Prince2012IonicSystems,Prince2015MolecularNanodrops}. They demonstrated agreement between quantum mechanical models of cluster energies with MD simulations of cluster energies. They determined fragmentation rates under some temperature and electric field conditions using exponential fit models, however, the number of samples and fragmentation percentage of the samples was not high enough to get statistically accurate results. Clusters at 298 K were simulated with electric fields between $5\times10^8$ V/m and $2.5\times10^9$ V/m. They briefly examined the geometry of the clusters during fragmentation under the effect of an electric field, noting the stretching of the cluster that occurred. They also determined that the addition of the electric field changed the energy evolution during the fragmentation process. They investigated the effects of cluster size on fragmentation rate and the way in which the cluster fragmented. They noted that larger clusters fragmented faster under the influence of an electric field and that large clusters were most likely to evaporate smaller charged clusters instead of neutrals. They also reported fragmentation rates for different size neutral clusters under the influence of an electric field, noting that neutral clusters do not fragment as quickly as similarly sized charged clusters under the same conditions. 

In the 2017 and 2019 works Prince also calculated fragmentation rates using quantum mechanical models and the Rice-Ramsperger-Kassel-Marcus (RRKM) model. RRKM states that some transition state must be reached in order for fragmentation to occur. When the energy of the cluster is larger than the energy needed to reach this transition state then the fragmentation rate will be nonzero. In 2017 and 2019 Prince calculated the binding energies of EMI-Im clusters under different temperature conditions \cite{Prince2017ADissociation,Prince2019SolvatedPropellants}. These results agreed with the CID experiments performed under similar conditions. These MD simulations followed a process similar to that used by Coles with the notable addition of an annealing step \cite{Coles2012InvestigatingBeams,Coles2013InvestigatingBeams}. In 2020 Roy et al. presented fragmentation analysis of EMI-BF4, BMI-BF4, HMI-BF4, and OMI-BF4 using the AMBER force field \cite{Roy2020Gas-PhaseClusters}. The fragmentation rates agreed with those collected from CID experiments. They also reported on the existence of various molecule configurations for each type of liquid and the effect on the enthalpies and free energies of the clusters.

\section{Experimental Methods}


\subsection{Retarding Potential Analysis} \label{sec:RPA}

\begin{figure}
    \centering
    \includegraphics[width=5.5in]{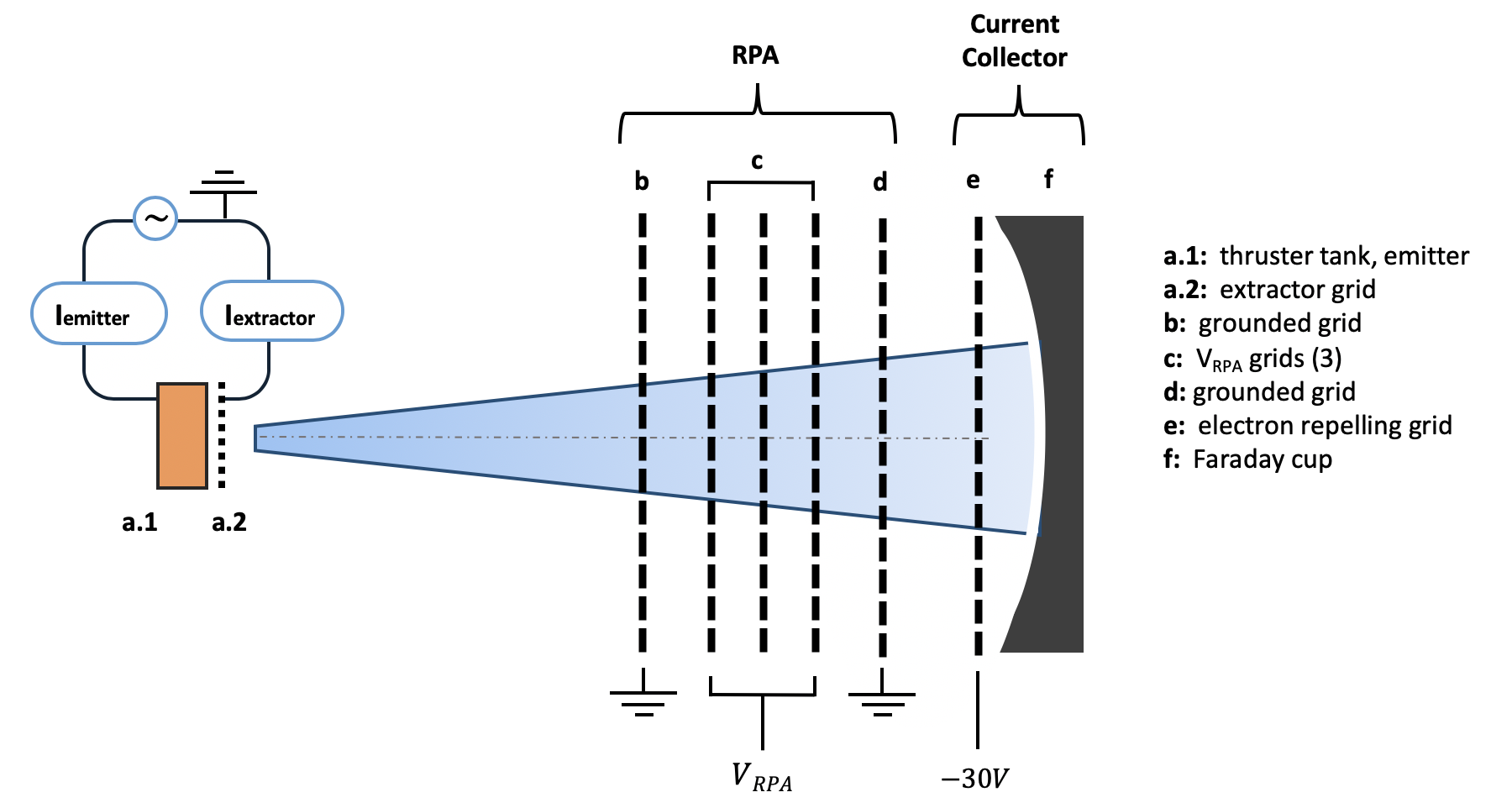}
    \caption{Retarding Potential Analyzer (RPA) Experimental Apparatus. Image created by Amelia Bruno.}
    \label{fig:RPAdiagram}
\end{figure}

A retarding potential analyzer (RPA) is an experimental apparatus used to determine the energy distribution of an ion beam. The RPA consists of several meshed grids in front of a Faraday cup or other current collector as shown in Figure \ref{fig:RPAdiagram}. The ion beam is aimed at the RPA while firing at a constant voltage $\phi_0$. A voltage $\phi_{st}$ called the stopping potential is applied between the high voltage grids and the grounded grids. The grid closest to the current collector is biased to -30 volts to repel any secondary electrons generated from the interaction of the beam with the grids or chamber. $\phi_{st}$ is controlled by a triangular waveform, which alternates between a voltage approximately 200 V below 0 V, and 200 V above the firing voltage of the ion source. Ions with kinetic energies less than $q \phi_{st}$ will be retarded by the high voltage grids and will not reach the current collector, while ions with kinetic energies greater than $q \phi_{st}$ will reach the current collector. Measuring the collected current as a function of the stopping potential thus yields the energy distribution of the ion beam. The stopping potential for different ions and ion clusters are given in the next sections.

\subsubsection{Monoenergetic Ions}

Ions that do not fragment before reaching the current collector are called monoenergetic ions. The kinetic energy of a monoenergetic ion is given by \ref{ionKEexit}. These ions are accelerated by the full potential of the accelerating potential field. Assuming the initial velocity of the ions is negligible the stopping potential is given by the firing voltage of the thruster.

\begin{equation}\label{monoenergeticStoppingPotential}
    \phi_{st} = \phi_0
\end{equation}

\subsubsection{Fragmented Dimers}

Ion clusters that fragment before reaching the current collector have energies lower than the firing potential of the ion source. Dimers can fragment once before reaching the current collector. The kinetic energy of the monomer resulting from the fragmentation of an emitted dimer is given by Equation \ref{ionEmonf}. Assuming the initial ion velocity is negligible the stopping potential for the monomer resulting from the fragmentation of a dimer can be written as

\begin{equation}\label{dimerStoppingPotentialAccelRegion}
    \phi_{st} = \phi_{di} + \frac{m_{mon}}{m_{di}} (\phi_{0} - \phi_{di})
\end{equation}

\subsubsection{Fragmented Trimers}

Trimers can fragment twice before reaching the current collector, once into dimers, and again into monomers. The kinetic energy of a dimer resulting from one fragmentation of a trimer is given by Equation \ref{XionKEdif}. Assuming the initial ion velocity is negligible the stopping potential for a dimer resulting from one fragmentation of a trimer is given by

\begin{equation}\label{trimerStoppingPotentialDimerOnly}
    \phi_{st} = \phi_{tri} + \frac{m_{di}}{m_{tri}} (\phi_0 - \phi_{tri})
\end{equation}

The kinetic energy of a monomer that results from two fragmentations of a trimer is given by Equation \ref{XionKEmonf}. Assuming the initial ion velocity is negligible the stopping potential for a monomer resulting from two fragmentations of a trimer is given by

\begin{equation}\label{trimerStoppingPotentialFull}
    \phi_{st} = \phi_{di} + \frac{m_{mon}}{m_{di}} (\phi_{tri} - \phi_{di}) + \frac{m_{mon}}{m_{tri}} (\phi_0 - \phi_{tri})
\end{equation}

\subsubsection{Idealized RPA}

This section shows some examples of RPA curves for different limiting cases of the previous equations for stopping potential as a function of fragmentation location. The data presented is the normalized current collected at the detector plotted against the normalized applied retarding voltage. When the retarding voltage is 0 all of the current emitted from the beam reaches the detector. When the retarding voltage is greater than or equal to the firing voltage of the ion source all of the beam will be retarded by the grids and no current will be collected. 

\begin{figure}
    \centering
    \includegraphics[width=0.7\linewidth]{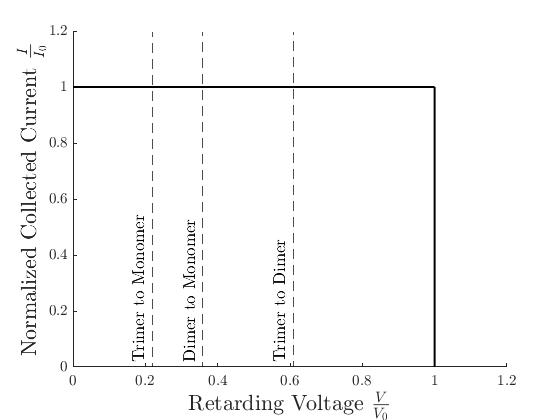}
    \caption{Monoenergetic RPA Example.}
    \label{fig:EXAMPLE_MonoRPA}
\end{figure}

\begin{figure}
    \centering
    \includegraphics[width=0.7\linewidth]{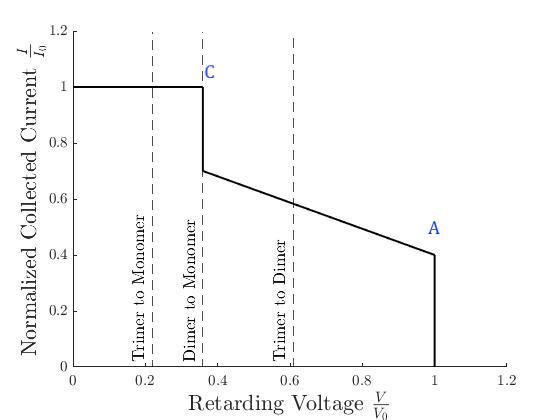}
    \caption{Monomer and Dimer RPA Example.}
    \label{fig:EXAMPLE_SimpleRPA}
\end{figure}

Figure \ref{fig:EXAMPLE_MonoRPA} shows an idealized RPA consisting only of monoenergetic ions. These monoenergetic ions appear at the normalized retarding potential because they have not lost any of the acceleration energy due to fragmentation. Figure \ref{fig:EXAMPLE_SimpleRPA} shows the results of an idealized RPA with only monomers and dimers in the beam. The vertical step at point A shows the monoenergetic portion of the beam composed of monomers emitted directly and dimers that did not fragment before reaching the current collector. The slope between points A and C shows monomers with a spread of kinetic energies resulting from fragmentation of dimers in the acceleration region. The stopping potential for these cases is given by the full Equation \ref{dimerStoppingPotentialAccelRegion}. The slope here is constant, which corresponds to fragmentation occurring at the same rate throughout the acceleration region \cite{Miller2020MeasurementSources}. The step at point C shows the portion of the beam composed of monomers resulting from the fragmentation of dimers in the field free region. These monomers all have the same energy because their parent dimer clusters were accelerated by the full potential of the electric field before fragmenting. The stopping potential of these monomers is the limiting case of Equation \ref{dimerStoppingPotentialAccelRegion} when the fragmentation potential of the dimer $\phi_{di}$ is taken to be 0. This is given by Equation \ref{dimerStoppingPotentialFF}.

\begin{equation}\label{dimerStoppingPotentialFF}
    \phi_{st, di to mon FF} = \frac{m_{mon}}{m_{di}} \phi_{0}
\end{equation}

\begin{figure}
    \centering
    \includegraphics[width=5in]{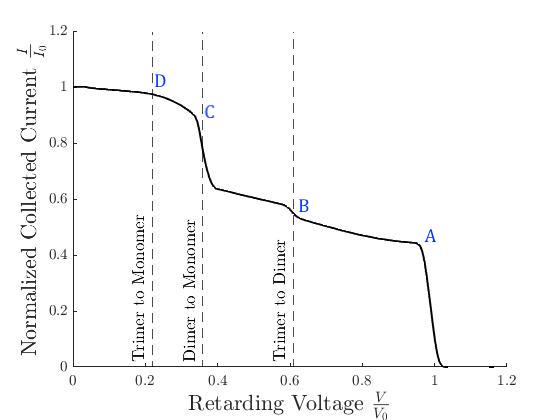}
    \caption{Experimental RPA Example.}
    \label{fig:EXAMPLE_RealRPA}
\end{figure}

Figure \ref{fig:EXAMPLE_RealRPA} shows an example of an experimental RPA curve. Again, the vertical step at point A shows the monoenergetic portion of the beam composed of monomers emitted directly and dimers and trimers that did not fragment before reaching the current collector. The step at point B represents the portion of the beam consisting of trimers that fragmented into dimers in the field free region and reached the detector without fragmenting into monomers. This corresponds to the limiting case of Equation \ref{trimerStoppingPotentialDimerOnly} when $\phi_{tri}, v_0 = 0$

\begin{equation}\label{TtoDFF}
    \phi_{st,tri to di FF} =  \frac{m_{di}}{m_{tri}} \phi_0
\end{equation}

Again, the step at point C represents emitted dimers that fragmented into monomers in the field free region as given by Equation \ref{dimerStoppingPotentialFF}. A step at point D would represent the portion of the beam composed of monomers resulting from two fragmentations of a trimer both occurring in the field free region. This corresponds to the limiting case of Equation \ref{trimerStoppingPotentialFull} where $\phi_{di} = \phi_{tri} = 0$

\begin{equation}\label{TtoMFF}
    \phi_{st,tri to mon FF} =  \frac{m_{mon}}{m_{tri}} \phi_0
\end{equation}

As discussed previously monomers resulting from fragmentation of emitted dimers in the acceleration region contribute to the collected current between points A and C with the stopping potential given by Equation \ref{dimerStoppingPotentialAccelRegion}. Trimers that fragment to dimers in the acceleration region and reach the detector as dimers contribute to the collected current between points A and B. Their stopping potential is given by the full expression of Equation \ref{trimerStoppingPotentialDimerOnly}. Trimers that fragment into dimers and then into monomers in the acceleration region contribute to the collected current between points A and C. Their potential is given by the full expression of Equation \ref{trimerStoppingPotentialFull}. Trimers that fragment into dimers in the acceleration region and then into monomers in the field free region contribute to the collected current between points C and D. Their stopping potential is given by the limiting case of Equation \ref{trimerStoppingPotentialFull} where $\phi_{di} = 0$

\begin{equation}\label{TtoDA,toMF}
    \phi_{st,tri to di A, di to mon FF} = \phi_{di} + \frac{m_{mon}}{m_{di}} \phi_{tri} + \frac{m_{mon}}{m_{tri}} (\phi_0 - \phi_{tri})
\end{equation}

\section{Computational Methods}

\subsection{Molecular Dynamics}\label{sec:MD}

Molecular dynamics simulations model the interactions of atoms and molecules to determine their response to external conditions. Intra-molecular and inter-molecular forces are simplified to represent the underlying physics of the interactions while being able to quickly calculate the resulting force and motion of each atom. Harmonic terms are used to represent the angles and bonds between atoms while cosine series are used to represent dihedrals. The Coulomb force between atoms is modeled by placing partial charges on each atom. The van der Waals force is modeled with the Lennard-Jones approximation. 

The dominant terms for each of these forces are combined into a parametrization known as a force field. Force field parameters are found by fitting the approximations to the physical forces to quantum mechanical calculations and experimental data for properties such as heat of vaporization, density, and heat capacity. Most force fields are based on two force fields OPLS-AA and AMBER, which contain force field parametrizations for the basic building blocks of many organic molecules \cite{Cornell1995AMolecules,Jorgensen1996DevelopmentLiquids}. Force fields such as the force field from Andrade et al. have been developed specifically for use with ionic liquids \cite{DeAndrade2002ComputationalValidation}. However, this field did not account for asymmetries due to the imidazolium ring. More recently, force fields from Liu et al. and Canogia Lopes and Padua have been improved to include asymmetries from the imidazolium ring and to include parameters for a larger range of ionic liquids \cite{CanongiaLopes2004ModelingField,Chaban2015SystematicLiquids,Liu2004ALiquids}. In particular the CLP field has parameterizations for the common ionic liquids such as EMI-FAP, EMI-Im, and EMI-BF$_4$ discussed previously.

\subsection{Approximate Bayesian Computation}\label{sec:ABClitRev}

Bayes rule gives the probability of a hypothesis H being true given that there is some evidence E. For example, we might want to determine the probability that the temperature of an ionic liquid cluster is 1000 K given the evidence provided by an experimental RPA curve.

\begin{equation}
    P(H|E) = \frac{P(E|H) P(H)}{P(E)}
\end{equation}

The term P(H) is called the prior, which describes the probability distribution for hypothesis H that is known before accounting for evidence E. P(H|E) is the posterior probability, which describes the probability of hypothesis H that is known after accounting for evidence E. P(E|H), or the likelihood, is the probability of obtaining evidence E given that the hypothesis H is true. P(E) is the marginal likelihood. 

Bayesian inference is a method that uses Bayes rule to update the probability of a given hypothesis using some input data as the evidence. Various sampling methods are employed to sample the different distributions to obtain the likelihood function and determine the posterior distribution. However, in many circumstances it is difficult to evaluate the likelihood function. In these cases, approximate Bayesian computation (ABC) can be used to determine the posterior given the evidence E. ABC is a method that uses the basic principle of Bayesian inference without needing to evaluate the likelihood function. It has been used extensively for studying genetic development and categorization of species \cite{Sunnaker2013ApproximateComputation,Csillery2010ApproximatePractice,Beaumont2002ApproximateGenetics, Beaumont2009AdaptiveComputation}. 

In ABC the posterior distribution for a parameter is estimated from the results of repeated simulations. For example, consider a simulation S with the input parameters $\theta$ that reproduces experimental data Z. The result of this simulation is given by $y = S(y|\theta)$. The goal is to determine the value of $\theta$ that produced the experimental data Z. Let $\pi(\theta)$ be the prior distribution of the parameters that is built from existing knowledge about the parameter values. Now produce many samples $\theta' \sim \pi(\theta)$, and for each sample simulate $x \sim S(x|\theta')$. If you accept sampled values $\theta' \sim \pi(\theta)$ into the posterior distribution when x = Z, then the posterior distribution will give you an exact representation of the distribution of the parameters $\theta$ conditioned on the experimental data Z. In practice, finding $\theta'$ for which x=y will be impossible. Instead, a sample should be accepted if the distance between x and Z is below some predetermined threshold $\delta$. Varying the value of $\delta$ will result in different uncertainties about the posterior value of $\theta$. In the limiting case where $\delta$ = 0 we will be able to determine $\theta$ exactly.

\chapter{Cluster Fragmentation Simulation}

This section details the methodology and results of molecular dynamics simulations performed to determine the behavior of electrosprayed ionic liquid clusters during fragmentation\footnote{Thank you to Kyle Sonandres and David Hernandez for simulating the EMI-Im, EMI-FAP, and EMI-BF$_4$ trimers}. As explained in section \ref{sec:MD}, MD simulations solve for the forces between the components of the cluster to determine the dynamic behavior of molecules in the cluster. MD simulations were performed on different ionic liquid clusters under different energy and electric field conditions to determine their fragmentation rates.

\section{Methodology}

This section explains the details of the MD simulation process. The focus of this work was repeated simulation of different ionic liquid cluster types. The effect of the initial energy as well as the applied electric field strength on fragmentation behavior were investigated by varying these conditions. The initial cluster temperatures simulated ranged between 300 K and 3000 K as previous work indicates that the temperature of clusters emitted from electrospray likely falls in this range \cite{Miller2019CharacterizationSources,Miller2020MeasurementSources,Coles2012InvestigatingBeams}. Electric field strengths between $5\times 10^{5} $ V/m and $1\times 10^{10} $ V/m were simulated. The high end of this range is approximately an order of magnitude greater than the electric field experienced by clusters immediately after emission from an electrospray source. The lower bound is approximately the electric field strength experienced by ion clusters after they have passed through the extractor grid while they are traveling through the final 5\% of the potential field. Space charge results in the electric field spreading further past the extractor at strengths of approximately $1\times 10^{4} $ V/m, however, simulating fragmentation at electric fields of this magnitude was beyond the capabilities of the computational resources used. More on this limitation is included in section \ref{sec:MDsimProc}. 10,000 samples of each type of cluster were simulated under each set of energy and electric field conditions. The results from each set of conditions were averaged over all of the samples. More details on the data post processing are given in sections \ref{sec:postLifetime} and \ref{sec:postGeometry}.

\subsection{Molecular Dynamics Cluster Simulation Procedure}\label{sec:MDsimProc}

This section details the process of simulating a single ionic liquid cluster. This process is based on previous work by Coles and Prince et al. \cite{Coles2012InvestigatingBeams,Coles2013InvestigatingBeams, Prince2017ADissociation,Prince2019SolvatedPropellants}. All MD simulations used LAMMPS, an open-source MD software that has been used previously for ionic liquid simulations \cite{Coles2012InvestigatingBeams,Coles2013InvestigatingBeams,Takahashi2008ComputationalThrusters,Takahashi2009AtomisticThrusters,Plimpton1995FastDynamics,Mehta2018MolecularEMIM-BF4}. The simulations use the Canongia Lopes and Padua force field as described in section \ref{sec:MD} \cite{CanongiaLopes2004ModelingField}. Lennard-Jones forces were cutoff at 10 
$\mathring{\mathrm{A}}$ and Coulomb forces were cut off at 100 $\mathring{\mathrm{A}}$. The timestep for the simulation was 0.5 fs. The procedure for simulating a single cluster was as follows:

\begin{enumerate}
    \item Determine coordinates of atoms in the cluster from stable emitted clusters. This data was taken from emission simulations performed by Coles \cite{Coles2012InvestigatingBeams,Coles2013InvestigatingBeams}
    \item Randomize the velocities of each atom of the cluster at 298 K.
    \item Apply a Nose-Hoover thermostat to equilibrate the temperature of the cluster to the desired temperature \cite{Evans1985TheThermostat}. The damping parameter of the thermostat was 1000 fs.
    \item Apply a constant electric field strength until the separation between the molecules of the cluster reaches the fragmentation threshold $\delta_{f(sim)}$. More about $\delta_{f(sim)}$ is described in section \ref{sec:FragDef}
\end{enumerate}

\begin{figure}
    \centering
    \includegraphics[width=6.25in]{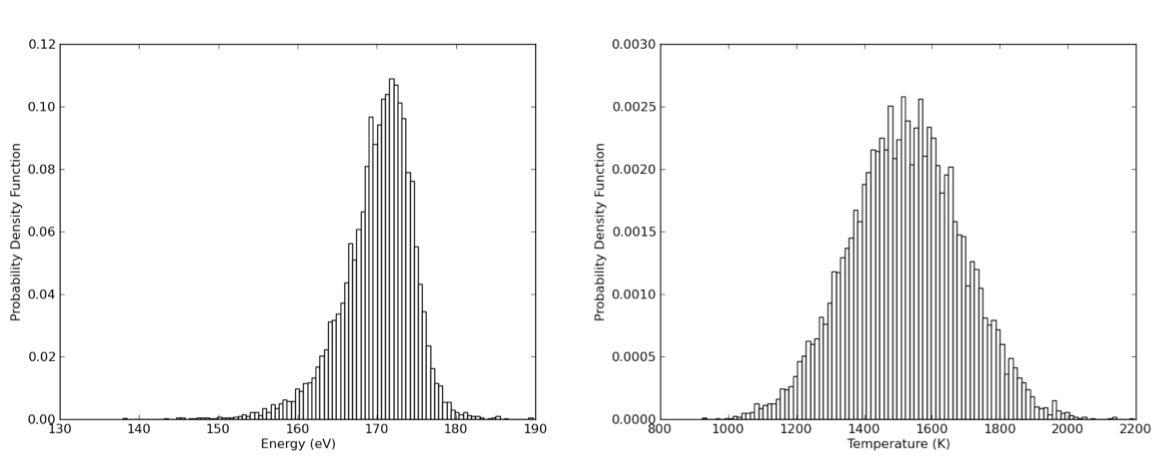}
    \caption{Comparison of energy and temperature distributions for 10,000 samples of EMI-BF$_4$ dimers at 1500 K.}
    \label{fig:CompareTempEnergy}
\end{figure}

The goal of the thermostat in the second step is to set the internal energy of the cluster. Unfortunately, current MD methods do not provide a simple way to set internal energy, so the temperature of the cluster was used instead. However, results show that equilibration to a selected temperature results in samples with a narrow band of internal energies. Figure \ref{fig:CompareTempEnergy} shows an example of the temperature and energy distributions for EMI-BF$_4$ dimers simulated at 1500 K. Tables of the temperature and energy distributions for each set of simulated conditions is included in Appendix \ref{app:appa}. The standard deviation in initial internal energy for each group of samples was much smaller than the difference in internal energies between each simulated temperature. Additionally, efforts to reduce the temperature variation in the samples in post-processing produced a negligible change in the post-processing statistics. Thus, the spread of sample temperatures shown in Figure \ref{fig:CompareTempEnergy} was deemed acceptable. Unless otherwise noted, 10,000 samples of each cluster type were simulated at each chosen set of temperature and electric field conditions.

\subsection{Defining Fragmentation}\label{sec:FragDef}

\begin{figure}
    \centering
    \includegraphics[width=4in]{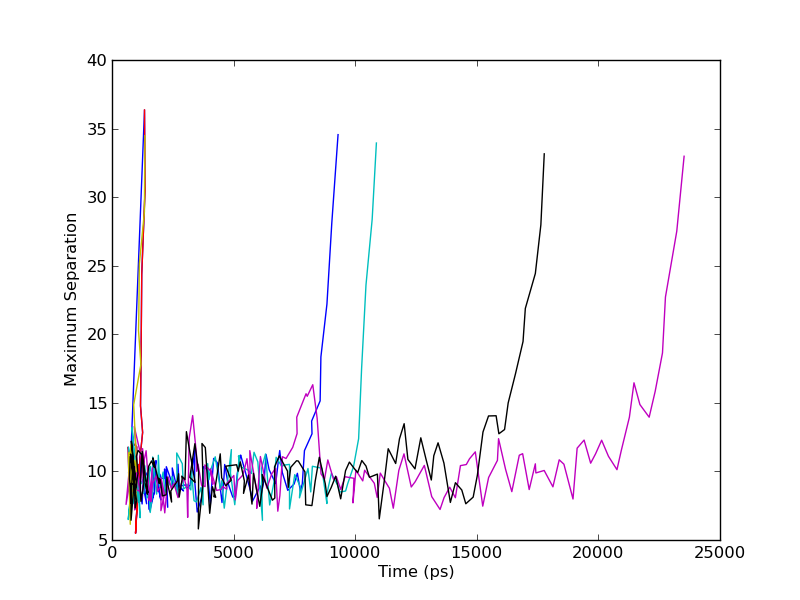}
    \caption{Maximum separation of molecules in EMI-BF$_4$ dimer clusters at 1000K before fragmentation.}
    \label{fig:recombinationTrajectories}
\end{figure}

Ionic liquid clusters simulated with electric fields often develop large separations between the molecules in the cluster without fragmenting \cite{Coles2012InvestigatingBeams}. Figure \ref{fig:recombinationTrajectories} shows an example of the maximum separation distances within an EMI-BF$_4$ dimer cluster during the simulation process for several dimer samples. The arches in this figure occur when a cluster stretches out but does not fragment. In order to determine when fragmentation occurs $\delta_{f(sim)}$ must be larger than the largest maximum separation that could occur in one of these arches. Thus, a $\delta_{f(sim)}$ value of 40 $\mathring{\mathrm{A}}$ was selected. This value is larger than the expected maximum separations that could occur without fragmentation for each of the simulated ionic liquids for the simulated temperatures and electric field strengths. Each cluster was simulated until this value of $\delta_{f(sim)}$ was reached. 

\begin{figure}
    \centering
    \includegraphics[width=6.25in]{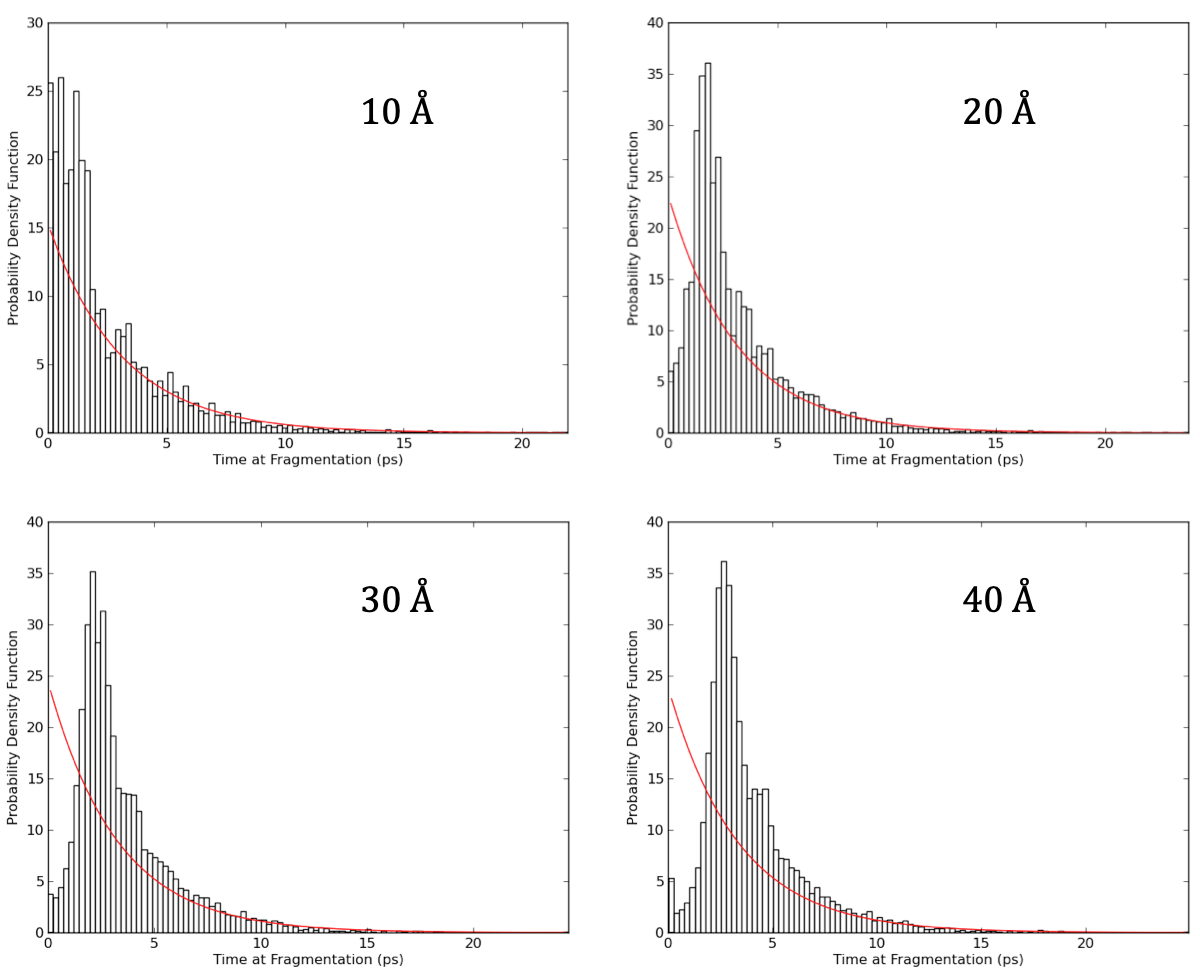}
    \caption{Lifetime distributions for positive EMI-BF$_4$ dimers at 2500K with an electric field strength of $1\times10^9$ V/m. The red line shows the exponential decay fit to the data.}
    \label{fig:reduceDeltaPost}
\end{figure}

While a separation of 40 $\mathring{\mathrm{A}}$ is appropriate for determining whether a cluster has fragmented, the time at which that separation is reached does not necessarily represent the time at which the cluster began to fragment. Particularly for the high electric field and high temperature cases, the time it takes for the maximum separation to reach 40 $\mathring{\mathrm{A}}$ is large compared to the time it takes for the cluster to begin to fragment. To account for this, the time at fragmentation was determined during post-processing using the value $\delta_{f(post)}$. Figure \ref{fig:reduceDeltaPost} shows the distribution of dimer lifetimes for EMI-BF$_4$ dimers simulated at 2500 K with an electric field strength of $1\times10^9$ V/m for different values of $\delta_{f(post)}$. For large values of $\delta_{f(post)}$ the distribution is somewhat Gaussian, which is thought to be a result of the $\delta_{f(post)}$ value determining the time it takes for the cluster to reach the large separation distance. For smaller values of $\delta_{f(post)}$ the distribution is closer to exponential decay, which is expected for a constant rate process. This shows that a smaller value of $\delta_{f(post)}$ can be used to determine the time it takes for the cluster to fragment. For some of the high temperature and electric field conditions, reducing $\delta_{f(post)}$ to 10 $\mathring{\mathrm{A}}$ had a large effect on the fragmentation statistics calculated in post-processing. In particular, the mean lifetime of the clusters decreased by nearly 50 \% for some of the simulation conditions when $\delta_{f(post)}$ was reduced from 20 $\mathring{\mathrm{A}}$ to 10 $\mathring{\mathrm{A}}$. The change in mean lifetime for lower electric field and temperature conditions was closer to 5 \% because the time it takes for the cluster to reach $\delta_{f(post)}$ was much smaller than the time it took the clusters to fragment. 

While reducing $\delta_{f(post)}$ improved the shape of the lifetime distributions for the higher electric field and temperature conditions, a $\delta_{f(post)}$ value of 10 $\mathring{\mathrm{A}}$ results in up to 30 \% of the high electric field and temperature condition samples being considered fragmented before the electric field is applied or within the first timestep of applying the electric field. Additionally, for these conditions the distribution of the lifetimes for samples that were not fragmented before the electric field was applied was Gaussian. Even if the samples that had fragmented before the simulation began were included in the data, the lifetime distribution would not be the exponential decay that is expected for a constant rate process. 

It is possible that the behavior seen for these conditions is a result of the frequency at which the fragmentation condition was checked during the simulation. Simulations with mean lifetimes greater than 15 ps were checked for fragmentation every 0.5 ps while simulations with mean lifetimes below 15 ps were checked for fragmentation every 0.1 ps. Simulations with Gaussian mean lifetime distributions and mean lifetimes less than 1ps were checked for fragmentation every 0.01 ps. Even with the higher fragmentation checking frequency the lifetime distributions remained Gaussian for some conditions. As such, this data is not considered to be quantitatively accurate and could be overestimating the lifetime by a as much as a factor of 4. However, these conditions occurred with applied electric fields greater than the highest electric field experienced by electrospray clusters immediately after emission and further work was not done to fix this problem. Future work should be performed to determine the cause of this behavior at the highest simulated electric fields and determine whether fragmentation under these conditions is still a constant rate process as is observed for lower electric fields and temperatures. Additionally, work should be done to characterize the simulations in which fragmentation behavior is observed before the temperature equilibration is complete. It is possible that adding further constraints on the molecule motion during the temperature equilibration step could reduce the number of clusters being considered fragmented before the electric field is applied.

The final chosen $\delta_{f(post)}$ value for dimers was 20 $\mathring{\mathrm{A}}$. All results included in the next sections were determined using this value. This was greater than two standard deviations above the largest separation average before fragmentation for all simulated dimers, resulting in fewer than 15 \% of the dimers being considered fragmented before the start of the simulations. Tables of the mean and standard deviation of the maximum separation of the cluster before fragmentation are given in Appendix \ref{app:appa} for each of the simulation conditions. Tables of the post-processed statistics for other values of $\delta_{f(post)}$ are also included in Appendix \ref{app:appa}.

The final chosen $\delta_{f(post)}$ value for dimers was 30 $\mathring{\mathrm{A}}$. While the mean maximum separation before fragmentation is less than 15 $\mathring{\mathrm{A}}$ for the trimer simulations a 30 $\mathring{\mathrm{A}}$ $\delta_{f(post)}$ was necessary to reach above 95 \% fragmentation with the computational resources available. Further work is required to better understand the effect of the $\delta_{f(post)}$ value on the high electric field trimer simulations.

\subsection{Determining Mean Lifetime}\label{sec:postLifetime}

This section explains how the mean lifetime was calculated from the results of the simulations as well as how the results of the simulations were used to extrapolate to other simulation conditions. The mean lifetime, $\tau$, was determined for a set of samples simulated at the same conditions by averaging the lifetimes. Mean lifetimes were also calculated by fitting a line of exponential decay to the lifetime graph and calculating $\tau$ from the expressions for exponential decay given in equation \ref{expDecay}. 

\begin{equation}\label{expDecay}
    y = \frac{1}{\tau} e^{-\frac{t}{\tau}}
\end{equation}

where y is the number of clusters fragmenting at that time, t is the time, and $\tau$ is the mean lifetime. A linear fit of the form $y = ax+b$ can be found to the datapoints given by (t, ln(y)). For this fit $a = -\frac{1}{\tau}$ and $b = ln(\frac{1}{\tau})$. Thus, the mean lifetime can be calculated as $\tau = -\frac{1}{a}$ or $\tau = \frac{1}{e^b}$. The results of determining the mean lifetime through an exponential fit are given in tables in Appendix \ref{app:appa}. The exponential fits were performed by splitting the lifetimes into 100 bins and finding a linear fit to (t, ln(y)). Error between the exponential fit and the mean lifetime decay were calculated as

\begin{equation}\label{ErrorExpFit}
    E = \sum_{0}^{100} (y_i - {y_{fit}})^2
\end{equation}

where $y_i$ is the number of clusters fragmenting at the times put into bin i and $y_{fit}$ is the number of clusters predicted to fragment at the times put into bin i by the exponential fit. The mean lifetimes calculated using the exponential fit slope were higher than those calculated using the fit constant. As expected, the simulation conditions that resulted in lifetime distributions that were closer to Gaussian than exponential decay had large errors in the exponential fits. The mean lifetimes determined from the exponential fits for these conditions did not match the mean lifetimes found by averaging the lifetimes. In cases where the error in the exponential fit was low the difference between the methods of calculating the mean lifetime were also low. 



\subsection{Analyzing Geometry and Fragmentation Pathway}\label{sec:postGeometry}

The sections below explain how the geometry of the clusters during fragmentation was determined from the results of MD simulations.

\subsubsection{Maximum Separation}

One aspect of cluster geometry that is of interest is the average and standard deviation of the maximum separation between the components of the cluster before fragmentation begins. The mean maximum separation roughly indicates the diameter of the cluster before fragmentation. The standard deviation indicates how much the cluster diameter fluctuates before fragmentation. The mean maximum separation before fragmentation was found by averaging the maximum separation at each timestep at which the fragmentation condition was checked before $\delta_{f(post)}$ was reached. The standard deviation was calculated from the same values. 




\subsubsection{Fragmentation Pathway}\label{MemoryMethods}


Another aspect of cluster geometry that is of interest is the fragmentation pathway, or the way in which the cluster splits up during fragmentation. This influences the types of ions and neutrals left in the ion beam after fragmentation, which can have effects on surfaces surrounding the electrospray emitter. Dimers have two fragmentation pathways, neutral evaporation, or total fragmentation. Neutral evaporation occurs when a single ion escapes from the cluster, leaving behind a neutral ion pair that is not accelerated further by the applied electric field. Total fragmentation occurs when the cluster fragments into three monomers, two charged in the firing polarity and one charged opposite of the firing polarity. The two monomers charged in the firing polarity are accelerated away from the emission site while the charge opposite of the firing polarity is accelerated towards the emission site and likely recombines with the emission meniscus or collides with other emitted ions. 

The fragmentation pathway for each sample is determined in post-processing when the cluster reaches $\delta_{f(sim)}$. If the separation between each of the three pairings of ions in the cluster is greater than 35 $\mathring{\mathrm{A}}$ the cluster is considered to have taken the total fragmentation pathway. If the separation of one of the pairs of ions in the cluster is less than 35 $\mathring{\mathrm{A}}$ the cluster is considered to have taken the neutral evaporation pathway. The smaller distance between one pair of ions indicates that the neutral remains intact during the fragmentation process. Future work is needed to simulate the behavior of neutral clusters after neutral evaporation fragmentation to determine if the criteria described above accurately predicts whether the neutral will remain intact after fragmentation.



\begin{figure}
    \centering
    \includegraphics[width=5in]{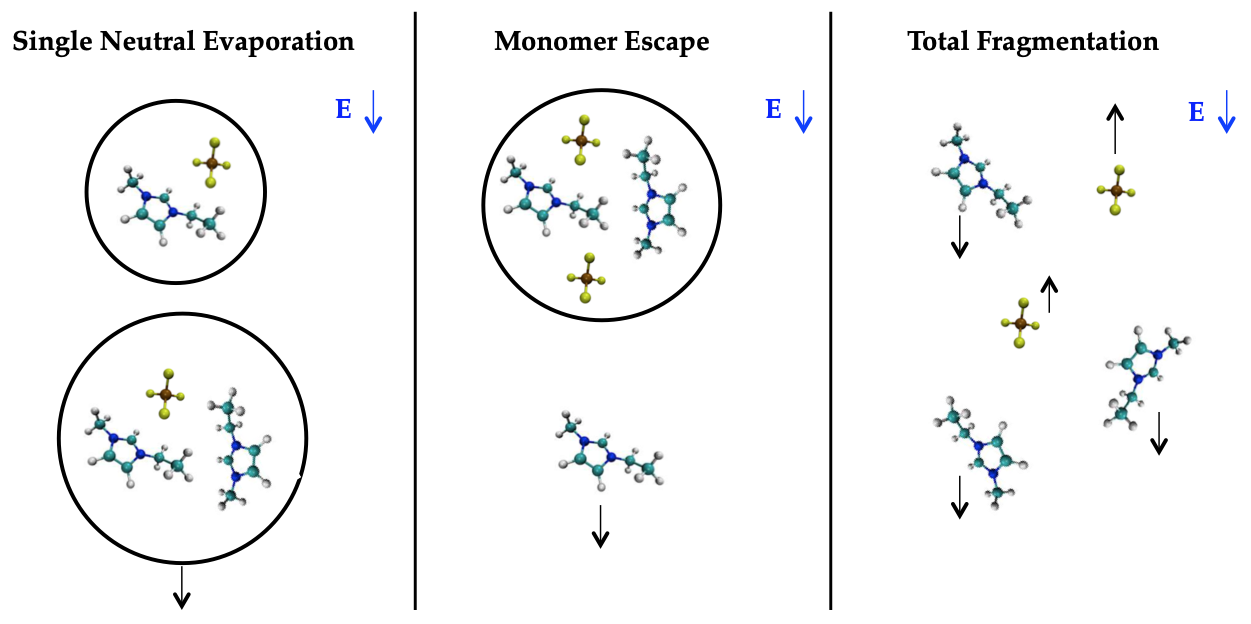}
    \caption{Different fragmentation pathways for positive trimers.}
    \label{fig:TrimerPathwayDiagram}
\end{figure}

According to previous work and observations of the data collected for this work, trimers have the following fragmentation pathways: single neutral evaporation, monomer escape, and total fragmentation \cite{Prince2015MolecularNanodrops}. Figure \ref{fig:TrimerPathwayDiagram} shows diagrams of each of these pathways. The black arrows give the direction that each cluster or ion will be accelerated in the electric field given by the blue arrow. The neutral clusters will continue at the same velocity that the parent cluster had before fragmentation in the direction of the electric field. Single neutral evaporation occurs when fragmentation results in a dimer and a single neutral ion pair. Monomer escape occurs when fragmentation results in a single ion monomer and a cluster of two neutrals. Total fragmentation occurs when fragmentation results in 5 separate ions, three charged in the firing polarity and two charged opposite of the firing polarity. 

During each trimer simulation the distances between the 10 different ion pairings in the trimer cluster were calculated. The simulations were then categorized into the three different pathways by finding the number of the ion pairs that were larger than a desired cutoff. Total fragmentation was recorded when all of the ion pair distances were greater than 30 $\mathring{\mathrm{A}}$, single neutral evaporation was recorded when 6 of the 10 pairs were larger than 30 $\mathring{\mathrm{A}}$, and monomer escape was recorded when 4 of the 10 pairs were larger than 30 $\mathring{\mathrm{A}}$. Any of the simulations that did not fit these specifications were labelled as unknown. More discussion about these cases is given in Section \ref{sec:PathwayResults}.

\subsection{Analyzing Electric Field Memory}\label{sec:postMemory}

This section details how MD results were used to determine whether there is a memory effect of the electric field on the fragmentation behavior of the clusters. Determination of the mean lifetime of clusters from fitting an exponential decay equation assumes that clusters are fragmenting according to a constant rate process. If fragmentation is a constant rate process then the distribution of the lifetimes of the clusters as a function of time should be given by an exponential distribution as given in Equation \ref{expDecay}. The exponential distribution is memoryless, which is a quality described by equation \ref{memoryless}

\begin{equation}\label{memoryless}
    P(X>x+a | X>a) = P(X>x)
\end{equation}

If we choose any point in time along the graph of lifetime distribution vs time, remove all the data to the left of that point, consider that point to be the new t=0, and rescale the data to the right to account for the different total number of samples being examined, all of the data to the right of the point should fit the same exponential decay equation. Now consider the case in which the lifetime distribution is not memoryless. For example, suppose exposing the cluster to the electric field changes the energy distribution in the ion cluster. In this case it is possible that ions exposed to the electric field for a longer time fragment at a rate higher than those exposed to the electric field for a shorter amount of time. If this were the case, then the memoryless quality of the exponential distribution would not hold for the distribution of lifetimes of clusters. 

To determine whether this is the case for ion cluster fragmentation the 2500 K, $5\times10^8$ V/m conditions were examined more closely. 90,000 simulations were run for this set of conditions. More samples were needed for the memory investigation because higher resolution of the lifetime distribution decay behavior was necessary for the calculations explained below. The $\delta_{post}$ value was taken to be 15 $\mathring{\mathrm{A}}$. The mean lifetime of the cluster was calculated first using the average of all of the samples. Time dependent mean lifetimes were then determined using three methods.

Method 1: The histogram bins were divided evenly into 7 separate sections by the fragmentation time. Exponential fits were found for each of the sections and mean lifetimes calculated from the slope of the fit as described in Section \ref{sec:postLifetime}.

Method 2: The mean lifetime was determined for the center fragmentation time of each histogram bin by calculating the probability of fragmentation from the number of clusters that fragmented in a given bin divided by the total number of clusters left to fragment after taking into account the fragmentation in previous bins. 

Method 3: An exponential function was fit to the full histogram. The mean lifetime was determined for the center fragmentation time of each histogram bin by calculating the probability of fragmentation from the number of clusters that fragmented at a given point on the exponential fit curve divided by the total number of clusters left to fragment after taking into account the fragmentation at previous points on the exponential fit curve. 

The same three methods were also used to characterize the change in mean lifetime of clusters exposed to a time dependent electric field. In particular, 60,000 clusters were simulated by an electric field that decreased linearly from $5\times10^8$ to $5\times10^7$ over 200 ps. If fragmentation is a constant rate process the mean lifetimes should increase as the electric field decreases. If fragmentation is not a constant rate process then it is possible that the fragmentation rates will not increase as much during the simulation as would be predicted using the instantaneous electric field.

\section{Results}

\subsection{Mean Lifetime}\label{sec:resultLifetime}

\begin{figure}
    \centering
    \includegraphics[width=5in]{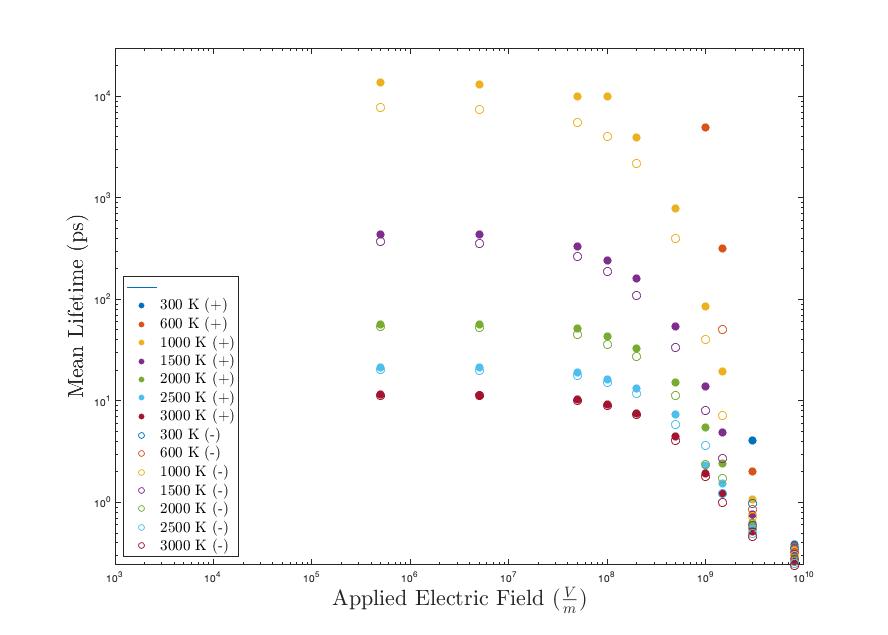}
    \caption{Mean lifetimes for positive and negative EMI-BF$_4$ dimers.}
    \label{fig:EMIBF4DimersMeanLifetimes}
\end{figure}

\begin{figure}
    \centering
    \includegraphics[width=5in]{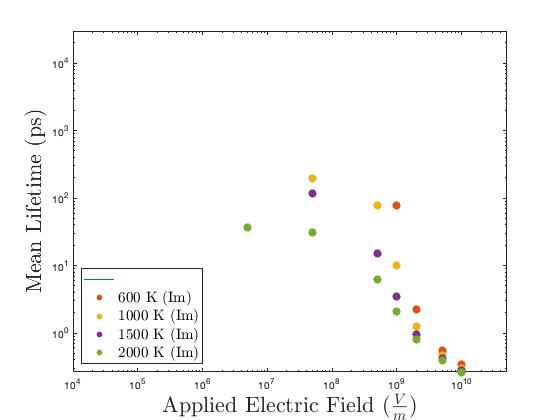}
    \caption{Mean lifetimes for positive EMI-Im dimers.}
    \label{fig:EMIIMDimersMeanLifetimes}
\end{figure}

\begin{figure}
    \centering
    \includegraphics[width=5in]{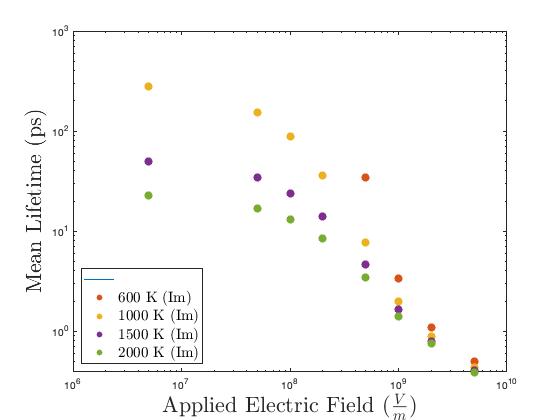}
    \caption{Mean lifetimes for positive EMI-FAP dimers.}
    \label{fig:EMIFAPDimersMeanLifetimes}
\end{figure}

Figures \ref{fig:EMIBF4DimersMeanLifetimes}, \ref{fig:EMIIMDimersMeanLifetimes}, \ref{fig:EMIFAPDimersMeanLifetimes}, and \ref{fig:EMIBF4DimersMeanLifetimes} show the mean lifetime results for positive and negative mode EMI-BF$_4$ dimers, EMI-Im dimers, EMI-FAP dimers, and EMI-BF$_4$ trimers respectively. Results are plotted in logarithmic scales to visualize all results on the same plot. Full tables with the mean lifetime results are included in Appendix \ref{app:appa}. The sections below discuss the effects of electric field, cluster temperature, polarity, cluster size, and ionic liquid type on the mean lifetime. 

\subsubsection{Electric Field}\label{sec:Efieldresults}
Mean lifetime decreases with increasing electric field. This is because the application of the electric field reduces the energy barrier to fragmentation by applying forces on the positive and negative components of the cluster in opposite directions. As seen in the graphs there is a change in the dependence of the fragmentation rate on the strength of the electric field between $2\times 10^7$ and $2\times 10^8 $ V/m. For electric fields lower than $2\times 10^7 $ V/m the dependence on the electric field is small. The fragmentation rate is nearly constant with less than 25 \% change in the fragmentation rate for an order of magnitude change in the electric field between $5\times 10^{7} $ V/m and $5\times 10^6 $ V/m for all temperatures simulated. For higher temperatures the change in mean lifetime between these two electric field is even smaller, around 10\% for the 3000 K case. For electric fields greater than $2\times 10^8 $ V/m the dependence on the electric field is much stronger. Increases in electric field strength by an order of magnitude from $2\times 10^8 $ V/m to $2\times 10^9 $ V/m results in decreases in mean lifetime of at least an order of magnitude for most simulated conditions for EMI-BF$_4$. 

Decreases in mean lifetime in response to increasing electric field are larger for lower temperature conditions than for higher temperature conditions. This is likely because the energy in the ions is larger for the higher temperatures, which means that the clusters are already closer to the energy barrier for fragmentation, and only a small increase in electric field will provide the necessary extra energy to fragment.

\subsubsection{Temperature}
Mean lifetime decreases with increasing cluster temperature. This is because the cluster temperature is related to the cluster energy. When the cluster has more energy the probability that it overcomes the activation barrier for fragmentation is higher. Thus the time for which the cluster survives unfragmented, and thus the mean lifetime, is lower. The decrease in mean lifetime for 500 degree increases in temperature is smaller at higher temperatures. For example, the decrease in mean lifetime between 1000 K and 1500 K for EMI-BF$_4$ dimers at $1\times 10^{9} $ V/m is 83\% and the decrease in mean lifetime between 1500 K and 2000 K for EMI-BF$_4$ dimers $5\times 10^{7} $ V/m is 57\%.

Additionally, there appear to be two limiting cases for the dependence of mean lifetime on temperature. As the electric field becomes small the temperature dependence is large while the electric field dependence is low. This corresponds to cases when the internal energy from the temperature of the ion is larger than the reduction in activation energy provided by the electric field. When the electric field becomes large, the temperature dependence is small. This corresponds to cases when the internal energy from the temperature of the ion cluster is smaller than the reduction in activation energy provided by the electric field. In the limiting case when the electric field goes to 0 the mean lifetime depends only on the temperature of the cluster. Presumably at a high enough electric field the fragmentation rate of the cluster would depend only on the electric field strength and would not be affected by the ion temperature. This would likely happen with electric field strengths several orders of magnitude greater than simulated here as the results at the highest electric fields still show differences of up to 50\% in the mean lifetimes for different temperature conditions. 

\subsubsection{Polarity}

Data is only available for the negative polarity of the EMI-BF$_4$ dimer. For all simulation conditions the negative polarity fragmentation rates are lower than for the positive polarity. This is at least partly due to the energy content of both polarities at the same temperature. For example, at 1000 K the positive polarity mean energy is 82 eV while the negative polarity mean energy is 192 eV. The mean lifetimes for the positive and negative polarities at an electric field of $1.5\times 10^9$ V/m are 20. To conclusively determine whether there is a difference in the energy barrier for fragmentation between the two polarities simulations would need to be performed with the same total internal energy

\subsubsection{Cluster Size}

The effect of cluster size on fragmentation behavior can be determined by comparing the results for EMI-BF$_4$ dimers and trimers at the same temperature. For dimers and trimers at the same temperature trimer mean lifetimes at each electric field strength are shorter than those of dimers. A similar change in the dependence of the mean lifetime as a function of electric field to the one discussed in Section \ref{sec:Efieldresults} occurs for trimers at approximately $1\times10^8$ V/m. Further work would be needed to determine how the fragmentation rates compare for clusters of different size with the same total energy as equilibrating clusters to the same temperature yields different energies for clusters with different numbers of molecules. 

\subsubsection{Ionic Liquid}

Fragmentation rates also depend on the type of ionic liquid studied. For the same cluster temperature EMI-FAP had the shortest mean lifetimes, followed by EMI-BF$_4$, and EMI-Im had the longest mean lifetimes. This held true for 600 K, 1000 K, 1500 K, and 2000 K. This is interesting considering Miller's results, which showed that the mean lifetimes for EMI-FAP were the longest, EMI-Im, the second longest, and EMI-BF$_4$ the shortest. This discrepancy is possibly due to the energy content of the clusters differing between Miller's experimental characterization and these simulations. Miller calculated that the temperatures of EMI-Im dimers were approximately 600 K while the temperatures of the EMI-FAP dimers were approximately 550 K. The smaller temperature for the EMI-FAP dimers may have led to them having lower fragmentation rates than would be expected at the same temperature as seen in these simulations. Additionally, the temperatures calculated by Miller relied on fragmentation rates calculated in the field free region. It is possible that the fragmentation rates with no electric field have a different dependence on ionic liquid than do the fragmentation rates with an electric field. More simulations would be necessary to distinguish between the effect of the temperature of the clusters measured by Miller and the effects of the ionic liquid with no electric field. 

Another explanation for this disagreement is the definition of temperature used in this work. The energy content of EMI-Im dimers at 1000 K is 309 eV while the energy content of EMI-FAP dimers at 1000 K is 801 eV. It is possible that the larger amount of total energy results in shorter mean lifetimes for EMI-FAP than would be expected if they were at the same energy. However, EMI-BF$_4$ dimers at 1000 K have an energy of 192 eV and EMI-BF$_4$ dimers were found to have shorter mean lifetimes than EMI-Im at this temperature even though EMI-Im has a larger total energy content. This requires further investigation. Future simulations will be performed with clusters at the same total energy content similar to simulations performed by Coles to determine whether the difference in mean lifetimes is due to a difference in energy content at each temperature or if it is a result of the geometry of the clusters of different ionic liquids \cite{Coles2013InvestigatingBeams}.


\begin{figure}
    \centering
    \includegraphics[width=5in]{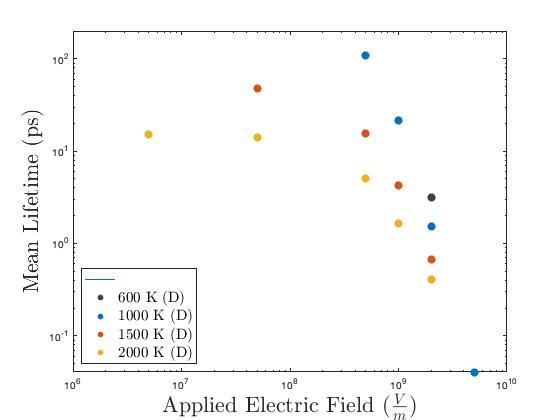}
    \caption{Mean lifetimes for positive EMI-BF$_4$ trimers.}
    \label{fig:EMIBF4TrimersMeanLifetimes}
\end{figure}



\subsection{Geometry and Fragmentation Pathway}\label{sec:resultGeometry}

\subsubsection{Separation Before Fragmentation}

Full tables of the maximum separation before fragmentation for each ionic liquid under different electric field and temperature conditions are included in Appendix \ref{app:appa}. Typical maximum separations range between 7 and 13 $\mathring{\mathrm{A}}$ for EMI-BF$_4$ dimers with slight but statistically insignificant differences between the polarities. For EMI-Im dimers the range is 9 to 13 $\mathring{\mathrm{A}}$ and for EMI-FAP dimers the range is 9 to 15 $\mathring{\mathrm{A}}$. EMI-BF$_4$ trimers max separation ranges between 10 and 16 $\mathring{\mathrm{A}}$, which is a result of the larger number of molecules in the cluster.

For all ionic liquids simulated the mean separation before fragmentation increases with temperature. This is because the random motions from the internal energy are larger. This is similar to results seen previously by Prince et al. \cite{Prince2015MolecularNanodrops}. For the same reason the standard deviation of the separation before fragmentation increases with temperature. Interestingly the mean maximum separation is larger for lower electric fields than for higher electric fields. This is contrary to what is expected from the stretching of the cluster by the electric field that was seen by Prince \cite{Prince2015MolecularNanodrops}. One possible explanation for this discrepancy is that the mean max separation takes into account the time that the cluster takes to fragment. It is possible that over the longer time spent in the electric field the clusters have more time to stretch under its force, resulting in a larger mean max separation.

\subsubsection{Fragmentation Pathway}\label{sec:PathwayResults}

\begin{figure}
    \centering
    \includegraphics[width=5in]{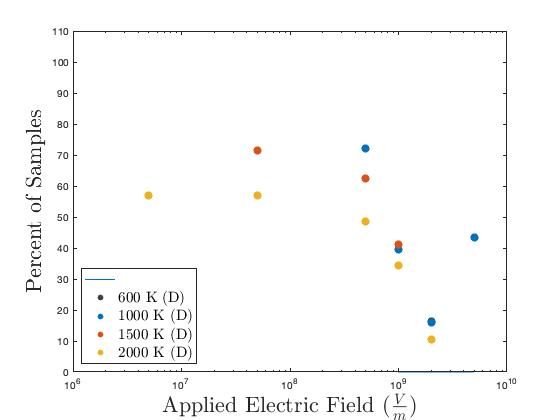}
    \caption{Percentage of positive EMI-BF$_4$ trimers that fragment according to the neutral evaporation pathway.}
    \label{fig:NeutralEvap}
\end{figure}

\begin{figure}
    \centering
    \includegraphics[width=5in]{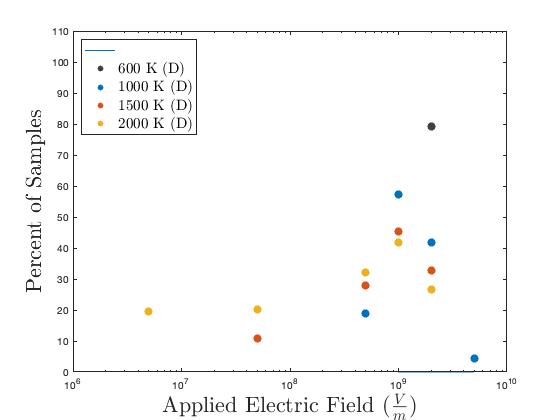}
    \caption{Percentage of positive EMI-BF$_4$ trimers that fragment according to the monomer escape pathway.}
    \label{fig:MonomerEsc}
\end{figure}

\begin{figure}
    \centering
    \includegraphics[width=5in]{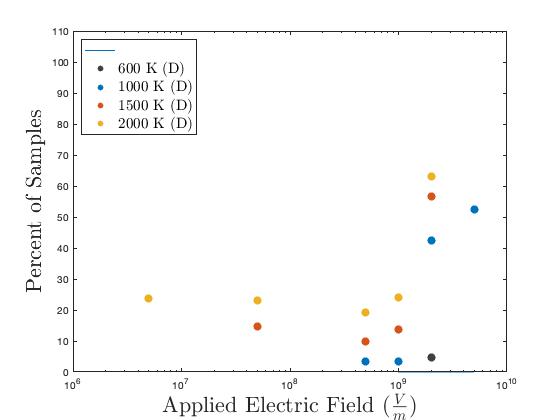}
    \caption{Percentage of positive EMI-BF$_4$ trimers that fragment according to an unknown separation pathway.}
    \label{fig:UnknownSep}
\end{figure}

Tables with the number of total separations recorded for each ionic liquid and set of conditions are included in Appendix \ref{app:appa}. Only one total separation was recorded for a trimer sample at the 1000 K and $5\times 10^9$ V/m conditions. Figures \ref{fig:NeutralEvap}, \ref{fig:MonomerEsc}, and \ref{fig:UnknownSep} show the proportion of EMI-BF$_4$ trimer samples that fragmented according to the neutral evaporation, monomer escape, and unknown pathways respectively. At lower electric field strengths neutral evaporation was more likely than monomer escape, however at electric fields higher than $1\times10^9$ V/m the opposite was true. For lower temperatures the likelihood of neutral evaporation was higher than for higher temperatures. Both of these trends indicate that the energy barrier for monomer escape is higher than that of neutral evaporation. For higher temperatures the likelihood that the separation data was not clearly one of the three defined pathways was higher. Simulations of trimers would need to be performed for a longer time to determine the full fragmentation pathway in these cases. One possibility is that these cases will eventually go on to fragment totally into 5 separate ions but that this process takes longer than the other fragmentation pathways and so was not seen in this data set.

Another possibility is that there are other pathways such as double neutral evaporation that may also occur beyond the timescale of this dataset. Double neutral evaporation occurs when the fragmentation products are a single ion monomer and two separate, two ion neutral clusters. It is possible that for some conditions the single neutral evaporation pathway is followed, but immediately after this fragmentation event the child dimer cluster fragments into a neutral cluster and a monomer. If the dimer has a high enough energy this may happen so soon after the initial fragmentation that it is nearly indistinguishable from the double neutral evaporation pathway. Future work will include simulations of trimer products for a longer time period to characterize the fragmentation behavior after the initial trimer fragmentation.


\subsection{Electric Field Memory}\label{sec:resultMemory}

\begin{figure}
    \centering
    \includegraphics[width=4.5in]{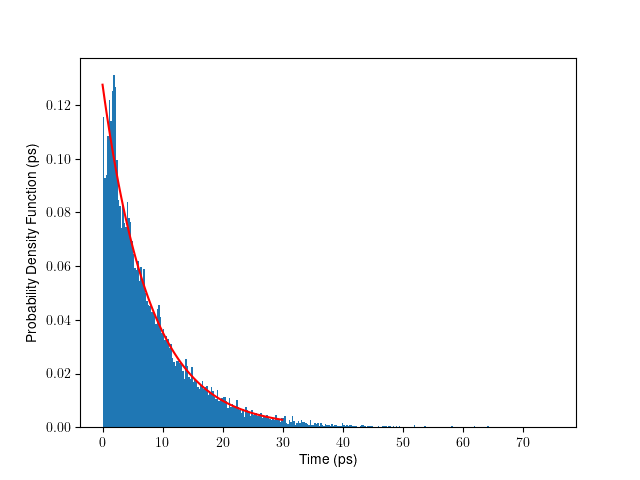}
    \caption{Histogram of fragmentation times for 87,000 2500 K EMI-BF$_4$ dimers simulated at an electric field of $5\times10^8$ V/m.}
    \label{fig:MemoryConstantEHistWide}
\end{figure}

\begin{figure}
    \centering
    \includegraphics[width=4.5in]{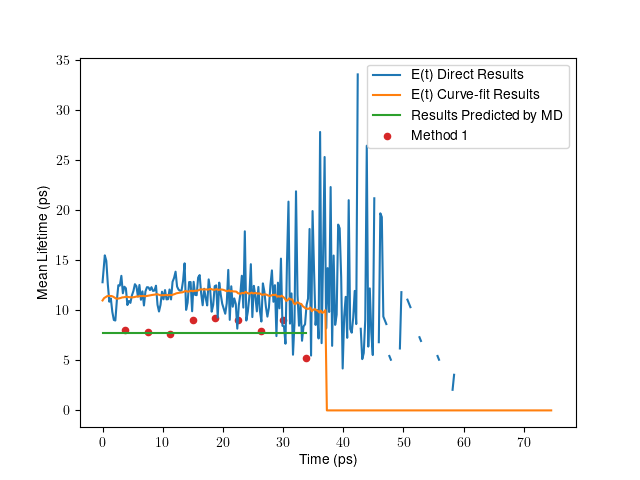}
    \caption{Mean lifetimes calculated for 87,000 2500 K EMI-BF$_4$ dimers simulated at an electric field of $5\times10^8$ V/m using three different methods.}
    \label{fig:MemoryConstantElifeWide}
\end{figure}

Figure \ref{fig:MemoryConstantEHistWide} shows the histogram with the exponential curve fit for the 2500 K constant electric field results with 300 bins and temperatures limited to +- 10\% of 2500 K. Figure \ref{fig:MemoryConstantElifeWide} shows the mean lifetimes calculated using the three methods described in Section \ref{MemoryMethods} from this data as a function of the time the clusters have been in the electric field. For method one it appears that the first three fitted exponential curves have decreasing mean lifetime. However, for the next 6 mean lifetimes calculated by method one the trend is unclear. The results for method two and three are similar to method one, with small decreases in mean lifetime as a function of the amount of time that the clusters have been exposed to the electric field. This is what would be expected if the application of the electric field changed the energy of the cluster enough to make the constant rate assumption invalid. One explanation for the disagreement of the mean lifetimes calculated using method one at the longer simulation times is that results calculated near the very end of the fragmentation times are likely to be inaccurate. At the end of the simulation there are not enough clusters left to represent the low probability of fragmenting. This is why the variance in the mean lifetimes calculated using method two increases as the time gets larger. 

\begin{figure}
    \centering
    \includegraphics[width=4.5in]{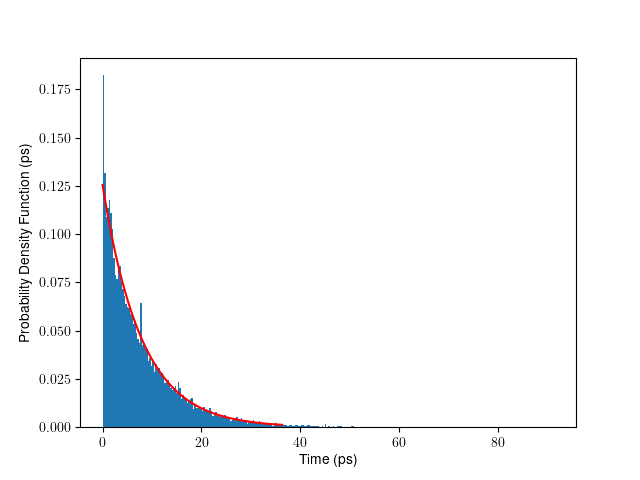}
    \caption{Histogram of fragmentation times for 60,000 2500 K EMI-BF$_4$ dimers simulated with a varying electric field.}
    \label{fig:MemoryVariableEHistWide}
\end{figure}

\begin{figure}
    \centering
    \includegraphics[width=4.5in]{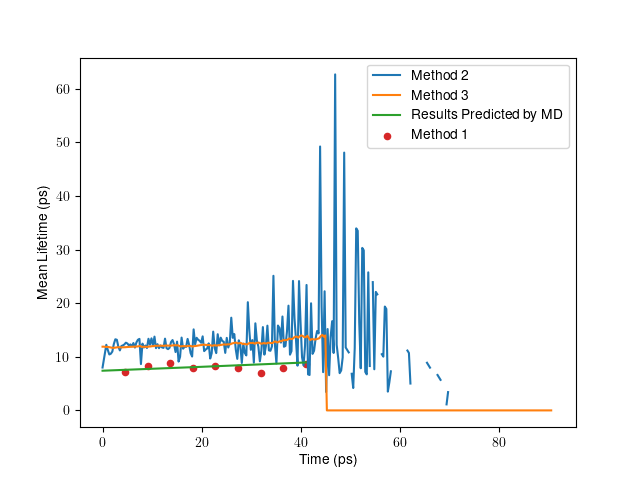}
    \caption{Mean lifetimes calculated for 60,000 2500 K EMI-BF$_4$ dimers simulated with a varying electric field using three different methods.}
    \label{fig:MemoryVariableElifeWide}
\end{figure}

Figure \ref{fig:MemoryVariableEHistWide} shows the histogram with the exponential curve fit for the 2500 K EMI-BF$_4$ dimers simulated with a time varying electric field. Figure \ref{fig:MemoryVariableElifeWide} shows the mean lifetimes calculated using the three methods described in Section \ref{MemoryMethods} from this data as a function of the time the clusters have been in the electric field. The mean lifetimes calculated using all three methods appear to follow the trend in the mean lifetimes predicted by MD assuming that fragmentation is a memoryless constant rate process.

While these results are not conclusive, they suggest that the memory effect of the electric field on fragmentation behavior is negligible. However, more simulations would need to be performed and the post-processing methods would need to be improved to ensure the data is precise enough to detect any memory effect. More tests would also need to be done with other electric field conditions, particularly at higher electric fields, to see if this assumption holds. Additionally, simulations would need to be performed in the same way for other ionic liquids and clusters sizes to ensure this result holds for other cases.
\chapter{Cluster Fragmentation Physical Model}

This section explains the derivation of the dipole point model and comparison of the model to the MD results. The dipole point model is an alternative to the Schottky model for determining fragmentation rates for ionic liquid dimers under the influence of electric fields. This model is similar to the Schottky method in that it uses a physical model of the geometry of the fragmenting cluster over time to determine the reduction in the fragmentation activation energy provided by the electric field. This model is only valid for the neutral evaporation pathway. Evaluations of the model are compared to MD data for various temperatures and electric fields to determine the parameters that are needed to match the model to the MD results. 

\section{Derivation}

\begin{figure}
    \centering
    \includegraphics[width=2in]{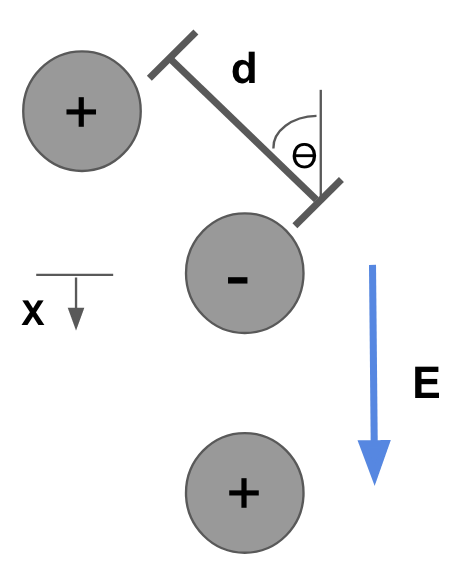}
    \caption{Geometry of dipole and escaping ion used for the dipole point model.}
    \label{fig:Dipolegeometry}
\end{figure}

Figure \ref{fig:Dipolegeometry} shows the geometry assumed for the dipole point model. The dipole distance, d, is the distance between the molecules that make up the neutral pair that is left behind after fragmentation. For this basic dipole point model, the dipole distance is assumed to remain constant during fragmentation. This model does not apply to the total fragmentation pathway in which the neutral breaks up into two ions during fragmentation. The dipole angle, $\theta$, is the angle that the dipole makes with respect to the direction of the electric field during fragmentation. The dipole angle is assumed to remain constant during the fragmentation process. While this assumption does not conserve angular momentum, it results in a simple expression for the mean lifetime as a function of the electric field, and thus is used for this preliminary analysis. Future work will include MD simulations to improve our understanding of the geometry of the cluster during fragmentation and to improve the assumptions made in the physics-based model. The E vector shows the direction of the electric field which is assumed to be the same direction as the motion of the escaping ion. The escaping ion is assumed to remain on this axis during fragmentation. For all derivations the ion with the polarity opposite that of the cluster is taken to be located at (0,0) during the fragmentation process. The following equations are in the reference frame of this opposite polarity ion. 

This model assumes that each of the molecules in the cluster behaves like a point charge. The only force between the point charges is the Coulomb force given by equation \ref{Coulomb}. 

\begin{equation}\label{Coulomb}
    F_C = \frac{q^2}{4 \pi \epsilon_0 r^2} \hat{r} = \frac{q^2 k}{r^2} \hat{r}
\end{equation}

where q is the elementary charge on each ion, $\epsilon_0$ is the permittivity of free space, r is the distance between the two ions, and $\hat{r}$ is the vector pointing from one ion to the other. This model ignores all other forces between the molecules. 

Like the Schottky model, the goal of the dipole point model is to find the minimum work needed to remove the escaping ion from the cluster. This can then be included in the Arrhenius rate law as a reduction in the activation energy of the process. To find the minimum work to remove the ion we first find the total force on the escaping ion

\begin{equation}\label{FtotalBasicDipole}
    F_{total} = q E - \frac{q^2 k}{x^2} + \frac{q^2 k (x+ d cos(\theta))}{((x+dcos(\theta))^2 + (dsin(\theta))^2)^{(\frac{3}{2})}}
\end{equation}

This is a superposition of the forces from the electric field, the ion of opposite polarity in the neutral, and the ion of the same polarity in the neutral. This equation only accounts for the x direction component of the force as we are assuming the ion stays on the axis of the electric field during the fragmentation process. This can be integrated from $\infty$ to x to find the work required to move the escaping ion from $\infty$ to the location x as a function of x. This is given by Equation \ref{WtotalBasicDipole}

\begin{equation}\label{WtotalBasicDipole}
    W_{total} = q E x + \frac{q^2 k}{x^2} - \frac{q^2 k}{((x+dcos(\theta))^2 + (dsin(\theta))^2)^{(\frac{1}{2})}} 
\end{equation}

This is the same as the work required to move the escaping ion from a location x to infinity. The minimum x location of this function will occur when the total force on the escaping ion is 0. This is the location at which the electric field force overcomes the attraction between the escaping ion and the neutral pair, resulting in fragmentation. This can be solved for numerically using Equation \ref{FtotalBasicDipole}.

\section{Model Evaluation}

\begin{figure}
    \centering
    \includegraphics[width=4.5in]{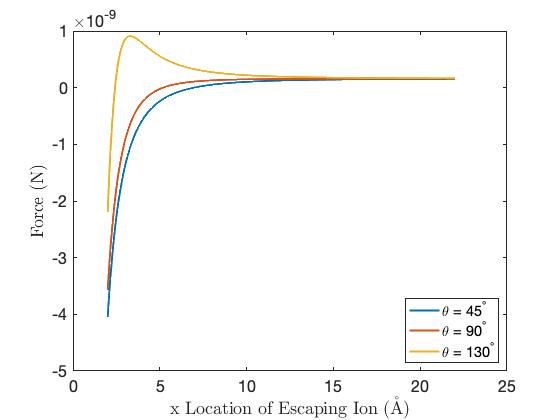}
    \caption{Force on the escaping ion for varying dipole angles. d = $2\mathring{\mathrm{A}}$, E = $1\times10^9$V/m}
    \label{fig:ForceTheta}
\end{figure}

\begin{figure}
    \centering
    \includegraphics[width=4.5in]{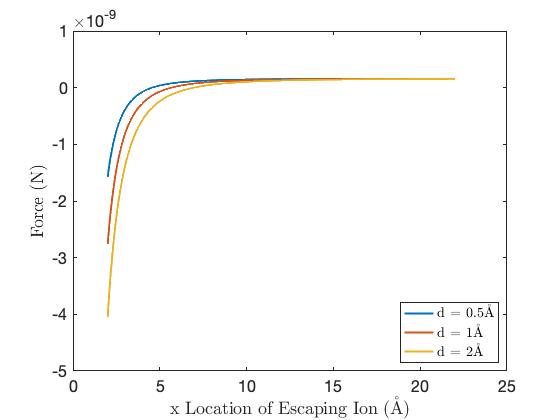}
    \caption{Force on the escaping ion for varying dipole separations. $\theta$ = 45, E = $1\times10^9$V/m}
    \label{fig:ForceD}
\end{figure}

This section details the evaluation of this model for various cluster geometries. Figures \ref{fig:ForceTheta} and \ref{fig:ForceD} shows the results of evaluating Equation \ref{FtotalBasicDipole} for various values of $d$ and $\theta$. As the angle of the dipole increases the total positive force on the escaping ion increases. This is because the same polarity ion in the cluster is closer to the escaping ion. When the angle of the dipole is greater than 90$^{\circ}$ a positive bump appears in the force. This is because the same polarity ion is between the escaping ion and the opposite polarity ion, shielding the escaping ion. For a dipole angle of 45$^{\circ}$, as the distance within the dipole increases the force on the escaping ion is more negative. This is because the same polarity ion in the cluster is further away and thus shields the escaping ion less. The opposite is true when the dipole angle is greater than 90$^{\circ}$ as the same polarity ion would be closer to the escaping ion and thus shield it more.

\begin{figure}
    \centering
    \includegraphics[width=4.5in]{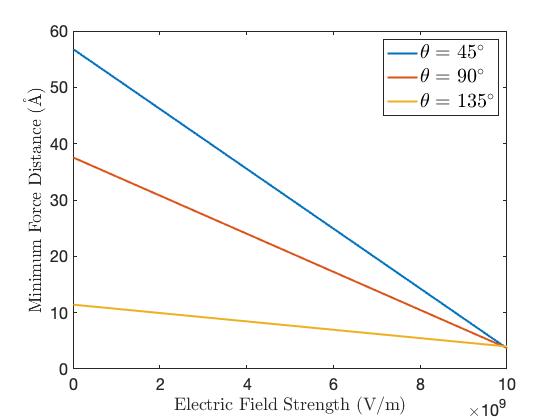}
    \caption{Location of 0 force on the escaping ion for varying dipole angles. d = $10\mathring{\mathrm{A}}$}
    \label{fig:XMINTheta}
\end{figure}

\begin{figure}
    \centering
    \includegraphics[width=4.5in]{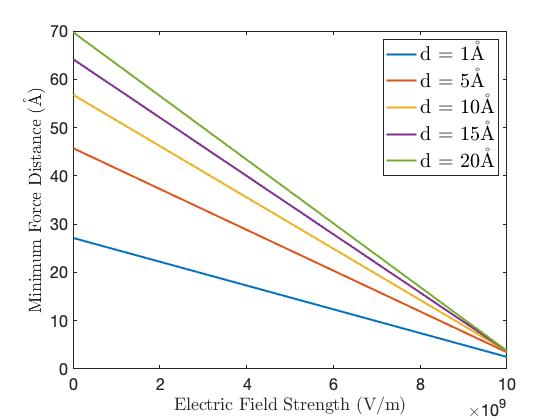}
    \caption{Location of 0 force on the escaping ion for varying dipole separations. $\theta$ = 45}
    \label{fig:XMIND}
\end{figure}

\begin{figure}
    \centering
    \includegraphics[width=4.5in]{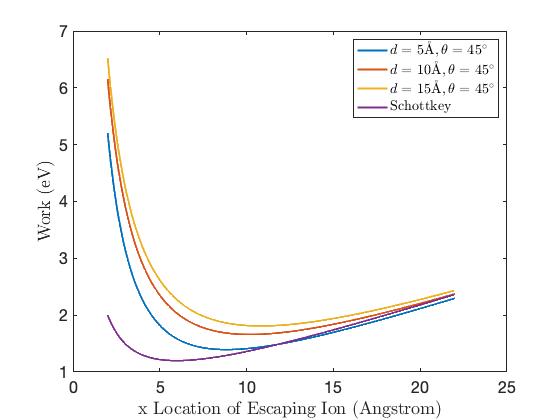}
    \caption{Contour plot of minimum work value for different values of dipole angle and dipole distance for the stationary dipole model. E = $1\times10^9$V/m}
    \label{fig:CompareWorkDipoleSchottky}
\end{figure}

\begin{figure}
    \centering
    \includegraphics[width=4.5in]{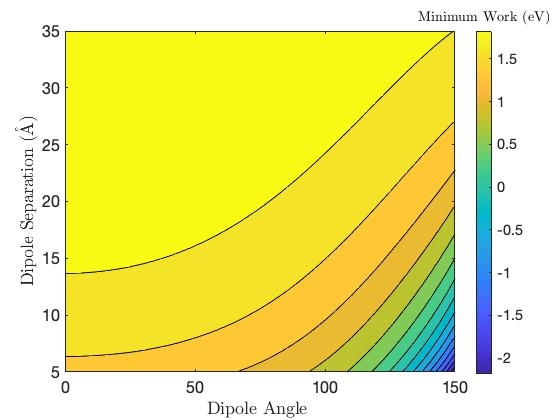}
    \caption{Contour plot of minimum work value for different values of dipole angle and dipole distance for the stationary dipole model. E = $1\times10^9$V/m}
    \label{fig:StationaryDipoleMinWorkContour}
\end{figure}

Figures \ref{fig:XMINTheta} and \ref{fig:XMIND} show the x location at which the force on the escaping ion is minimized. As the angle of the dipole increases the distance at which the force is 0 decreases. As the dipole separation increases for $\theta = 45 ^{\circ}$ the distance at which the force is 0 increases. Figure \ref{fig:CompareWorkDipoleSchottky} shows a comparison of the work to remove the escaping ion to point x as a function of x for various dipole angles and dipole distances. The minimum work for the Schottky model given by Equation \ref{WSchottky} is also shown. For the specific parameters shown here the work to remove the escaping ion is lower for the Schottky model than for the stationary dipole model. This is because in the dipole model some of the force of the opposite polarity ion is shielded by the similar polarity ion. Figure \ref{fig:StationaryDipoleMinWorkContour} shows the resulting minimum work values for a range of dipole angles and dipole distances. A negative minimum work occurs when the force from the same polarity ion completely shields the force of the opposite polarity ion. This happens at very high dipole angles where the same polarity ion is between the escaping ion and the opposite polarity ion. The largest positive work values result when the dipole separation is large and the dipole angle is small.

\section{Fit to MD Results}

\begin{figure}
    \centering
    \includegraphics[width=4.5in]{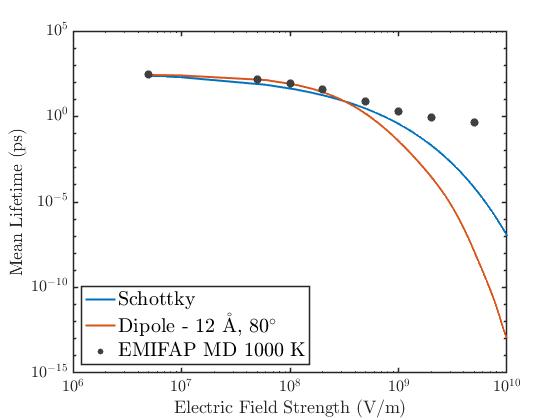}
    \caption{Best fits for the stationary dipole and Schottky models for EMI-FAP at 1000 K.}
    \label{fig:EMIFAP1000fits}
\end{figure}

\begin{figure}
    \centering
    \includegraphics[width=4.5in]{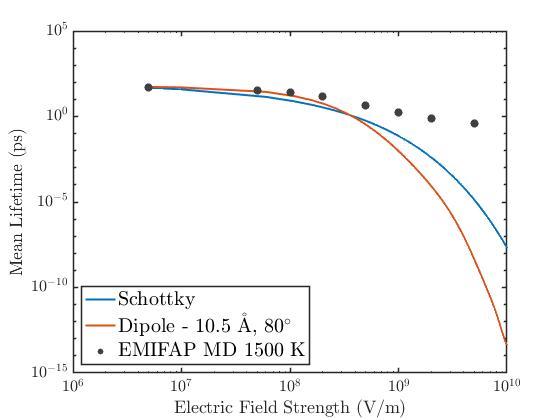}
    \caption{Best fits for the stationary dipole and Schottky models for EMI-FAP at 1500 K.}
    \label{fig:EMIFAP1500fits}
\end{figure}

\begin{figure}
    \centering
    \includegraphics[width=4.5in]{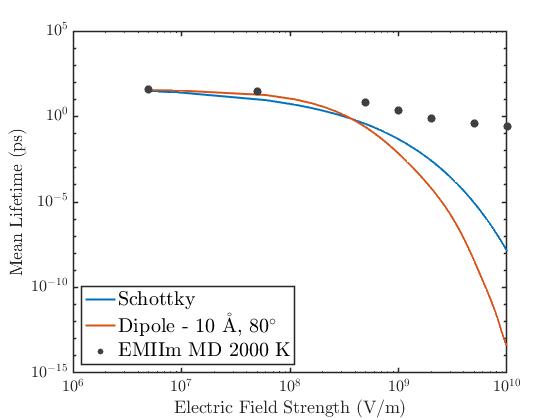}
    \caption{Best fits for the stationary dipole and Schottky models for EMI-Im at 2000 K.}
    \label{fig:EMIIm2000fits}
\end{figure}

Figures \ref{fig:EMIFAP1500fits}, \ref{fig:EMIIm2000fits}, \ref{fig:EMIFAP1000fits} show the best fits for the MD data for EMI-Im at 2000 K and EMI-FAP at 1000 K and 1500 K using the stationary dipole model and the Schottky model. The mean lifetimes for the dipole model were calculated using Equation \ref{DipoleWmin}. 

\begin{equation}\label{DipoleWmin}
    K = A exp(-\frac{1}{k T}(E_a + C - W_{min})) = \frac {1}{\tau}
\end{equation}

The constant C was used to match the fragmentation rate of the model to the lowest electric field case investigated with MD. The constant coefficient, A and the activation energy, $E_a$ for EMI-Im and EMI-FAP were taken from Reference \cite{Miller2019CharacterizationSources}. $W_{min}$ was calculated by numerically minimizing Equation \ref{WtotalBasicDipole}. The dipole distance parameters for the fits are 12 $\mathring{\mathrm{A}}$, 10.5 $\mathring{\mathrm{A}}$, and 10 $\mathring{\mathrm{A}}$ respectively. These chosen dipole separations were calculated using the MD data for mean max separation before fragmentation assuming that the maximum separation measured using MD occurred between the two ions with the same polarity in the cluster and that the distance between each of these ions and the ion of opposite polarity is the same. The dipole angle was varied to find the best fit to the data.

For EMI-FAP at 1000 K the dipole model fits the MD data better than the Schottky model between $5\times10^5$ V/m and $2\times10^8$ V/m. At electric fields above $2\times10^8$ V/m the Schottky model is a closer fit to the data than the dipole model. However, above $2\times10^8$ V/m the error in both of the fits is more than 100\%. For EMI-FAP at 1500 K the dipole model is again a better fit than the Schottky model between $5\times10^5$ V/m and $2\times10^8$ V/m however, there are larger errors between the dipole model and the MD results at electric fields above $2\times10^8$ V/m. For EMI-Im at 2000 K the dipole model fits better than the Schottky model between $5\times10^5$ V/m and $2\times10^8$ V/m, however, the difference between the MD results and the dipole predictions are larger than for EMI-FAP at 1500 K at lower electric fields. 

These results suggest that at higher temperatures the effects of geometry or forces not accounted for by the simple point charge model are larger. More simulations with lower temperatures would need to be run and fit with the dipole model to understand this better. Additional energy characterization could also help explain the differences seen between the different temperature conditions. Additionally, the results show that the dipole model in its current form results in too high a fragmentation rate for electric fields around $1\times10^9$ V/m. This electric field strength is typically experienced by emitted clusters for a short period immediately after emission. It is possible that some of the differences observed between the model predictions and the MD results are due to the inability of the MD to represent the fragmentation rates of clusters at high electric fields. As discussed in Section \ref{sec:FragDef} the four highest electric fields simulated resulted in many clusters fragmenting before the first check of the separation condition was performed. It is possible that the current method of detecting separations was not able to accurately determine fragmentation rates below 1 ps, and that the mean lifetimes for the highest four electric field conditions have been overestimated. However, if clusters fragmented in the region according to the rates predicted by the dipole model, then nearly all of the clusters would fragment immediately after emission. This indicates that while the dipole model better represents the fragmentation rates in regions of electric field between $5\times10^5$ V/m and $2\times10^8$ V/m it does not explain fragmentation behavior at electric fields greater than $2\times10^8$ V/m.


\chapter{Simulating Experimental RPA Data}

This section details the methodology and results of simulations of experimental RPA data. Simulations were performed using an N-body code developed by the Space Propulsion Lab SOLVEiT team as well as a lower fidelity, lower computational cost RPA simulator. The goal of these simulations was to determine the temperatures of the different clusters in the beam that result in fragmentation behavior most similar to that seen in experiments. Initial steps were taken to develop and test an approximate Bayesian computation routine to infer cluster temperature from experimental RPA curves using the lower fidelity RPA simulator. The next steps to validating this routine are also described.

\section{N-Body Simulation}

The N-body code used in this work to produce high-fidelity simulations of experimental RPA curves was developed by the Space Propulsion Lab SOLVEiT team \footnote{I would like to thank the SOLVEiT team for their work. In particular, Ximo Gallud Cidoncha, for his development of the EHD model that was used to generate electric fields used for the RPA simulator, Summer Hoss, for her development of the initial RPA post-processing routine, and Elaine Petro and Sebastian Hampl, for the development of the N-body simulator that was used to simulate RPA curves to compare to experimental results.}. This section explains the basic implementation of the N-body procedure as well as how it was applied to simulate RPA curves. Simulated RPA curves are compared to experimental RPA curves.

\subsection{N-Body Implementation}

The N-body simulation used for this work is a code that determines the evolution of an electrospray beam after emission by propagating individual particles in space and time. This propagation is performed by calculating the forces on each of the particles from the other particles in the beam as well as the force of the electric field applied between the emitter and the extractor. The initial conditions for the injection of the particles are based on electrohydrodynamic (EHD) models of electrospray menisci developed by Coffman and Gallud-Cidoncha \cite{Cidoncha2019InformingEvaporation,Coffman2016StructureIons}. The EHD model determines the shape of the liquid meniscus, the Laplacian electric field around the meniscus, the total current emitted from the meniscus, and the current density of emitted ions at different places on the meniscus. Particle injection is randomly selected from the probability density function defined by the EHD model. Once particles are injected they are propagated according to Newton's second law as given in Equation \ref{NewtonPropagation}. 

\begin{equation}\label{NewtonPropagation}
m_i\ddot{\mathbf{r}}_i=q_i\left(\mathbf{E}_{l}+\mathbf{E}_p\right)=q_i\mathbf{E}_{l}\left(\mathbf{r}_i\right) + \sum_{j} \frac{q_i q_j \left(\mathbf{r}_i - \mathbf{r}_j\right)}{4 \pi \varepsilon_0 |\mathbf{r}_i - \mathbf{r}_j|^3}
\end{equation}

The electric field in this case is the summation of the Laplacian electric field that comes from the application of voltage between the emitter and the extractor ($\mathbf{E}_l$), and the Poisson electric field that comes from the presence of the other charges in the ion beam ($\mathbf{E}_p$). The particle trajectories are integrated using a leapfrog scheme, which is further described in Reference \cite{Petro2019DevelopmentModeling}. The N-body code does not take into account any potential collisions that may occur between ions in the beam or collisions that may occur between ions or neutrals and the extractor grid surface. Development of these features is currently underway.

Cluster fragmentation is modeled as discrete events occurring each timestep in the simulation. The probability of a given ion cluster fragmenting during a timestep of duration $\delta t$ is given by the integration of the constant rate probability over the length of the timestep

\begin{equation}\label{dtfragProb}
p = 1 - e^{\frac{-\delta t}{\tau}}
\end{equation}

where $\tau$ is the mean lifetime of the cluster \cite{Petro2019DevelopmentModeling}. At the beginning of each timestep the probability of fragmentation is calculated from the mean lifetime of the cluster and the electric field at the location of the cluster. Some error in the fragmentation model is introduced by calculating the probability of fragmentation using the electric field at the beginning of each timestep. This error results in changes of less than 1\% in the final beam energy distribution when the timestep is smaller than 2 ps.

The N-body simulation accounts for fragmentation in dimers, trimers, and dimers formed from the fragmentation of trimers, also called trimer products. All clusters of one type are currently assumed to have uniform temperature. Trimers and trimer products were assumed to have the same temperature and electric field dependence as dimers because the MD data presented previously is not sufficient to predict trimer fragmentation behavior over the electric field range needed for these simulations. Simulations were repeated with different portions of the beam composed of each type of emitted species, monomers, dimers, and trimers. Different temperatures were also tested for each of the species. RPA curves were simulated for each of these tests to determine the fragmentation rates for each species that best matched experimental data. The details of the RPA post-processing are given in Section \ref{sec:RPApostProcess}. Results of this process are given in section \ref{sec:RPApostProcessresults}.

\subsection{Post-Processing RPA Implementation}\label{sec:RPApostProcess}

RPA curves were simulated by post processing the data from the N-body simulation. At the final timestep of the simulation the positions and velocities of all of the particles were recorded. Particles were then filtered by the distance travelled from the emission site. Particles in regions with electric fields higher than $10^7$ V/m were removed because the fragmentation behavior of these particles would depend heavily on the electric field strength. The $10^7$ V/m threshold was chosen because below this point the fragmentation rates of the clusters do not change more than 10 \% with a change in electric field. 

The remaining particles were propagated in space and time using Newton's second law assuming that the electric field is negligible. This assumption is supported by the low density of the beam resulting in negligible space charge effects and the decay of the Laplacian field that occurs by this point in the simulation. More data about the electric field as a function of distance from the emitter is given in section \ref{sec:RPAsim}. The detector geometry used was identical to that of the spherical RPA detector used to collect the experimental data. The radius of curvature was 8.85 cm and the solid angle was 90$^{\circ}$. The distance between the source and the center of the detector was taken to be 8.3 cm. The particles are propagated until they reached the detector surface. Any particles that did not reach the detector surface because their angle with respect to the ion beam center axis is too large are removed. The probability of field free fragmentation is calculated for the rest of the particles using Equation \ref{dtfragProb} using the time to reach the detector from the original location as the value of $\delta t$. Particles were selected randomly according to these probabilities to fragment. The neutral and charged products of the fragmentation were assumed to continue in the same direction and at the same velocity as the parent ion cluster \cite{Petro2019DevelopmentModeling}. 

The kinetic energy required to retard the particles at the detector is calculated using Equation \ref{retardingEnergy}

\begin{equation}\label{retardingEnergy}
    KE_{retard} = \frac{1}{2} m v_{\perp}^2
\end{equation}

where $v_{\perp}$ is the velocity perpendicular to the surface of the detector. More details about the effect of the detector geometry on the RPA curves is included in Appendix \ref{app:appc}. Particle energies are binned and normalized by the voltage applied to the source. The bins are summed such that the height of the RPA curve at each energy represents the number of particles in the beam with retarding energy less than that energy. The height of the curve is normalized by the number of particles reaching the detector. 

\subsection{Post-Processing RPA Results}\label{sec:RPApostProcessresults}

\begin{figure}
    \centering
    \includegraphics[width = 4in]{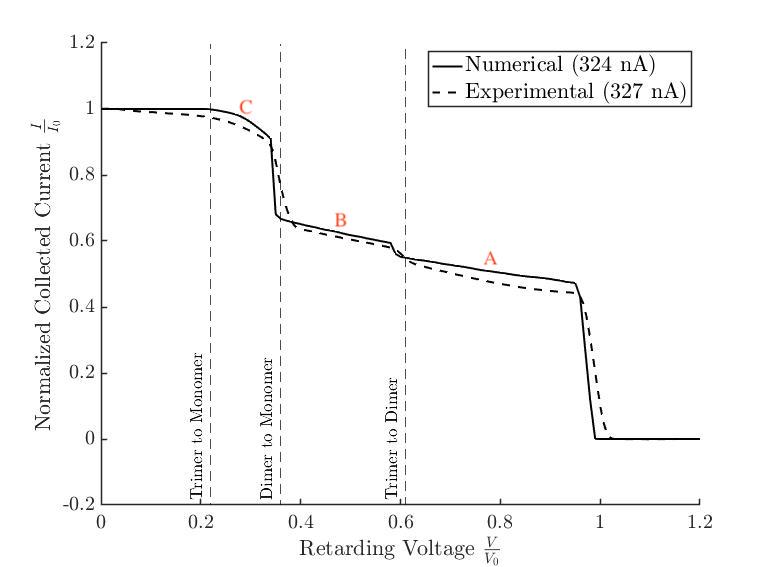}
    \caption{Comparison of numerical and experimental RPA curves for 324 nA and 327 nA respectively. Labeled points on the graph used for calculating error for ABC computation described further in Section \ref{sec:DistMetrics}}
    \label{fig:BESTSOLVEITRPA}
\end{figure}

\begin{figure}
    \centering
    \includegraphics[width = 4in]{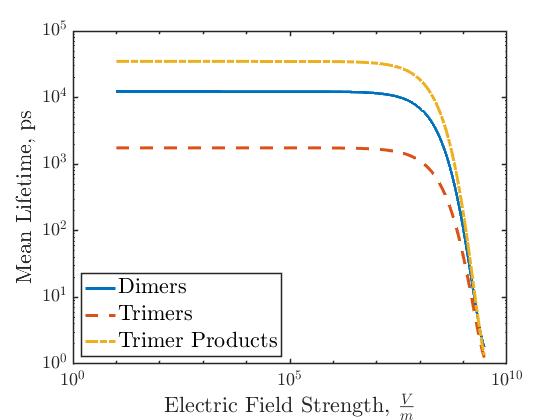}
    \caption{Optimal mean lifetimes as a function of electric field for different cluster types for 327 nA.}
    \label{fig:OPTMEANLIFESOLVEIT}
\end{figure}

Figure \ref{fig:BESTSOLVEITRPA} shows the numerical and experimental RPA curves for 324 nA and 327 nA respectively. Figure \ref{fig:OPTMEANLIFESOLVEIT} shows the mean lifetimes of each of the different cluster types that produced the simulated RPA that best matched the experimental data. This simulation was performed with 40\% monomers, 40\% dimers, and 20\% trimers.


\section{Approximate Bayesian Computation}\label{sec:ABC}

This section explains how approximate Bayesian computation is applied to determine the temperatures of different ionic liquid clusters and the percentage of the beam composed of each species using experimental RPA data.

\subsection{Bayesian Problem Formulation}

In this formulation of ABC the parameters that are being estimated are the temperatures of the different species present in the beam, $T_d$, $T_t$, and $T_{tp}$ for dimers, trimers, and trimer products. For this work the clusters of the same species are assumed to all have the same temperature. Section \ref{sec:massDist} presents results from inference routines developed for inferring the beam mass composition as well as the temperatures of the clusters. The forward model for the ABC routine is the simulation of an RPA curve similar to the post-processing done for SOLVEiT data. Details about the implementation of the RPA simulator are explained in Section \ref{sec:RPAsim}. The posterior distributions of the temperatures are investigated by calculating the average and standard deviation of the cluster temperatures as well as plotting the joint distribution. The L2 error as given in Section \ref{sec:DistMetrics} is calculated between the experimental RPA curve and the RPA curve simulated at the mean temperature values. These statistics attempt to represent the ability of the posterior distribution to represent the experimental data.

\subsubsection{Prior Distribution}

The prior distribution of the temperatures contains all of the information known about the temperatures of the clusters in the beam. Uniform and independent Gaussian priors are tested to determine which yields the estimated temperatures that result in the lowest L2 error. The uniform prior is given by U(300,1500). The lower bound of temperatures is given by the room temperature ions in the liquid before emission. Emitted clusters will not have energies lower than this. The higher bound is defined by investigation of the MD results. Temperatures higher than 1500 degrees result in fragmentation rates that would result in nearly all clusters fragmenting immediately after emission, which is not observed in experimental data. The Gaussian prior is given by three independent Gaussian distributions with means given by [1000,1000,500] and variance given by [1225,1225,100]. The mean temperatures are given by estimated temperature of dimer clusters from experimental data assuming an Arrhenius rate law for the dissociation of the clusters. The trimer product mean temperature is found by assuming equipartition of energy during fragmentation.

\subsubsection{Distance Metrics}\label{sec:DistMetrics}

The distance metric used to determine whether a given sample should be accepted has a large effect on the final posterior distribution. This work uses two different distance metrics, the custom point metric, and the L2 error metric. For the custom point distance metric the distance between the simulated and experimental RPA curves is defined as 

\begin{equation}
    \delta_{Custom} = (RPA_E(A) - RPA_i(A))^2 + (RPA_E(B) - RPA_i(B))^2 + (RPA_E(C) - RPA_i(C))^2
\end{equation}

where $RPA_i(A)$ is the RPA simulated from temperature sample i evaluated at point A. This metric selects for samples that have the same beam mass composition as the RPA curve value at points A, B, and C. The points A, B, and C were chosen to be the middle of the slopes between the monoenergetic, trimer, and dimer steps. These points are shown on the simulated RPA curve shown in Figure \ref{fig:BESTSOLVEITRPA}. The L2 error is defined as 

\begin{equation}
    \delta_{L2} = \sum_{k = 1}^{N} (RPA_E(k) - RPA_i(k))^2
\end{equation}

where N is the number of samples, k is the point on the RPA curve that is being evaluated, and $RPA_E$ is the experimental RPA curve.

\subsection{Forward Model: RPA Simulation}\label{sec:RPAsim}

\begin{figure}
    \centering
    \includegraphics[width=4.5in]{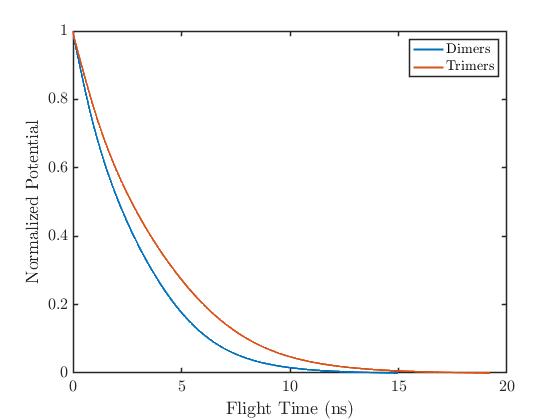}
    \caption{SOLVEiT potential field as a function of time for dimers and trimers.}
    \label{fig:SOLVEITpotential}
\end{figure}

\begin{figure}
    \centering
    \includegraphics[width=4.5in]{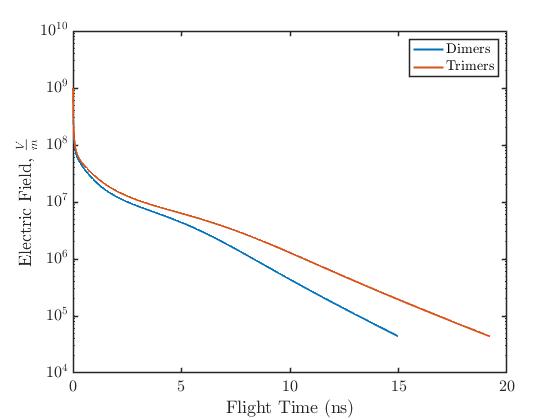}
    \caption{SOLVEiT electric field as a function of time for dimers and trimers.}
    \label{fig:SOLVEITEfield}
\end{figure}

The forward model for this ABC implementation is an RPA simulator. The inputs for the simulator that are fixed are the electric and potential field as a function of time for each species and the fragmentation rates of each cluster type. Figures \ref{fig:SOLVEITpotential} and \ref{fig:SOLVEITEfield} show the potential and electric field for dimers and trimers as a function of time as calculated by the SOLVEiT N-body code.

Fragmentation rates are calculated from the MD results that provide mean lifetimes as a function of temperature and electric field presented in Section \ref{sec:resultLifetime}. Comprehensive data is only available for dimers and as such this data was used to calculate fragmentation probability for all cluster types. Further work is needed to implement this simulator using MD results specific to each type of cluster. The inputs that are varied for each simulation in the ABC routine are the temperatures of each type of cluster and the proportion of the beam composed of each type of cluster.

The simulator first initializes lists of particles based on the beam mass composition input. Each cluster is given a temperature according to the species temperature inputs. The simulator propagates the clusters in time. At each timestep the probability of fragmentation is calculated using the temperature for each cluster and the electric field experienced at the beginning of the timestep as given by the SOLVEiT electric field input. The probability of fragmentation is calculated from the MD results. The potential at which the cluster fragments is recorded. Any trimers that fragment are added to the list of trimer products and are allowed to fragment again in subsequent timesteps according to the fragmentation rates for trimer products. Once all clusters have reached the end of the electric field region the probability of fragmentation in the field free region is calculated using the mean lifetimes of 1.49 $\mu s$ \cite{Miller2020MeasurementSources}. Further work must be done to characterize the field free fragmentation rates of different cluster types with MD. Clusters are randomly selected to fragment according to these probabilities and the potential of fragmentation is recorded as 0. 

Trial and error iterations with the RPA simulator indicated that in order to match the curvature of the RPA curve between the dimer to monomer and trimer to monomer steps the temperature of the trimers had to be between 1100 and 1400 K. This region corresponds to the second fragmentation of trimer products in the field free region. However, at these temperatures all of the trimers fragmented in the acceleration region. In contrast, experimental RPA curves show a well-defined step of trimers fragmenting into dimers in the field free region but not subsequently fragmenting into monomers. This could be due to very low energy trimers resulting from rapid fragmentation of low energy larger clusters very close to the emission site. To account for this step in the simulation 5\% of the beam was designated as trimers that will fragment into dimers in the field free region but not again into monomers. These trimers were separated from the trimers allowed to fragment in the acceleration region and added to the simulation at the end of the rest of the fragmentation location calculations.

Once all fragmentation locations have been determined the kinetic energies are calculated using Equations \ref{dimerStoppingPotentialAccelRegion}, \ref{trimerStoppingPotentialDimerOnly}, and \ref{trimerStoppingPotentialFull} depending on what the species is and what potentials it fragmented at. The kinetic energies are then binned. The bin heights are normalized by the total number of particles injected into the simulation and the kinetic energy values are normalized by the total source voltage applied multiplied by the elementary charge q. The cumulative sum of the bin heights is then plotted against kinetic energy to yield the RPA plot.

The part of the RPA curve to the left of the trimer to monomer step consists of the products of fragmentation of clusters larger than trimers at multiple locations in the acceleration and field free regions. To account for fragmentation of larger clusters not included in the simulation the part of the RPA curve to the left of the trimer to monomer step is artificially added at the end as a straight line from the trimer to monomer step to the zero-potential location. These simulations assume a large cluster beam percentage of 2\%, which is consistent with the experimental data used for comparison. Future work is necessary to include larger clusters in the simulation.



\section{Results}

\subsection{Prior Shape and Distance Metric}

\begin{figure}[htbp]
\begin{minipage}{.5\linewidth}
\centering
\captionsetup{width=.8\linewidth}
\includegraphics[width=1.1\linewidth]{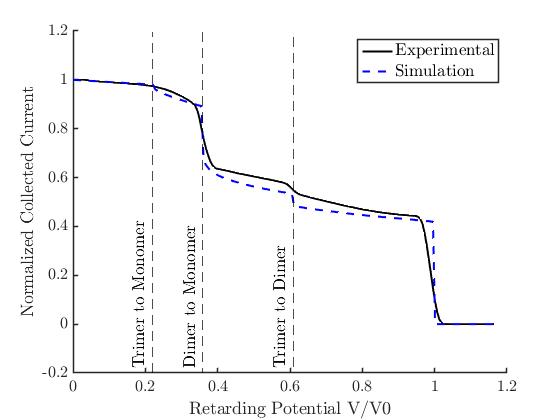}
\caption{Custom distance - Gaussian prior}
\label{fig:Custom_Gaussian_meanRPA}
\end{minipage}%
\begin{minipage}{.5\linewidth}
\centering
\captionsetup{width=.8\linewidth}
\includegraphics[width=1.1\linewidth]{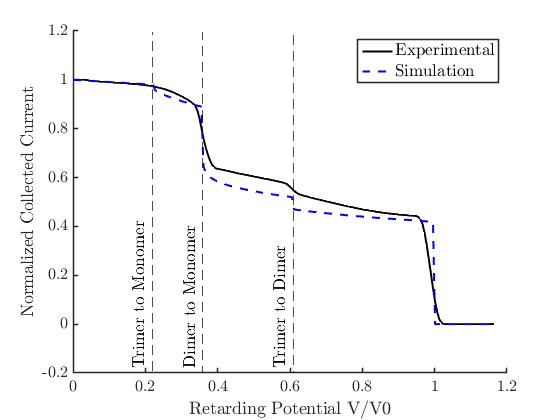}
\caption{Custom distance - uniform prior}
\label{fig:Custom_Uniform_meanRPA}
\end{minipage}\par\medskip
\end{figure}

\begin{figure}[htbp]
\begin{minipage}{.5\linewidth}
\centering
\captionsetup{width=.8\linewidth}
\includegraphics[width=1.1\linewidth]{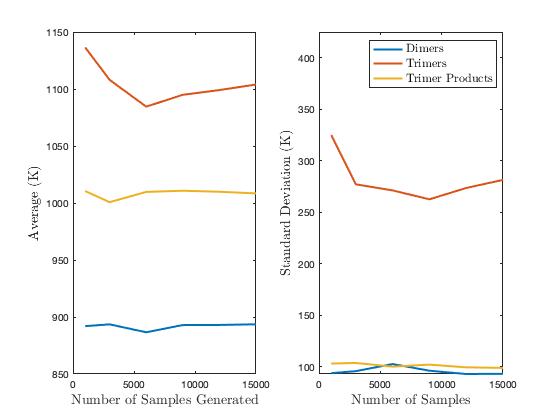}
\caption{Custom distance - Gaussian prior}
\label{fig:Custom_Gaussian_meanSTD}
\end{minipage}%
\begin{minipage}{.5\linewidth}
\centering
\captionsetup{width=.8\linewidth}
\includegraphics[width=1.1\linewidth]{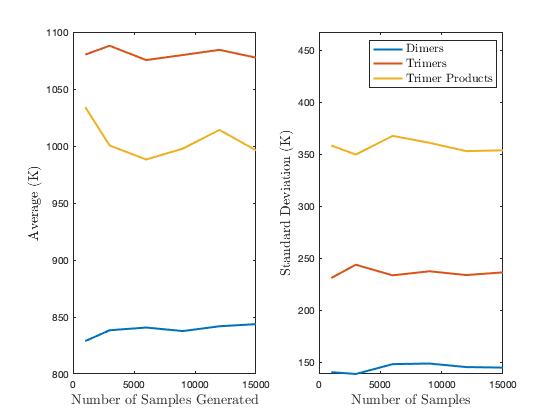}
\caption{Custom distance - uniform prior}
\label{fig:Custom_Uniform_meanSTD}
\end{minipage}\par\medskip
\end{figure}

\begin{figure}[htbp]
\begin{minipage}{.5\linewidth}
\centering
\captionsetup{width=.8\linewidth}
\includegraphics[width=1.1\linewidth]{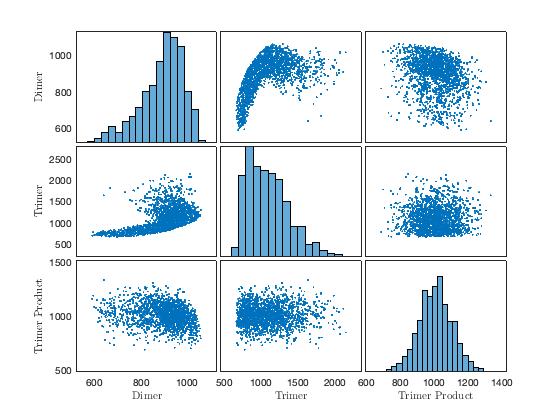}
\caption{Custom distance - Gaussian prior}
\label{fig:Custom_Gaussian_postDist}
\end{minipage}%
\begin{minipage}{.5\linewidth}
\centering
\captionsetup{width=.8\linewidth}
\includegraphics[width=1.1\linewidth]{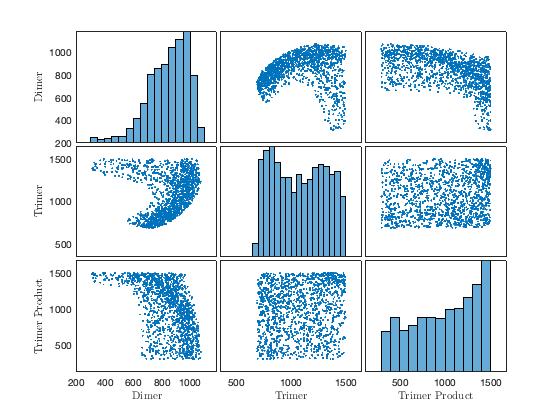}
\caption{Custom distance - uniform prior}
\label{fig:Custom_Uniform_postDist}
\end{minipage}\par\medskip
\end{figure}

\begin{figure}[htbp]
\begin{minipage}{.5\linewidth}
\centering
\captionsetup{width=.8\linewidth}
\includegraphics[width=1.1\linewidth]{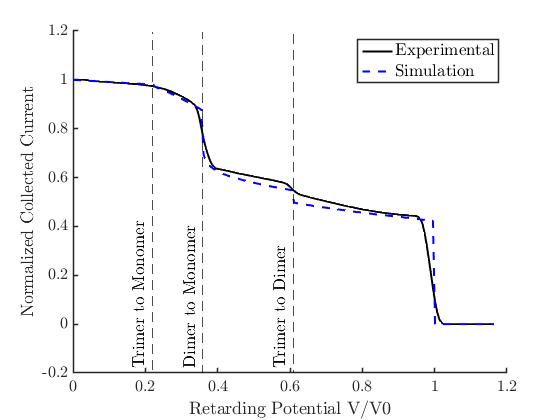}
\caption{L2 distance - Gaussian prior}
\label{fig:L2_Gaussian_meanRPA}
\end{minipage}%
\begin{minipage}{.5\linewidth}
\centering
\captionsetup{width=.8\linewidth}
\includegraphics[width=1.1\linewidth]{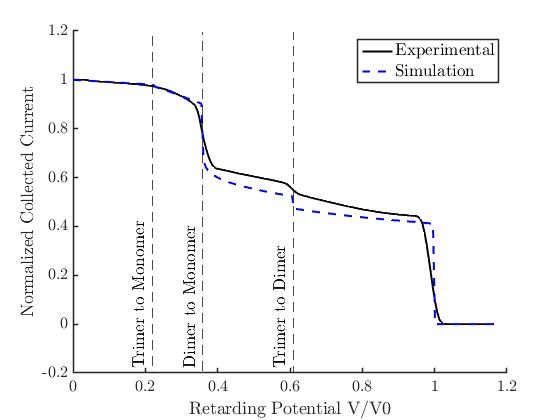}
\caption{L2 distance - uniform prior}
\label{fig:L2_Uniform_meanRPA}
\end{minipage}\par\medskip
\end{figure}

\begin{figure}[htbp]
\begin{minipage}{.5\linewidth}
\centering
\captionsetup{width=.8\linewidth}
\includegraphics[width=1.1\linewidth]{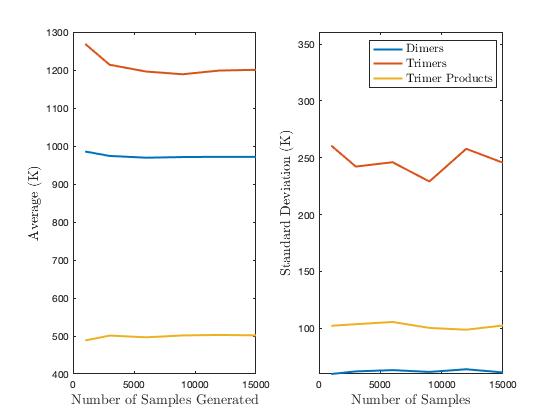}
\caption{L2 distance - Gaussian prior}
\label{fig:L2_Gaussian_meanSTD}
\end{minipage}%
\begin{minipage}{.5\linewidth}
\centering
\captionsetup{width=.8\linewidth}
\includegraphics[width=1.1\linewidth]{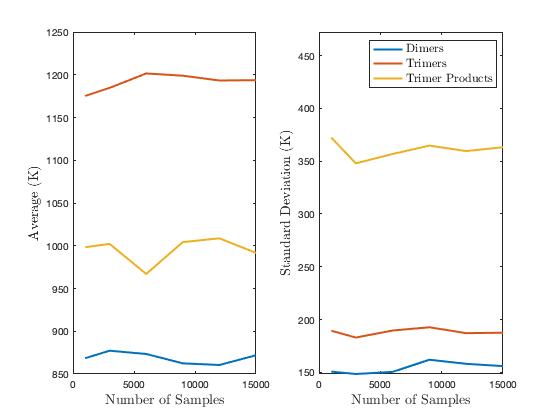}
\caption{L2 distance - uniform prior}
\label{fig:L2_Uniform_meanSTD}
\end{minipage}\par\medskip
\end{figure}

\begin{figure}[htbp]
\begin{minipage}{.5\linewidth}
\centering
\captionsetup{width=.8\linewidth}
\includegraphics[width=1.1\linewidth]{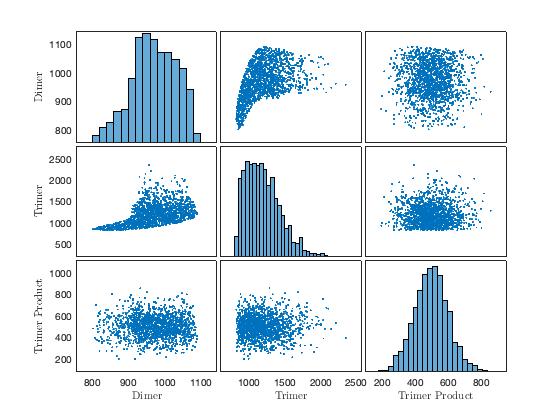}
\caption{L2 distance - Gaussian prior}
\label{fig:L2_Gaussian_postDist}
\end{minipage}%
\begin{minipage}{.5\linewidth}
\centering
\captionsetup{width=.8\linewidth}
\includegraphics[width=1.1\linewidth]{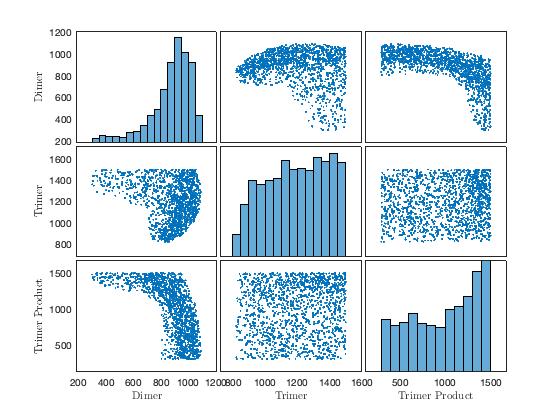}
\caption{L2 distance - uniform prior}
\label{fig:L2_Uniform_postDist}
\end{minipage}\par\medskip
\end{figure}

Figures \ref{fig:Custom_Gaussian_meanRPA} through \ref{fig:L2_Uniform_postDist} show the mean and standard deviation of the cluster temperatures, the simulated RPA curve for the mean temperatures, and the posterior distribution of the temperatures for various combinations of prior shape and distance metric. Plots next to each other have the same distance metric with a different prior. The custom distance prior selects for agreement between simulated and experimental RPA curves at the center of the slopes between the vertical steps. While this distance metric may work for completely accurate RPA simulations, the error introduced into this RPA simulation due to the on-axis emission assumption mean that selecting for the curves to coincide at these points yields very dissimilar curves. When using the L2 distance metric the Gaussian prior results in better informed posteriors. The L2 error for the mean of the posterior resulting from using a Gaussian prior is lower than for the uniform prior and the RPA curve can be seen to agree with the experimental curve at more than just points A, B, and C. 

It is possible but not likely that with a more accurate RPA simulation the custom distance metric might approach the behavior of the L2 distance metric. This would only happen if specifying the exact height of the RPA curve at points A, B, and C also specified the slopes of the curve between those points. From trial-and-error attempts at fitting temperatures it was observed that both temperature and beam composition have effects on both the height of the curves in each location and the slopes. The beam mass composition has a larger effect on the height of the curve because of the presence of the monoenergetic step composed entirely of monomers. The slopes are determined more by the temperatures and thus the fragmentation rates of the different types of clusters. The standard deviation of the posterior is smaller for the Gaussian prior, which is a result of the shape of the prior and the influence it has on the shape of the posterior. 

One interesting result from testing these two priors is evidence that large differences between the mean temperatures of the clusters can result in a small difference in the simulated RPA curves. This suggests that there may be multiple groups of temperatures for the clusters that yield similarly low L2 errors in the simulated RPA. More investigation with less specific prior distributions would be needed to determine whether this is actually the case. One potential test would be to look at the L2 error at more specific locations in the space of temperatures of the posteriors for the Gaussian and uniform priors and compare them.

\subsection{Acceptance Probability}

\begin{figure}[htbp]
\begin{minipage}{.5\linewidth}
\centering
\captionsetup{width=.8\linewidth}
\includegraphics[width=1.1\linewidth]{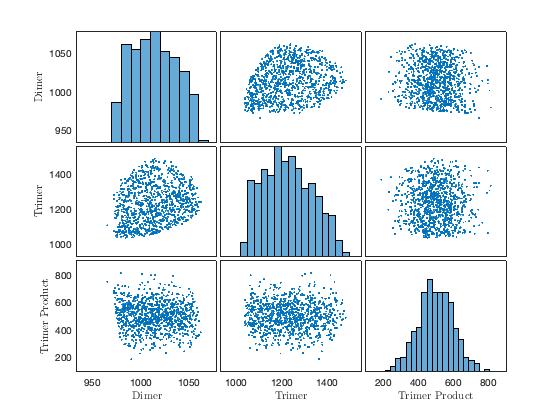}
\caption{2.3\% acceptance}
\label{fig:L2_Gaussian_postDist_AcceptA}
\end{minipage}%
\begin{minipage}{.5\linewidth}
\centering
\captionsetup{width=.8\linewidth}
\includegraphics[width=1.1\linewidth]{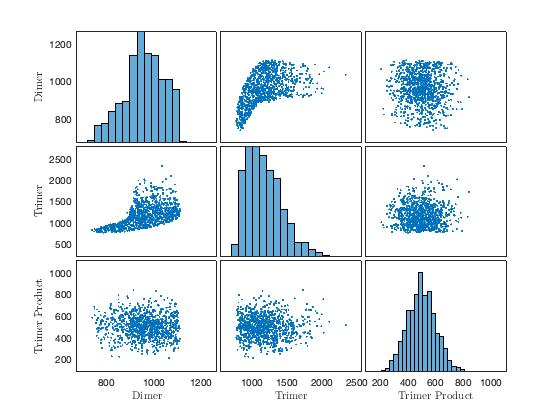}
\caption{14.7\% acceptance}
\label{fig:L2_Gaussian_postDist_AcceptF}
\end{minipage}\par\medskip
\end{figure}

\begin{figure}[htbp]
\begin{minipage}{.5\linewidth}
\centering
\captionsetup{width=.8\linewidth}
\includegraphics[width=1.1\linewidth]{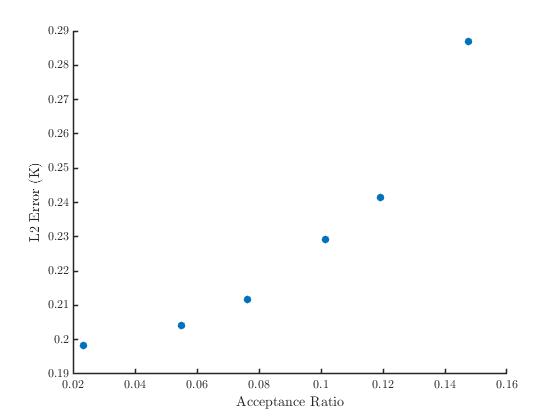}
\caption{L2 Error for different acceptance probability}
\label{fig:L2_Gaussian_Accept_L2Error}
\end{minipage}%
\begin{minipage}{.5\linewidth}
\centering
\captionsetup{width=.8\linewidth}
\includegraphics[width=1.1\linewidth]{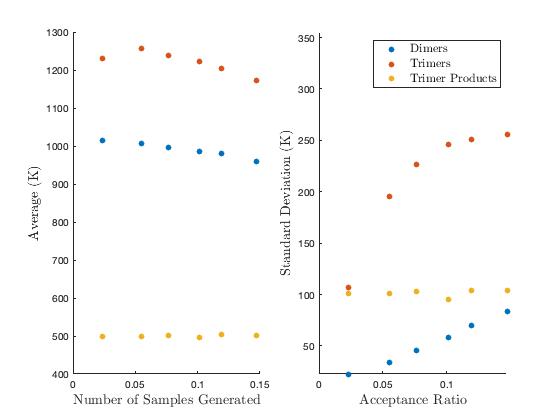}
\caption{Mean and standard deviation for different acceptance probability}
\label{fig:L2_Gaussian_Accept_meanSTD}
\end{minipage}\par\medskip
\end{figure}

\begin{figure}[htbp]
\begin{minipage}{.5\linewidth}
\centering
\captionsetup{width=.8\linewidth}
\includegraphics[width=1.1\linewidth]{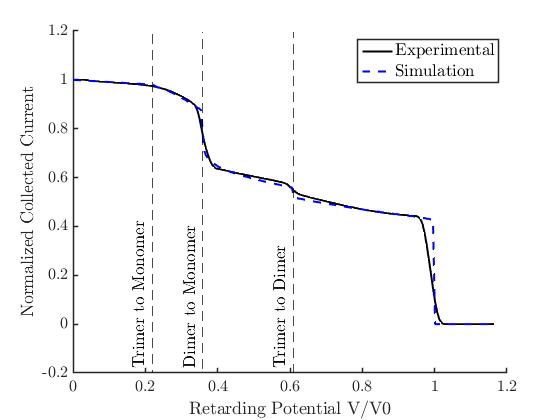}
\caption{2.3\% acceptance}
\label{fig:L2_Gaussian_meanRPA_AcceptA}
\end{minipage}
\begin{minipage}{.5\linewidth}
\centering
\captionsetup{width=.8\linewidth}
\includegraphics[width=1.1\linewidth]{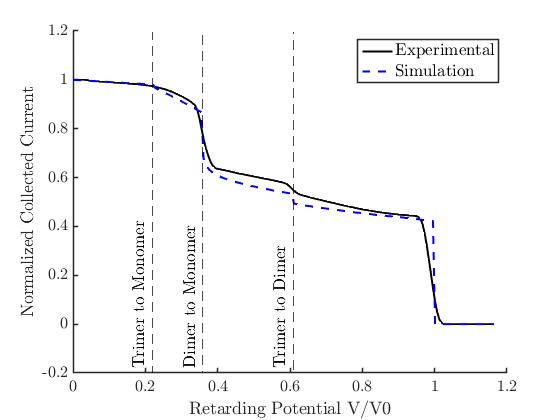}
\caption{14.7\% acceptance}
\label{fig:L2_Gaussian_meanRPA_AcceptF}
\end{minipage}\par\medskip
\end{figure}

\ref{fig:L2_Gaussian_postDist_AcceptA} and \ref{fig:L2_Gaussian_postDist_AcceptF} show the posterior distributions of the cluster temperatures for the L2 distance metric with a Gaussian prior for acceptance probabilities of 2.3\% and 14.7\%. \ref{fig:L2_Gaussian_Accept_L2Error} shows the L2 error between the experimental and simulated RPA curves for the varying acceptance probabilities. \ref{fig:L2_Gaussian_Accept_meanSTD} shows the mean and standard deviation of the cluster temperatures for the varying acceptance probabilities.  \ref{fig:L2_Gaussian_meanRPA_AcceptA} and \ref{fig:L2_Gaussian_meanRPA_AcceptF} show the simulated RPA curves for the mean cluster temperatures for acceptance probabilities of 2.3\% and 14.7\%. Appendix \ref{app:appb} includes posterior distributions and mean RPA curves for the other acceptance probabilities. 

Having a lower acceptance probability yields smaller L2 errors in the mean simulated RPA. However, the decrease in the error gets smaller as the acceptance probability gets smaller. This is because a significant portion of the error is due to inaccuracies in the forward model such as the effects of energy spreading. Thus, reducing the acceptance probability below 2 \% is unlikely to result in a further decrease in the L2 error. The reduction in the shape of the joint distributions is not uniform across the clusters. As seen in Figure \ref{fig:L2_Gaussian_meanRPA_AcceptA} the 2.3\% acceptance results in simulated results almost identical to the experimental results. Differences primarily occur at the corners of the monoenergetic and field free fragmentation steps, which is a result of the forward model not including energy spreading effects.

\subsection{Inferring Beam Mass Composition}\label{sec:massDist}

\begin{figure}
    \centering
    \includegraphics[width=1\linewidth]{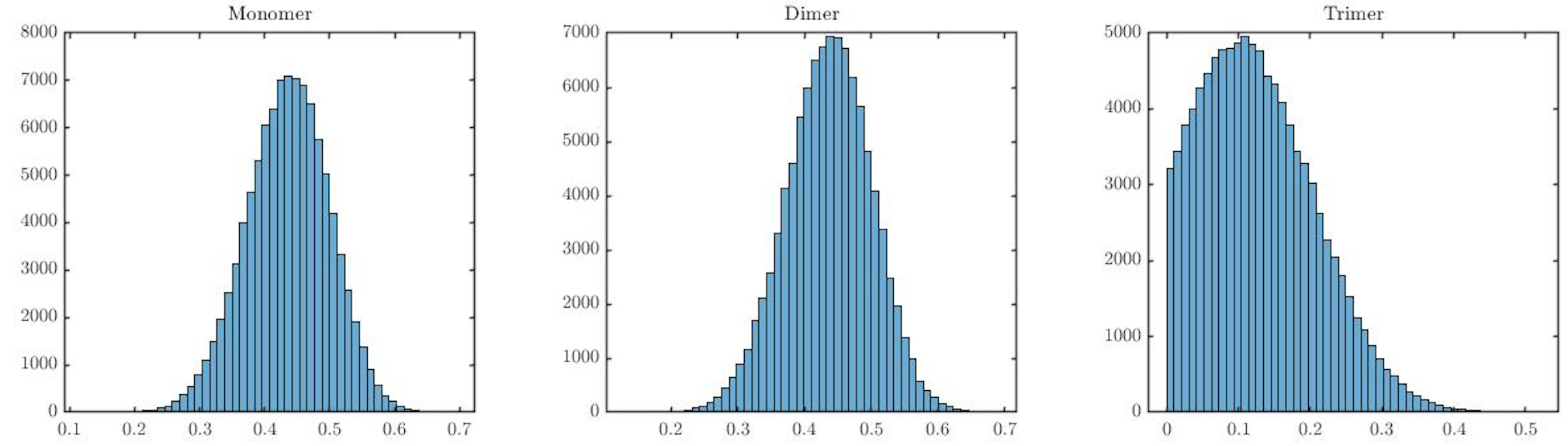}
    \caption{Prior distributions for the beam percentages of each of the three species accounted for in the RPA simulation.}
    \label{fig:massDist_PRIOR}
\end{figure}

Time of flight data can be collected immediately after RPA data to obtain the beam mass composition to use for inferring the cluster temperatures from the RPA curve. However, high resolution RPA data can only be collected with a partial beam apparatus. Results from SOLVEiT simulations as well as experimental data indicate that the beam mass composition varies strongly with the angle that the collected part of the beam makes with the central axis of the beam \cite{Perez-Martinez2015IonMicrotips}. The beam mass distribution from the previous section was determined using trial and error to demonstrate that the inference method could be used to determine the temperatures that minimized the L2 error. However, it is also possible to use the RPA data to infer the most likely beam mass composition. To do this two independent Gaussian distributions with mean 0.45 and variance 0.005 were used as the prior distribution for the monomer and dimers populations. The trimer population was taken to be the portion of the beam left. If samples of the monomer and dimer prior were taken and they added up to more than 1 a new set of samples was taken. Figure \ref{fig:massDist_PRIOR} shows the resulting prior distribution of the monomer, dimer, and trimer populations. 

\begin{figure}
    \centering
    \includegraphics[width=4.5in]{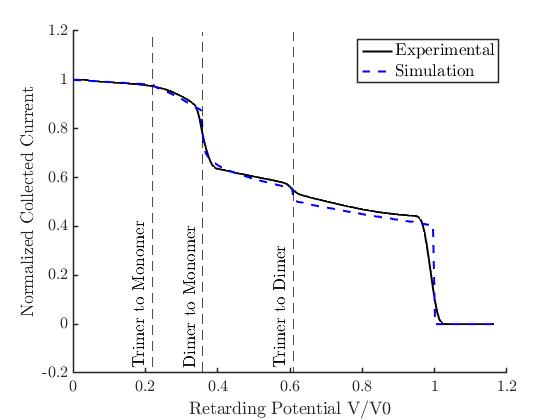}
    \caption{Mean RPA graph resulting from inferring the beam mass distribution and the cluster temperatures simultaneously.}
    \label{fig:meanDist_meanRPA}
\end{figure}

\begin{figure}
    \centering
    \includegraphics[width=5in]{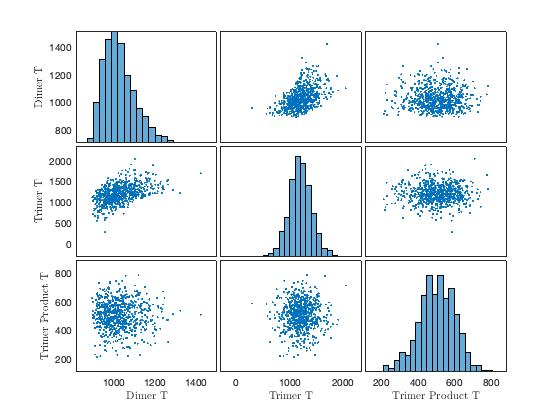}
    \caption{Temperature posterior resulting from inferring the beam mass distribution and the cluster temperatures simultaneously.}
    \label{fig:meanDist_postDistT}
\end{figure}

\begin{figure}
    \centering
    \includegraphics[width=5in]{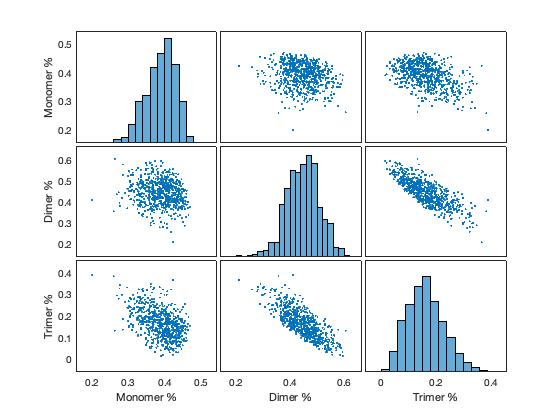}
    \caption{Beam percentages resulting from inferring the beam mass distribution and the cluster temperatures simultaneously.}
    \label{fig:meanDist_postDistM}
\end{figure}

Figure \ref{fig:meanDist_meanRPA} shows the mean RPA resulting from the simultaneous inference of the beam mass distribution and the cluster temperatures. The temperature prior distribution was independent Gaussian and the distance metric was the L2 error.  The acceptance level was 2.5\% and 1000 samples in the posterior were generated. Figure \ref{fig:meanDist_postDistT} shows the posterior distribution of the temperatures and \ref{fig:meanDist_postDistM} shows the posterior distribution of the beam percentages of the different emitted species. 

The resulting mean temperatures for the dimers, trimers, and trimer products are 1028 K, 1203 K, and 505 K. The mean beam percentages for monomers, dimers, and trimers are 39\%, 45\%, and 16\%. For similar acceptance levels the routine that inferred both the beam mass distribution and the cluster temperatures had a higher L2 error than the routine that used the constant beam mass distribution determined by trial and error. However, the routine inferring all 6 parameters of the problem was able to represent the majority of the behavior in the RPA curve. It is possible that decreasing the acceptance level further would result in better agreement between the simulated and experimental RPA curves.


\chapter{Conclusions and Future Work}

\section{Conclusions}

Ionic liquid electrospray emission is a promising technology that can be used for applications from space propulsion to microetching. Fragmentation of ionic liquid clusters has significant effects on the performance of electrospray emission for all applications. Characterization of fragmentation behavior in regions with electric field is especially important for understanding the effects of fragmentation on electrospray performance. Previous investigations of ionic liquid cluster fragmentation have focused on regions with no electric field or clusters with temperatures well below those expected from electrospray emission sources. The goal of this work was to use various computational methods to simulate important characteristics of ionic liquid cluster fragmentation. This included determining the effect of the electric field and temperature on fragmentation and finding the electrospray beam properties such as cluster temperature and beam mass distribution that best matched experimental data.

Molecular dynamics simulations were performed to characterize fragmentation of various ionic liquid clusters under different temperature and electric field conditions. Positive EMI-FAP and EMI-Im dimers, positive and negative EMI-BF$_4$ dimers, and positive EMI-BF$_4$ trimers were simulated for temperatures between 300 K and 2500 K and electric field strengths between $5\times10^5$ V/m and $1\times10^10$ V/m. The mean lifetime of the clusters was determined for each of the sets of conditions. Mean lifetime decreased for increasing temperature and electric field. In the limit of large electric field, the effect of the electric field dominated the effect of the temperature. In the limit of low electric field, the effect of the electric field was negligible compared to the effect of the temperature. Larger clusters dissociated faster at the same temperature conditions than did the dimers. Negative polarity EMI-BF$_4$ dimers dissociated faster than positive polarity dimers at the same temperature, however, it was unclear if this was due to a difference in the energy barrier for fragmentation for the two polarities or due to the different energy content the two polarities have at the same temperature. At the same temperature conditions EMI-BF$_4$ fragmented the fastest, EMI-Im the second fastest, and EMI-FAP the slowest. This was contrary to the previous evidence presented by Coles and Miller. Further work must be done to determine whether this effect was due to the different energy content of the ionic liquids at the same temperature or a difference in the energy barrier to fragmentation depending on the size of the anion of the liquid. 

The fragmentation pathways were determined for each set of conditions. Results show that the percentage of clusters undergoing total fragmentation is negligible for all clusters. For trimers, the percentage of clusters undergoing single neutral evaporation is higher when the electric field and temperature are lower, however more work is needed to characterize the separation behavior of clusters that did not match any predefined pathway. Simulations were performed with time varying electric fields, which support the conclusion that ionic liquid cluster fragmentation is a constant rate process. 

A physics-based model was developed to approximate the fragmentation behavior of dimers as a stationary dipole and an escaping ion. The model was fit to the MD results by varying the geometrical parameters of the dipole. While the model was able to fit the MD results more accurately than the Schottky model could for electric fields lower than $2\times10^8$ V/m, the fit to the data at the higher electric fields was worse than for the Schottky model. Further work is necessary to determine the geometry of clusters during fragmentation and to compare this geometry to the dipole model. 

Approximate Bayesian computation was used to determine the temperatures of the different clusters and the portion of the beam composed of each type of cluster from a given experimental RPA curve and the MD results for mean lifetimes of EMI-BF$_4$ dimers. An RPA simulator was developed that uses a physics-based approach to generating experimental RPA curves similar to results from full N-body simulations but fast enough to be used in inference routines. An inference routine was developed and tested with various prior distributions and distance metrics. The independent Gaussian prior and the L2 error distance metric resulted in the lowest L2 error between the mean simulated RPA curve and the experimental data. The beam was determined to have 39\% monomers, 45\% dimers, and 16\% trimers and the temperatures of the dimers, trimers and trimer products were 1028 K, 1203 K, and 505 K respectively. Further work is necessary to expand these methods to determine beam properties based on trimer and trimer product MD simulations. Further work is also needed to expand the simulation to include non-uniform temperature distributions for each type of cluster. 

\section{Future Work}

This section details some of the future work that could be performed with MD simulations, computational methods, and experiments to further understand ionic liquid fragmentation behavior.

\subsection{Molecular Dynamics}

While the data provided in this work indicate general trends in ion cluster fragmentation rates under different conditions, more MD simulations are needed to complete this characterization. The following sections explain some of the improvements that could be made to the fragmentation simulations as well as emission simulations that could provide additional data about the energy distribution of electrospray ion beams.

\subsubsection{Fragmentation Simulations}\label{sec:FutureMDfrag}

More fragmentation simulations will be performed with additional ionic liquids that are of interest for electrospray applications including BMI-I and EMIFSI. Additionally, fragmentation simulations will be performed on various types of clusters that result from the addition of various salts to ionic liquids. For example, there is growing interest in the use of Lithium salts with various ionic liquids \cite{Smith2015InfluenceLiquids}. The use of these salts has been shown to increase the monomer population in an electrospray beam. Understanding the fragmentation behavior of these clusters would support the considerations on how to use these mixtures for propulsion as well as other applications.

More detailed investigations of the geometry of clusters during fragmentation will also be performed. In particular, the time dependent relative position of components of the clusters during the fragmentation process will be investigated to improve the analytical physical model developed in this work. This will included determining whether there is a temperature and electric field dependence of the orientation of the molecules in order to compare with the work done by Prince et al. and Roy et al. \cite{Prince2015MolecularNanodrops,Roy2020Gas-PhaseClusters}. In addition to improving the analytical physical model of dimers presented here, this geometry characterization will be used to extrapolate to behaviors for larger clusters, which could reduce the need to perform computationally expensive MD simulations for each type of cluster emitted. Additional energy characterization will also be performed. Similar to the work done by Prince et al. the energy changes of the clusters during the fragmentation process will be examined to determine the effect of initial fragmentations on the subsequent thermal stability of clusters \cite{Prince2015MolecularNanodrops}.

\subsubsection{Emission Simulations}

Emission simulations should be performed to better characterize the solvation energy of ion clusters. From experimental RPA data it is known that there is an approximately 3 to 10 V difference between the voltage applied to an electrospray emitter and the voltage to which the ions are accelerated. This difference, $\delta V$, is likely distributed between the solvation energy of the ions and ohmic heating of the liquid meniscus \cite{Miller2020MeasurementSources}. However, the exact values of the solvation energies of various ionic liquid clusters are not known. Emission simulations could be used to determine the solvation energies of ion clusters by measuring the energy of clusters before and after emission \cite{Prince2015MolecularNanodrops}. These simulations could also help determine the initial energy distribution of the clusters in the beam, providing an opportunity to compare to the results of the ABC algorithm. Finally, emission simulations could indicate the conditions that support various beam mass compositions. The proportion of the ion beam composed of each cluster size depends on the firing conditions. Previous work has shown that emitter geometry, firing voltage, emitted current, and ionic liquid temperature all have effects on the beam mass composition but the exact dependence on each of these factors is not known \cite{Miller2015OnSources,Miller2019CharacterizationSources}.

\subsection{Approximate Bayesian Computation}

The work with ABC presented previously demonstrates that it will be a valuable tool in determining the energies of emitted ions and their fragmentation behavior from otherwise difficult to interpret experimental data. However, many improvements could be made that would increase the capability of the routine. 

First, the RPA simulator must be improved to be more accurate. The future simulator should take into account the spherical geometry of the detector as the SOLVEiT post-processing does to improve the accuracy. The simulator could also be made faster by determining the RPA curve shape as a function of fragmentation probability using the results from Miller \cite{Miller2019CharacterizationSources}. Using these equations in addition to beam information from experiments and SOLVEiT simulations the computationally expensive simulation of fragmentation of individual particles could be eliminated. Further work must be done to determine the applicability of these equations for the different regions of the RPA curve.

Once the RPA simulator has improved accuracy and run time, it must be updated to account for distributions in cluster energies. The current simulator assumes that all clusters of the same species have the same constant temperature and fragmentation behavior as a function of electric field. However, MD emission simulation results from Coles show that emitted species likely have a Maxwellian energy distribution \cite{Coles2012InvestigatingBeams}. Thus, the RPA simulator should be updated to simulate RPA curves with each species having a different energy distribution with certain parameters such as temperature for Maxwellian distributions or mean and standard deviation for Gaussian distributions. Then, the goal of the ABC routine would be to determine the parameters of the energy distribution. The simulator should also be updated to condition the beam energy and species distribution based on both experimental RPA and TOF data. Appropriate distance metrics should be developed to use both RPA and TOF data to best predict the desired quantities. A TOF simulator must be designed with similar considerations as the current RPA simulator, in particular the effect of the partial beam TOF detector designs. 

Once the ABC routine building blocks have been improved, full validation of the method must be performed. The ability of the routine to converge on the correct energy and species distribution parameters should be tested using experimental data curves generated in SOLVEiT. The dependence of the convergence on the initial guess for the desired parameters, the number of samples taken, and the distance metric used should be investigated. The effect of reducing the accuracy of the simulator in favor of a faster run time should also be tested. Finally, the routine should be run for different ionic liquids at different conditions and compared to SOLVEiT simulations to understand the implications on the emission energies of different experimental configurations.

\subsection{Experiment}

While MD simulations provide insight into fragmentation rates of ionic liquid clusters under different conditions, experimental results are needed to confirm the simulation results. Additional experiments could also help determine the exact conditions of ion clusters emitted from electrospray sources. The following sections detail some of the experiments that could be performed to better understand ionic liquid fragmentation behavior.

\subsubsection{Simultaneous RPA and TOF}

As discussed in section \ref{sec:ABC} current data does not allow for validation of the ABC algorithm. In particular, current simultaneous TOF and RPA data uses a partial beam TOF design. As shown by the SOLVEiT simulations and experimental work, TOF curves that use only the center of the beam measure a smaller population of trimers and larger clusters than exists in the full beam \cite{Perez-Martinez2015IonMicrotips}. Simultaneous RPA and TOF measurements with full beam instruments would be needed to provide data to accurately determine the ability of the ABC algorithm to determine the ion cluster energy distribution given TOF and RPA data.

\subsubsection{Charge to Mass Ratio}

TOF and RPA curves give information about the beam mass and energy composition. However, resolving fragmentation in high electric field regions very close to the emission site is difficult. The energy resolution of the spherical RPA used by Miller to determine field free region fragmentation rates is 5\% \cite{Miller2019CharacterizationSources}. Given a startup voltage of 1500 V this can detect fragmentation potential within 75 volts. Thus, the monoenergetic step in the RPA curve could include products of fragmentations that occur within the first 75 volts of the emission site. Fragmentations in this region would produce very low energy neutrals as the clusters will have only been accelerated through the 75 V potential. A neutral resulting from dimer fragmentation in this region would travel at less than $6.8\times10^3$ m/s, almost an order of magnitude slower than the typical $3.0\times10^4$ m/s for a fully accelerated dimer. Neutrals with velocities this low could accumulate on the extractor grid, causing damage to the hardware or an electrical short.

Measuring the charge to mass ratio of the complete ion beam would indicate the total amount of fragmentation occurring in the beam, including the region very close to the emission site. The total mass flow of the beam includes both the ions that produce the emitted current as well as the neutrals that result from fragmentation. The current provides the total number of ions emitted from the source. Using RPA and TOF curves the amount of fragmentation occurring immediately after emission could be estimated by subtracting the neutrals that result from fragmentation that appears clearly in the RPA and TOF curves from the total neutral population. 

As mentioned previously determining the total charge emitted from an electrospray thruster is the same as measuring the output current. This could be resolved accurately by placing the emitter less than a centimeter from a Faraday cup. The current from the Faraday cup would be amplified and monitored. The mass flow of the emitter during the firing process has previously been measured by taking the mass of the emitter before and after firing. However, this method poses some difficulties. For a single emitter the change in mass is very small, even when firing for a long period of time, and is thus difficult to measure accurately. Mass measurements before and after firing a full thruster with a fuel tank have been performed. However, emission from multiple tips makes it difficult to characterize fragmentation as the behavior of different tips could result in different individual beam mass and energy compositions. A possible solution to this problem is measuring the mass flow of a single emitter by visually monitoring the fluid. A droplet of ionic liquid would be placed on a disc mounted on a cross post on the tungsten tip. The electrical connection would be made to the liquid using a small platinum wire placed just inside the meniscus but not touching the tungsten. This would eliminate any possible electrochemistry that could be induced by applying voltage directly to the tungsten tip. The size of the ionic liquid meniscus on the tip would be monitored using a camera outside of the chamber over time as the emitter is fired, providing the mass flow of the emitter over time. This could also be combined with periodic TOF and RPA analysis. This data could then be analyzed to determine the amount of fragmentation happening within 75 volts of the emission site.

\subsubsection{High Resolution RPA}

The amount of fragmentation occurring just after emission could also be determined by using an RPA with an energy resolution of 2.5\%. This would allow for resolution of fragmentation events within 38 volts of emission. Previous work has shown that RPA instruments with energy resolution of 2.5\% can be built using multiple thinly spaced tungsten grids to apply the retarding voltage and repel secondary electrons \cite{Lozano2006EnergySource}. Additionally, shielding the source using anodized aluminum has been shown to improve RPA results considerably \cite{Miller2019CharacterizationSources}. Unfortunately, designs that give the high energy resolution desired are often planar. To avoid the effects of beam spreading they must only sample a small portion of the beam. High energy resolution RPA data from multiple angular portions of the beam would need to be gathered and meshed together to understand the behavior of the entire beam. This process might also provide an opportunity to validate results of SOLVEiT simulations as it would provide fragmentation behavior as a function of the spatial location in the beam.

\subsubsection{High Resolution TOF}

High energy resolution TOF measurements could also help resolve fragmentation rates in high field regions. Miller provided the probability of fragmentation in the acceleration region as a function of the TOF data as given in Equation \ref{TOFresolution} \cite{Miller2019CharacterizationSources}. However, this requires information about the initial velocity of the clusters and the number of clusters per unit length at emission, which are not known. These equations could be used in tandem with SOLVEiT simulations to determine which fragmentation rates best match the experimental data. 

Distinguishing fragmentation in the high electric field region would require very high-resolution time of flight measurements. The resolution of a TOF curve as related to the potential at which a cluster fragments is given by Equation \ref{TOFresolution}

\begin{equation}\label{TOFresolution}
    \frac{\Delta t}{t} = \frac{1}{2}\frac{V_0-V_B}{V_0}
\end{equation}

where $\Delta t$ is the time resolution of the TOF apparatus, t is the flight time of the ion, $V_0$ is the voltage applied to the source, and $V_B$ is the potential at which the ion cluster fragments \cite{Miller2019CharacterizationSources}. To resolve fragmented species energy within 2.5\% the time resolution needs to be approximately 1.25\%. For an EMI monomer with a source voltage of 1.5 kV and a detector that is 1.1 meters from the source the flight time is 21 $\mu s$. Thus, the time resolution needed to differentiate this EMI monomer from the dimers that fragmented within 38 V of the emission site is 270 ns. The resolution of the TOF apparatus depends both on the geometry of the device and the response of the electronics. To meet this resolution the angular spreading effect must be small enough that the difference between the arrival times of clusters at the widest angle and the clusters on axis must be much less than the desired resolution time in ns. The bandwidth of the amplifier of the CEM signal must be large enough that the rise time is shorter than the desired time resolution in ns.

\subsubsection{Quadrupole RPA}

Finally, work will be done to develop a quadrupole RPA apparatus. One of the biggest difficulties in interpreting the fragmentation in high electric field regions is that data for different species shows up in the same place on the RPA curve. While this problem doesn't exist for TOF curves, data on the initial velocity of clusters and the number of clusters per unit length is not known. A possible experimental extension to this work would be to get RPA curves for each of the species individually. This has been done previously by Miller et al. with capillary electrosprays operating in the mixed ion droplet regime \cite{Miller2017OrthogonalProfile}.

\renewcommand{\thechapter}{A}
\renewcommand{\chaptername}{Appendix}
\chapter{Selected Supplemental Molecular Dynamics Data}

This section contains some selected supplemental MD data from various ionic liquids and cluster sizes. Any conditions for which the result is -1 were not simulated.

\section*{EMI-BF$_4$ Positive Dimers}

\subsection*{Mean Lifetime for Different $\delta_{post}$}

\begin{table}[H]
\caption{EMI-BF$_4$ positive dimers mean lifetimes for $\delta_{post} = 10 \mathring{\mathrm{A}}$ calculated by averaging.}
\label{tab:EMI-BF$_4$posDimMeanLife10}
\begin{adjustwidth}{-.65in}{-1in}
\begin{centering}
\begin{tabular}{|l|l|l|l|l|l|l|l|l|l|l|}
\hline
E (V/A) & 0.8               & 0.3               & 0.15             & 0.1              & 0.05             & 0.02             & 0.01             & 0.005            & 0.0005           & 0.00005          \\ \hline
T (K)   &    &    &    &    &      &     &    &      &    &  \\ \hline
300   & 0.387 & 4.080  & -1  & -1   & -1    & -1               & -1     & -1    & -1       & -1        \\ \hline
600                  & 0.370 & 2.010  & 300.182 & -1  & -1   & -1    & -1   & -1   & -1   & -1   \\ \hline
1000                 & 0.339 & 1.083  & 19.434 & 85.093 & -1  & -1   & -1   & -1   & -1  & -1   \\ \hline
1500                 & 0.311 & 0.762 & 4.857 & 14.100 & 52.930 & 161.565 & 243.135 & 330.973 & 434.016 & -1   \\ \hline
2000                 & 0.289 & 0.626 & 2.423 & 5.452 & 15.150 & 32.618 & 41.050 & 47.895 & 56.623 & 56.390 \\ \hline
2500                 & 0.275 & 0.556 & 1.490 & 2.269 & 7.371 & 13.378 & 16.390 & 19.078 & 21.197 & 21.520 \\ \hline
3000                 & 0.263 & 0.509 & 1.194 & 1.957 & 4.447 & 7.568 & 9.217 & 10.290 & 11.389 & 11.437 \\ \hline
\end{tabular}
\end{centering}
\end{adjustwidth}
\end{table}

\begin{table}[H]
\caption{EMI-BF$_4$ positive dimers mean lifetimes for $\delta_{post} = 20 \mathring{\mathrm{A}}$ calculated by averaging.}
\label{tab:EMI-BF$_4$posDimMeanLife20}
\begin{adjustwidth}{-.65in}{-1in}
\begin{centering}
\begin{tabular}{|l|l|l|l|l|l|l|l|l|l|l|}
\hline
E (V/A) & 0.8               & 0.3               & 0.15             & 0.1              & 0.05             & 0.02             & 0.01             & 0.005            & 0.0005           & 0.00005          \\ \hline
T (K)   &                   &                   &                  &                  &                  &                  &                  &                  &                  &                  \\ \hline
300                  & 0.582 & 4.739  & -1               & -1               & -1               & -1               & -1               & -1               & -1               & -1               \\ \hline
600                  & 0.571 & 2.644  & 301.638 & -1               & -1               & -1               & -1               & -1               & -1               & -1               \\ \hline
1000                 & 0.555 & 1.665  & 20.867 & 87.001 & -1               & -1               & -1               & -1               & -1               & -1               \\ \hline
1500                 & 0.538 & 1.289  & 5.950 & 15.521 & 54.426 & 162.737 & 244.584 & 333.153 & 433.944 & -1               \\ \hline
2000                 & 0.518 & 1.103  & 3.304 & 6.414 & 16.050 & 33.355 & 41.604 & 48.082 & 56.912 & 56.983 \\ \hline
2500                 & 0.498 & 0.986 & 2.234 & 3.202 & 7.969 & 13.663 & 16.540 & 18.992 & 20.923 & 21.236 \\ \hline
3000                 & 0.477 & 0.890 & 1.776 & 2.598 & 4.881 & 7.719 & 9.084 & 10.208 & 11.105 & 11.120 \\ \hline
\end{tabular}
\end{centering}
\end{adjustwidth}
\end{table}

\begin{table}[H]
\caption{EMI-BF$_4$ positive dimers mean lifetimes for $\delta_{post} = 30 \mathring{\mathrm{A}}$ calculated by averaging.}
\label{tab:EMI-BF$_4$posDimMeanLife30}
\begin{adjustwidth}{-.65in}{-1in}
\begin{centering}
\begin{tabular}{|l|l|l|l|l|l|l|l|l|l|l|}
\hline
E (V/A) & 0.8               & 0.3              & 0.15             & 0.1              & 0.05             & 0.02             & 0.01             & 0.005            & 0.0005           & 0.00005          \\ \hline
T (K)   &                   &                  &                  &                  &                  &                  &                  &                  &                  &                  \\ \hline
300                  & 0.712 & 5.081 & -1               & -1               & -1               & -1               & -1               & -1               & -1               & -1               \\ \hline
600                  & 0.701 & 2.972 & 302.388 & -1               & -1               & -1               & -1               & -1               & -1               & -1               \\ \hline
1000                 & 0.685 & 1.969 & 21.384 & 87.678& -1               & -1               & -1               & -1               & -1               & -1               \\ \hline
1500                 & 0.669 & 1.572 & 6.465 & 16.183 & 55.506 & 164.592 & 246.902 & 335.558 & 436.011 & -1               \\ \hline
2000                 & 0.650 & 1.369 & 3.791 & 7.037 & 16.985 & 34.819 & 43.323 & 49.682 & 58.363 & 58.391 \\ \hline
2500                 & 0.630 & 1.239 & 2.686 & 3.791 & 8.737 & 14.771 & 17.796 & 20.148 & 21.921 & 22.187 \\ \hline
3000                 & 0.607 & 1.128 & 2.175 & 2.332 & 5.531 & 8.595 & 10.064 & 11.085 & 11.847 & 11.841 \\ \hline
\end{tabular}
\end{centering}
\end{adjustwidth}
\end{table}

\begin{table}[H]
\caption{EMI-BF$_4$ positive dimers mean lifetimes for $\delta_{post} = 40 \mathring{\mathrm{A}}$ calculated by averaging.}
\label{tab:EMI-BF$_4$posDimMeanLife40}
\begin{adjustwidth}{-.65in}{-1in}
\begin{centering}
\begin{tabular}{|l|l|l|l|l|l|l|l|l|l|l|}
\hline
E (V/A) & 0.8               & 0.3              & 0.15             & 0.1              & 0.05             & 0.02             & 0.01             & 0.005            & 0.0005           & 0.00005          \\ \hline
T (K)   &                   &                  &                  &                  &                  &                  &                  &                  &                  &                  \\ \hline
300                  & 0.827 & 5.856 & -1               & -1               & -1               & -1               & -1               & -1               & -1               & -1               \\ \hline
600                  & 0.815 & 3.240 & 303.810 & -1               & -1               & -1               & -1               & -1               & -1               & -1               \\ \hline
1000                 & 0.800 & 2.213 & 22.301 & 88.199 & -1               & -1               & -1               & -1               & -1               & -1               \\ \hline
1500                 & 0.783 & 1.797 & 7.369 & 17.197 & 56.757 & 166.876 & 253.760 & 347.896 & 448.514 & -1               \\ \hline
2000                 & 0.766 & 1.584 & 4.231 & 8.038 & 18.222 & 36.529 & 45.490 & 52.124 & 65.417 & 65.472 \\ \hline
2500                 & 0.744 & 1.446 & 3.060 & 4.278 & 9.913 & 16.364 & 19.728 & 22.245 & 24.144 & 24.383 \\ \hline
3000                 & 0.712 & 1.312 & 2.486 & 2.221 & 6.604 & 9.979 & 11.661 & 12.794 & 13.595 & 13.551 \\ \hline
\end{tabular}
\end{centering}
\end{adjustwidth}
\end{table}

\subsection*{Mean Lifetime from Averaging and from Exponential Fit}

\begin{table}[H]
\caption{EMI-BF$_4$ positive dimers mean lifetimes for $\delta_{post} = 20 \mathring{\mathrm{A}}$ calculated using the constant from the exponential fit.}
\label{tab:EMI-BF$_4$posDimMeanLife20constant}
\begin{adjustwidth}{-.65in}{-1in}
\begin{centering}
\begin{tabular}{|l|l|l|l|l|l|l|l|l|l|l|}
\hline
E (V/A) & 0.8                  & 0.3              & 0.15             & 0.1              & 0.05             & 0.02             & 0.01             & 0.005            & 0.0005           & 0.00005          \\ \hline
T (K)   &                      &                  &                  &                  &                  &                  &                  &                  &                  &                  \\ \hline
300                  & 0.0002 & 20.530 & -1               & -1               & -1               & -1               & -1               & -1               & -1               & -1               \\ \hline
600                  & 0.001  & 3.263 & 370.948 & -1               & -1               & -1               & -1               & -1               & -1               & -1               \\ \hline
1000                 & 0.481    & 1.261 & 22.628 & 81.818 & -1               & -1               & -1               & -1               & -1               & -1               \\ \hline
1500                 & 81.538     & 4.966 & 9.323 & 18.770 & 46.379 & 214.858 & 208.675 & 415.225 & 754.473 & -1               \\ \hline
2000                 & 30.921     & 2.363 & 3.850 & 8.549 & 18.373 & 31.160 & 40.468 & 50.887 & 88.737 & 85.569 \\ \hline
2500                 & 9.995     & 1.656 & 2.405 & 4.736 & 11.853 & 24.230& 22.939 & 27.032 & 27.765 & 54.005 \\ \hline
3000                 & 4.068     & 1.233 & 1.832 & 3.113 & 7.633 & 12.194 & 16.263 & 17.480 & 19.353 & 20.754 \\ \hline
\end{tabular}
\end{centering}
\end{adjustwidth}
\end{table}

\begin{table}[H]
\caption{EMI-BF$_4$ positive dimers mean lifetimes for $\delta_{post} = 20 \mathring{\mathrm{A}}$ calculated using the slope from the exponential fit.}
\label{tab:EMI-BF$_4$posDimMeanLife20slope}
\begin{adjustwidth}{-.65in}{-1in}
\begin{centering}
\begin{tabular}{|l|l|l|l|l|l|l|l|l|l|l|}
\hline
E (V/A) & 0.8                & 0.3               & 0.15             & 0.1              & 0.05             & 0.02             & 0.01             & 0.005            & 0.0005           & 0.00005          \\ \hline
T (K)   &                    &                   &                  &                  &                  &                  &                  &                  &                  &                  \\ \hline
300                  & 0.097 & 8.601  & -1               & -1               & -1               & -1               & -1               & -1               & -1               & -1               \\ \hline
600                  & 0.114  & 2.572  & 343.540 & -1               & -1               & -1               & -1               & -1               & -1               & -1               \\ \hline
1000                 & -1.995  & 1.200  & 21.024 & 84.455 & -1               & -1               & -1               & -1               & -1               & -1               \\ \hline
1500                 & -0.153 & 1.235  & 6.632 & 15.854 & 49.224 & 175.551 & 222.358 & 360.002 & 537.996 & -1               \\ \hline
2000                 & -0.198 & 0.847 & 3.122 & 6.696 & 16.490 & 31.619 & 40.739 & 50.557 & 69.218 & 66.211 \\ \hline
2500                 & -0.287 & 0.696 & 1.953 & 4.176 & 8.777 & 16.677 & 18.376 & 21.486 & 22.962 & 29.568 \\ \hline
3000                 & -0.577 & 0.559 & 1.438 & 2.550 & 5.509 & 8.618 & 10.907 & 12.129 & 13.550 & 13.923 \\ \hline
\end{tabular}
\end{centering}
\end{adjustwidth}
\end{table}

\subsection*{Energy and Temperature}

\begin{table}[H]
\caption{EMI-BF$_4$ positive dimers temperature and energy data.}
\label{tab:EMI-BF$_4$posDimEnergy}
\begin{tabular}{|l|l|l|l|l|}
\hline
T (K)  & T STD (K) & Energy (eV)       & Energy STD (K)   & T Energy Correlation \\ \hline
322.375 & 33.773    & 24.752 & 4.475 & 0.515              \\ \hline
589.269 & 68.526    & 90.779 & 10.980 & 0.646              \\ \hline
998.566 & 87.458    & 192.096 & 2.630 & 0.039             \\ \hline
1522.825 & 134.826    & 320.935 & 4.971 & 0.074             \\ \hline
2057.246 & 183.897    & 451.756 & 7.662 & 0.079            \\ \hline
2597.133 & 229.011    & 583.566 & 10.222 & 0.083             \\ \hline
3137.435 & 276.405   & 716.821 & 13.027 & 0.062             \\ \hline
\end{tabular}
\end{table}

\subsection*{Geometry and Fragmentation Pathways}

\begin{table}[H]
\caption{EMI-BF$_4$ positive dimer mean maximum separation before fragmentation.}
\label{tab:EMI-BF$_4$posDimSepMean}
\begin{tabular}{|l|l|l|l|l|l|l|l|l|l|l|}
\hline
E (V/A) & 0.8              & 0.3              & 0.15             & 0.1              & 0.05             & 0.02             & 0.01             & 0.005            & 0.0005           & 0.00005          \\ \hline
T (K)   &                  &                  &                  &                  &                  &                  &                  &                  &                  &                  \\ \hline
300     & -1               & 7.673 & -1               & -1               & -1               & -1               & -1               & -1               & -1               & -1               \\ \hline
600     & 8.490 & 8.207 & 8.213 & -1               & -1               & -1               & -1               & -1               & -1               & -1               \\ \hline
1000    & 8.835 & 8.675 & 8.744 & 8.911 & -1               & -1               & -1               & -1               & -1               & -1               \\ \hline
1500    & 9.057 & 9.036 & 9.235 & 9.437 & 9.642 & 9.731 & 9.728 & 9.730 & 9.774 & -1               \\ \hline
2000    & 9.169 & 9.234 & 9.277 & 9.870 & 10.211& 10.489 & 10.617 & 10.728 & 10.804 & 10.787\\ \hline
2500    & 9.231 & 9.341 & 9.449 & 9.711 & 10.687 & 11.119 & 11.341 & 11.536 & 11.717 & 11.763 \\ \hline
3000    & 9.271 & 9.424 & 9.553 & 9.831 & 11.0433 & 11.612 & 11.906 & 12.097 & 12.337 & 12.411\\ \hline
\end{tabular}
\end{table}

\begin{table}[H]
\caption{EMI-BF$_4$ positive dimers total fragmentation counts.}
\label{tab:EMI-BF$_4$posDimPath}
\begin{tabular}{|l|l|l|l|l|l|l|l|l|l|l|}
\hline
E (V/A) & 0.8 & 0.3 & 0.15 & 0.1 & 0.05 & 0.02 & 0.01 & 0.005 & 0.0005 & 0.00005 \\ \hline
T (K)   &     &     &      &     &      &      &      &       &        &         \\ \hline
300                  & 0   & 0   & -1   & -1  & -1   & -1   & -1   & -1    & -1     & -1      \\ \hline
600                  & 0   & 0   & 0    & -1  & -1   & -1   & -1   & -1    & -1     & -1      \\ \hline
1000                 & 0   & 0   & 0    & 0   & -1   & -1   & -1   & -1    & -1     & -1      \\ \hline
1500                 & 0   & 0   & 0    & 1   & 0    & 0    & 0    & 0     & 0      & -1      \\ \hline
2000                 & 4   & 0   & 0    & 0   & 0    & 0    & 0    & 0     & 0      & 0       \\ \hline
2500                 & 6   & 7   & 11   & 10  & 1    & 1    & 0    & 0     & 0      & 0       \\ \hline
3000                 & 18  & 14  & 20   & 0   & 3    & 1    & 0    & 0     & 1      & 0       \\ \hline
\end{tabular}
\end{table}

\section*{EMI-BF$_4$ Negative Dimers}

\subsection*{Mean Lifetime from Averaging and Chosen $\delta_{post}$}

\begin{table}[H]
\caption{EMI-BF$_4$ negative dimers mean lifetimes for $\delta_{post} = 20 \mathring{\mathrm{A}}$ calculated by averaging.}
\label{tab:EMI-BF$_4$negDimLife}
\begin{adjustwidth}{-.65in}{-1in}
\begin{centering}
\begin{tabular}{|l|l|l|l|l|l|l|l|l|l|l|}
\hline
E (V/A) & 0.8               & 0.3               & 0.15             & 0.1              & 0.05             & 0.02             & 0.01             & 0.005            & 0.0005           & 0.00005          \\ \hline
T (K)   &                   &                   &                  &                  &                  &                  &                  &                  &                  &                  \\ \hline
300     & 0.552& 1.465  & -1               & -1               & -1               & -1               & -1               & -1               & -1               & -1               \\ \hline
600     & 0.544 & 1.342  & 43.169 & 2.506 & -1               & -1               & -1               & -1               & -1               & -1               \\ \hline
1000    & 0.535 & 1.204 & 8.463& 40.771 & 435.613 & -1               & -1               & -1               & -1               & -1               \\ \hline
1500    & 0.520 & 1.078  & 3.647 & 9.351 & 35.052 & 76.346 & 188.530 & 259.499 & 330.215 & 339.721 \\ \hline
2000    & 0.502 & 0.991 & 2.466 & 4.601 & 12.195 & 24.848 & 36.799 & 44.729 & 53.280 & 53.235 \\ \hline
2500    & 0.483 & 0.911 & 1.886 & 3.156 & 6.429 & 11.971 & 15.242 & 17.558 & 19.746 & 19.865 \\ \hline
3000    & 0.462 & 0.840 & 1.549 & 2.369 & 4.420 & 7.236 & 8.962 & 9.917 & 10.823 & 10.919 \\ \hline
\end{tabular}
\end{centering}
\end{adjustwidth}
\end{table}

\subsection*{Energy and Temperature}

\begin{table}[H]
\caption{EMI-BF$_4$ negative dimers temperature and energy data.}
\label{tab:EMI-BF$_4$negDimEnergy}
\begin{tabular}{|l|l|l|l|l|}
\hline
T (K)  & T STD (K) & Energy (eV)        & Energy STD (K)   & T Energy Correlation \\ \hline
332.136 & 51.676    & -23.751 & 6.195 & 0.710              \\ \hline
566.901 & 78.189    & 14.690  & 7.762 & 0.600              \\ \hline
978.672 & 112.923    & 81.973  & 6.299 & 0.364              \\ \hline
1522.780 & 165.358    & 170.062  & 4.520 & 0.087             \\ \hline
2048.764 & 223.218    & 256.064  & 8.002 & 0.103              \\ \hline
2585.963 & 276.801   & 342.646  & 12.142 & 0.138              \\ \hline
3118.145 & 337.876    & 429.820  & 15.810 & 0.138              \\ \hline
\end{tabular}
\end{table}

\subsection*{Geometry and Fragmentation Pathways}

\begin{table}[H]
\caption{EMI-BF$_4$ negative dimer mean maximum separation before fragmentation.}
\label{tab:EMI-BF$_4$negDimSepMean}
\begin{tabular}{|l|l|l|l|l|l|l|l|l|l|l|}
\hline
E (V/A) & 0.8              & 0.3              & 0.15             & 0.1              & 0.05             & 0.02             & 0.01             & 0.005            & 0.0005           & 0.00005          \\ \hline
T (K)   &                  &                  &                  &                  &                  &                  &                  &                  &                  &                  \\ \hline
300     & 8.555 & 8.295 & -1               & -1               & -1               & -1               & -1               & -1               & -1               & -1               \\ \hline
600     & 8.800 & 8.587 & 8.642 & 9.992 & -1               & -1               & -1               & -1               & -1               & -1               \\ \hline
1000    & 9.031 & 8.930 & 8.960 & 9.120 & 9.168 & -1               & -1               & -1               & -1               & -1               \\ \hline
1500    & 9.184 & 9.214 & 9.240 & 9.649 & 9.794 & 9.884 & 9.9147 & 9.941 & 9.977 & 9.988 \\ \hline
2000    & 9.245 & 9.337 & 9.460 & 10.133 & 10.372 & 10.653 & 10.828 & 10.928 & 11.109 & 11.074 \\ \hline
2500    & 9.283 & 9.429 & 9.599 & 9.861 & 10.887 & 11.303 & 11.568 & 11.773 & 11.992 & 11.944 \\ \hline
3000    & 9.293 & 9.478 & 9.694 & 9.966 & 10.625 & 11.845 & 12.116 & 12.397 & 12.723 & 12.658 \\ \hline
\end{tabular}
\end{table}

\begin{table}[H]
\caption{EMI-BF$_4$ negative dimers total fragmentation counts.}
\label{tab:EMI-BF$_4$negDimPath}
\begin{tabular}{|l|l|l|l|l|l|l|l|l|l|l|}
\hline
Electric Field (V/A) & 0.8 & 0.3 & 0.15 & 0.1 & 0.05 & 0.02 & 0.01 & 0.005 & 0.0005 & 0.00005 \\ \hline
Temperature          &     &     &      &     &      &      &      &       &        &         \\ \hline
300                  & 0   & 0   & -1   & -1  & -1   & -1   & -1   & -1    & -1     & -1      \\ \hline
600                  & 0   & 0   & 0    & 0   & -1   & -1   & -1   & -1    & -1     & -1      \\ \hline
1000                 & 0   & 0   & 0    & 0   & 0    & -1   & -1   & -1    & -1     & -1      \\ \hline
1500                 & 0   & 0   & 0    & 0   & 0    & 0    & 0    & 0     & 0      & 0       \\ \hline
2000                 & 6   & 1   & 5    & 0   & 0    & 0    & 0    & 0     & 0      & 0       \\ \hline
2500                 & 7   & 12  & 11   & 6   & 0    & 0    & 0    & 0     & 0      & 0       \\ \hline
3000                 & 38  & 29  & 28   & 27  & 10   & 0    & 0    & 0     & 1      & 0       \\ \hline
\end{tabular}
\end{table}

\section*{EMI-BF$_4$ Positive Trimers}

\subsection*{Mean Lifetime from Averaging and Chosen $\delta_{post}$}

\begin{table}[H]
\caption{EMI-BF$_4$ positive trimers mean lifetimes for $\delta_{post} = 20 \mathring{\mathrm{A}}$ calculated by averaging.}
\label{tab:EMI-BF$_4$posTri}
\begin{tabular}{|l|l|l|l|l|l|l|}
\hline
E (V/A) & 0.5               & 0.2              & 0.1              & 0.05             & 0.005            & 0.0005           \\ \hline
T (K)   &                   &                  &                  &                  &                  &                  \\ \hline
600     & -1                & 7.958 & -1               & -1               & -1               & -1               \\ \hline
1000    & 0.612 & 4.001 & 25.530 & 117.210 & -1               & -1               \\ \hline
1500    & -1                & 2.333 & 6.979 & 18.746 & 49.793 & -1               \\ \hline
2000    & -1                & 1.684 & 3.646 & 7.268 & 14.836 & 15.440 \\ \hline
\end{tabular}
\end{table}

\subsection*{Energy and Temperature}

\begin{table}[H]
\caption{EMI-BF$_4$ positive trimers temperature and energy data.}
\label{tab:EMI-BF$_4$posTriEnergy}
\begin{tabular}{|l|l|l|l|l|}
\hline
Te (K)  & T STD (K) & Energy (eV)       & Energy STD (eV)   & T Energy Correlation \\ \hline
575.776 & 42.292   & 123.221 & 4.777& 0.235              \\ \hline
1001.397 & 70.864    & 289.087 & 7.092 & 0.204              \\ \hline
1526.053 & 110.466    & 493.223 & 9.126 & 0.141              \\ \hline
2036.280 & 146.136    & 691.179 & 10.333 & 0.088             \\ \hline
\end{tabular}
\end{table}

\subsection*{Geometry and Fragmentation Pathways}

\begin{table}[H]
\caption{EMI-BF$_4$ positive trimer mean maximum separation before fragmentation.}
\label{tab:EMI-BF$_4$posTriSepMean}
\begin{tabular}{|l|l|l|l|l|l|l|}
\hline
E (V/A) & 0.5              & 0.2              & 0.1              & 0.05             & 0.005            & 0.0005           \\ \hline
T (K)   &                  &                  &                  &                  &                  &                  \\ \hline
600     & -1               & 12.416 & -1               & -1               & -1               & -1               \\ \hline
1000    & 9.931 & 11.055 & 13.264 & 13.880 & -1               & -1               \\ \hline
1500    & -1               & 10.308 & 13.358 & 14.223 & 15.151 & -1               \\ \hline
2000    & -1               & 10.093 & 12.873 & 14.251 & 15.545 & 15.755 \\ \hline
\end{tabular}
\end{table}

\begin{table}[H]
\caption{EMI-BF$_4$ positive trimers total fragmentation counts.}
\label{tab:EMI-BF$_4$posTriTotFrag}
\begin{tabular}{|l|l|l|l|l|l|l|}
\hline
E (V/A) & 0.5 & 0.2 & 0.1 & 0.05 & 0.005 & 0.0005 \\ \hline
T (K)   &     &     &     &      &       &        \\ \hline
600     & -1  & 0   & -1  & -1   & -1    & -1     \\ \hline
1000    & 1   & 0   & 0   & 0    & -1    & -1     \\ \hline
1500    & -1  & 0   & 0   & 0    & 0     & -1     \\ \hline
2000    & -1  & 0   & 0   & 0    & 0     & 0      \\ \hline
\end{tabular}
\end{table}

\begin{table}[H]
\caption{EMI-BF$_4$ positive trimers single neutral evaporation counts.}
\label{tab:EMI-BF$_4$posTriNeuEvap}
\begin{tabular}{|l|l|l|l|l|l|l|}
\hline
E (V/A) & 0.5  & 0.2  & 0.1  & 0.05 & 0.005 & 0.0005 \\ \hline
T (K)   &      &      &      &      &       &        \\ \hline
600     & -1   & 1620 & -1   & -1   & -1    & -1     \\ \hline
1000    & 4323 & 322  & 3941 & 7205 & -1    & -1     \\ \hline
1500    & -1   & 1062 & 4113 & 6234 & 7138  & -1     \\ \hline
2000    & -1   & 270  & 3421 & 4846 & 5692  & 5687   \\ \hline
\end{tabular}
\end{table}

\begin{table}[H]
\caption{EMI-BF$_4$ positive trimers monomer escape counts.}
\label{tab:EMI-BF$_4$posTriMonEscp}
\begin{tabular}{|l|l|l|l|l|l|l|}
\hline
E (V/A) & 0.5 & 0.2  & 0.1  & 0.05 & 0.005 & 0.0005 \\ \hline
T (K)   &     &      &      &      &       &        \\ \hline
600     & -1  & 7910 & -1   & -1   & -1    & -1     \\ \hline
1000    & 447 & 841  & 5720 & 1895 & -1    & -1     \\ \hline
1500    & -1  & 3265 & 4531 & 2791 & 1096  & -1     \\ \hline
2000    & -1  & 692  & 4182 & 3224 & 2002  & 1947   \\ \hline
\end{tabular}
\end{table}

\begin{table}[H]
\caption{EMI-BF$_4$ positive trimers unknown separation count types.}
\label{tab:EMI-BF$_4$posTriUnknown}
\begin{tabular}{|l|l|l|l|l|l|l|}
\hline
E (V/A) & 0.5  & 0.2  & 0.1  & 0.05 & 0.005 & 0.0005 \\ \hline
T (K)   &      &      &      &      &       &        \\ \hline
600     & -1   & 470  & -1   & -1   & -1    & -1     \\ \hline
1000    & 5229 & 853  & 336  & 330  & -1    & -1     \\ \hline
1500    & -1   & 5666 & 1356 & 975  & 1460  & -1     \\ \hline
2000    & -1   & 1642 & 2397 & 1930 & 2306  & 2366   \\ \hline
\end{tabular}
\end{table}

\section*{EMI-Im Positive Dimers}

\subsection*{Mean Lifetime from Averaging and Chosen $\delta_{post}$}

\begin{table}[H]
\caption{EMI-Im positive dimers mean lifetimes for $\delta_{post} = 20 \mathring{\mathrm{A}}$ calculated by averaging.}
\label{tab:EMI-ImposDimLife}
\begin{tabular}{|l|l|l|l|l|l|l|l|}
\hline
E (V/A) & 1                 & 0.5               & 0.2              & 0.1              & 0.05             & 0.005            & 0.0005           \\ \hline
T (K)   &                   &                   &                  &                  &                  &                  &                  \\ \hline
600     & 0.603 & 0.938 & 3.216 & 80.473 & -1               & -1               & -1               \\ \hline
1000    & 0.590 & 0.897 & 2.126 & 11.800 & 79.405 & 199.535 & -1               \\ \hline
1500    & 0.569 & 0.852 & 1.706 & 4.712 & 16.679 & 120.666 & -1               \\ \hline
2000    & 0.545 & 0.796 & 1.461 & 3.063 & 7.368 & 30.675 & 36.373 \\ \hline
\end{tabular}
\end{table}

\subsection*{Energy and Temperature}

\begin{table}[H]
\caption{EMI-Im positive dimers temperature and energy data.}
\label{tab:EMI-ImposDimEnergy}
\begin{tabular}{|l|l|l|l|l|}
\hline
T (K)  & T STD (K) & Energy (eV)       & Energy STD (eV)   & T, Energy Correlation \\ \hline
600.115 & 51.405    & 187.332 & 4.830 & 0.280             \\ \hline
993.336 & 81.191   & 309.735 & 3.363 & 0.113              \\ \hline
1496.611 & 121.023    & 476.335 & 4.560 & 0.085             \\ \hline
2019.525 & 157.922    & 624.842 & 6.325 & 0.051             \\ \hline
\end{tabular}
\end{table}

\subsection*{Geometry and Fragmentation Pathways}

\begin{table}[H]
\caption{EMI-Im positive dimer mean maximum separation before fragmentation.}
\label{tab:EMI-ImposDimSepMean}
\begin{tabular}{|l|l|l|l|l|l|l|l|}
\hline
E (V/A) & 1                & 0.5              & 0.2              & 0.1              & 0.05             & 0.005            & 0.0005           \\ \hline
T (K)   &                  &                  &                  &                  &                  &                  &                  \\ \hline
600     & 9.249 & 9.220 & 8.963 & 9.084 & -1               & -1               & -1               \\ \hline
1000    & 9.499 & 9.572 & 9.537 & 9.897 & 10.210 & 10.301 & -1               \\ \hline
1500    & 9.578 & 9.763 & 9.952 & 10.682 & 11.078 & 11.574 & -1               \\ \hline
2000    & 9.582 & 9.775 & 10.153 & 10.586 & 11.776 & 12.839 & 13.127 \\ \hline
\end{tabular}
\end{table}

\begin{table}[H]
\caption{EMI-Im positive dimers total fragmentation counts.}
\label{tab:EMI-ImposDimPath}
\begin{tabular}{|l|l|l|l|l|l|l|l|}
\hline
E (V/A) & 1  & 0.5 & 0.2 & 0.1 & 0.05 & 0.005 & 0.0005 \\ \hline
T (K)   &    &     &     &     &      &       &        \\ \hline
600     & 0  & 0   & 0   & 0   & -1   & -1    & -1     \\ \hline
1000    & 0  & 0   & 3   & 0   & 0    & 0     & -1     \\ \hline
1500    & 2  & 2   & 6   & 0   & 0    & 0     & -1     \\ \hline
2000    & 10 & 7   & 9   & 13  & 3    & 0     & 0      \\ \hline
\end{tabular}
\end{table}

\section*{EMI-FAP Positive Dimers}

\subsection*{Mean Lifetime from Averaging and Chosen $\delta_{post}$}

\begin{table}[H]
\caption{EMI-FAP positive dimer mean lifetimes for $\delta_{post} = 20 \mathring{\mathrm{A}}$ calculated by averaging.}
\label{tab:EMI-FAPposDimLife}
\begin{tabular}{|l|l|l|l|l|l|l|l|l|}
\hline
E (V/A) & 0.5               & 0.2              & 0.1              & 0.05             & 0.02             & 0.01             & 0.005            & 0.0005           \\ \hline
T (K)   &                   &                  &                  &                  &                  &                  &                  &                  \\ \hline
600     & 0.930 & 1.832 & 4.762 & 36.706 & -1               & -1               & -1               & -1               \\ \hline
1000    & 0.876 & 1.646 & 3.119 & 9.446 & 37.629 & 86.993 & 149.061 & 269.713 \\ \hline
1500    & 0.816 & 1.462 & 2.611 & 5.774 & 14.979 & 24.411 & 34.084 & 47.780 \\ \hline
2000    & 0.769 & 1.330 & 2.219 & 4.282 & 8.980 & 13.224 & 16.804 & 21.688 \\ \hline
\end{tabular}
\end{table}

\subsection*{Energy and Temperature}

\begin{table}[H]
\caption{EMI-FAP positive dimers temperature and energy.}
\label{tab:EMI-FAPposDimEnergy}
\begin{tabular}{|l|l|l|l|l|}
\hline
T (K)            & T STD (K)        & Energy (eV)       & Energy STD (eV)   & T Energy Correlation \\ \hline
600.400 & 64.472 & 655.327 & 16.437& 0.656    \\ \hline
1000.159 & 106.654 & 801.744 & 27.227 & 0.685    \\ \hline
1497.446 & 160.248 & 986.397 & 40.855 & 0.683    \\ \hline
2001.163 & 213.939 & 1173.137 & 54.117 & 0.673    \\ \hline
\end{tabular}
\end{table}

\subsection*{Geometry and Fragmentation Pathways}

\begin{table}[H]
\caption{EMI-FAP positive dimer mean maximum separation before fragmentation.}
\label{tab:EMI-FAPposMeanSep}
\begin{tabular}{|l|l|l|l|l|l|l|l|l|}
\hline
E (V/A) & 0.5              & 0.2              & 0.1              & 0.05             & 0.02             & 0.01             & 0.005            & 0.0005           \\ \hline
T (K)   &                  &                  &                  &                  &                  &                  &                  &                  \\ \hline
600     & 9.853 & 9.962 & 10.042 & 10.290 & -1               & -1               & -1               & -1               \\ \hline
1000    & 9.726 & 9.992 & 10.522 & 10.824 & 11.245 & 11.428 & 11.521 & 11.676 \\ \hline
1500    & 9.718 & 10.185 & 10.566 & 11.193 & 12.214 & 12.530 & 12.844 & 13.346 \\ \hline
2000    & 9.770 & 10.321 & 10.896 & 11.694 & 12.659 & 13.567 & 13.947 & 14.912 \\ \hline
\end{tabular}
\end{table}

\begin{table}[H]
\caption{EMI-FAP positive dimer total fragmentation count.}
\label{tab:EMI-FAPposTotFrag}
\begin{tabular}{|l|l|l|l|l|l|l|l|l|}
\hline
E (V/A) & 0.5 & 0.2 & 0.1 & 0.05 & 0.02 & 0.01 & 0.005 & 0.0005 \\ \hline
T (K)   &     &     &     &      &      &      &       &        \\ \hline
600     & 0   & 0   & 0   & 0    & -1   & -1   & -1    & -1     \\ \hline
1000    & 5   & 0   & 0   & 0    & 0    & 0    & 0     & 0      \\ \hline
1500    & 19  & 6   & 11  & 1    & 0    & 0    & 0     & 0      \\ \hline
2000    & 30  & 21  & 13  & 6    & 0    & 0    & 0     & 0      \\ \hline
\end{tabular}
\end{table}\label{app:appa}
\renewcommand{\thechapter}{B}
\renewcommand{\chaptername}{Appendix}
\chapter{Supplemental Simulated Experimental Data}

\section{Effect of Timestep on RPA Simulator}

The size of the timestep used in the RPA simulator affects the accuracy of the simulation process. During each timestep the probability of fragmentation during the timestep is calculated using the electric field at the beginning of the timestep. When the timestep is larger the electric field at the beginning of the timestep is larger than the average of the electric field over the timestep. Having a larger timestep overestimates the electric field during each timestep and thus overestimate the fragmentation rates during the timestep. In particular, during the first timestep where the electric field is highest overestimating fragmentation results in a large amount of clusters fragmenting so close to the emission site that they are indistinguishable from the monoenergetic step. 

\begin{figure}
    \centering
    \includegraphics[width=4.5in]{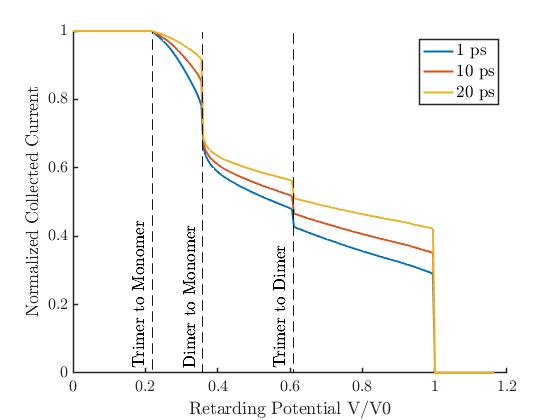}
    \caption{Simulated RPA curves for various timesteps with the electric field calculated at the beginning of the timestep.}
    \label{fig:ABCtimestepBegin}
\end{figure}

\begin{figure}
    \centering
    \includegraphics[width=4.5in]{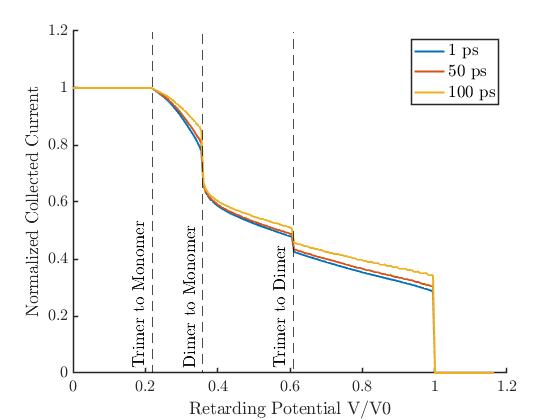}
    \caption{Simulated RPA curves for various timesteps with the electric field calculated as the average electric field over the timestep.}
    \label{fig:ABCtimestepMid}
\end{figure}

Figure \ref{fig:ABCtimestepBegin} shows simulated RPA curves for various timesteps with the electric field taken at the beginning of the timestep while \ref{fig:ABCtimestepMid} shows simulated RPA curves for various timesteps with the electric field taken as an average over the timestep. Using an averaged electric field allows for an increase in the simulation timestep of 50 ps with less than 6.5 \% change in the height of the monoenergetic step and thus the total fragmentation amount. The simulation time with a timestep of 50 ps is almost 300 times shorter than the simulation time with 1 ps, facilitating rapid simulation of many samples for the approximate Bayesian computation routine. Increasing the timestep further to 100 ps results in a further order of magnitude reduction in the simulation time however, the change in the height of the monoenergetic step for 100 ps is 20 \%. While increasing the timestep yields simulated RPA curves that are less smooth the kinks in the curve are unlikely to affect the overall function of the ABC routine as it does not contribute significantly to the distances calculated between the simulated and experimental curves.

\section{Supplemental Acceptance Probability Results}\label{sec:AcceptanceProbabilityResults} 

\begin{figure}[htbp]
\begin{minipage}{.5\linewidth}
\centering
\captionsetup{width=.8\linewidth}
\includegraphics[width=1.1\linewidth]{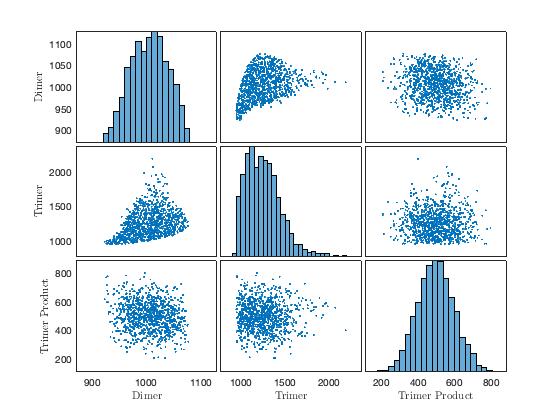}
\caption{5.5\% acceptance}
\label{fig:L2_Gaussian_postDist_AcceptB}
\end{minipage}%
\begin{minipage}{.5\linewidth}
\centering
\captionsetup{width=.8\linewidth}
\includegraphics[width=1.1\linewidth]{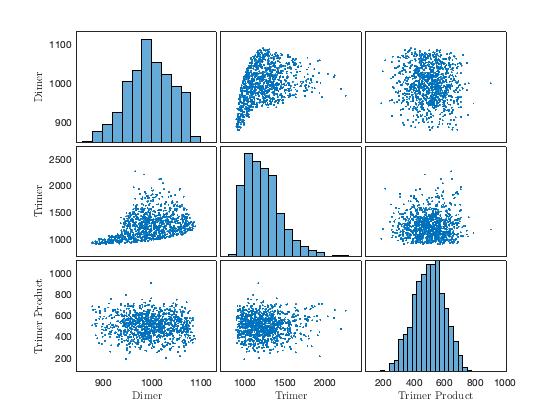}
\caption{7.6\% acceptance}
\label{fig:L2_Gaussian_postDist_AcceptC}
\end{minipage}\par\medskip
\end{figure}

\begin{figure}[htbp]
\begin{minipage}{.5\linewidth}
\centering
\captionsetup{width=.8\linewidth}
\includegraphics[width=1.1\linewidth]{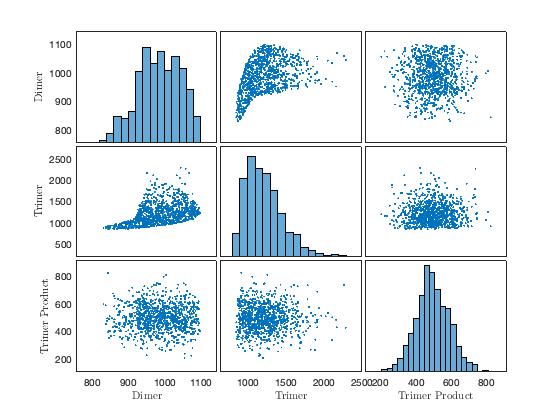}
\caption{10.1\% acceptance}
\label{fig:L2_Gaussian_postDist_AcceptD}
\end{minipage}%
\begin{minipage}{.5\linewidth}
\centering
\captionsetup{width=.8\linewidth}
\includegraphics[width=1.1\linewidth]{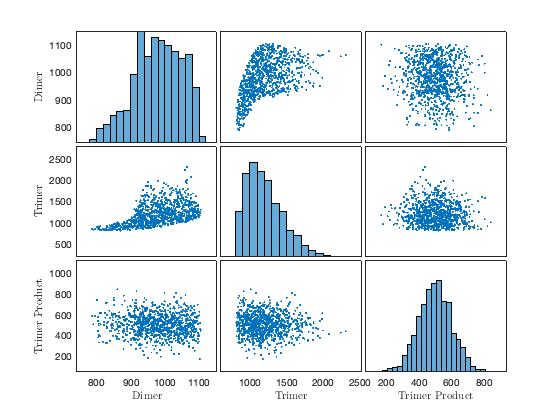}
\caption{11.9\% acceptance}
\label{fig:L2_Gaussian_postDist_AcceptE}
\end{minipage}\par\medskip
\end{figure}

\begin{figure}[htbp]
\begin{minipage}{.5\linewidth}
\centering
\captionsetup{width=.8\linewidth}
\includegraphics[width=1.1\linewidth]{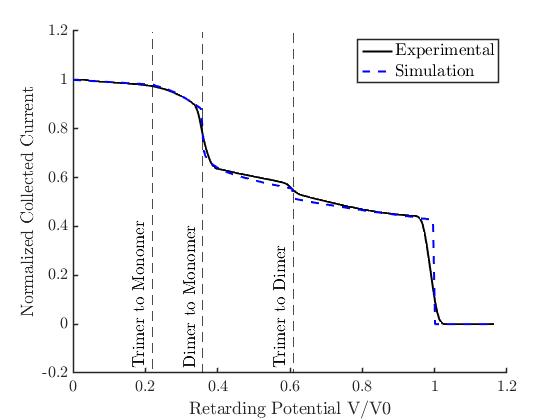}
\caption{5.5\% acceptance}
\label{fig:L2_Gaussian_meanRPA_AcceptB}
\end{minipage}
\begin{minipage}{.5\linewidth}
\centering
\captionsetup{width=.8\linewidth}
\includegraphics[width=1.1\linewidth]{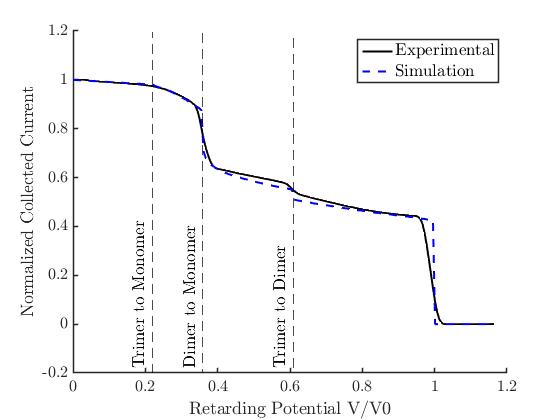}
\caption{7.6\% acceptance}
\label{fig:L2_Gaussian_meanRPA_AcceptC}
\end{minipage}\par\medskip
\end{figure}

\begin{figure}[htbp]
\begin{minipage}{.5\linewidth}
\centering
\captionsetup{width=.8\linewidth}
\includegraphics[width=1.1\linewidth]{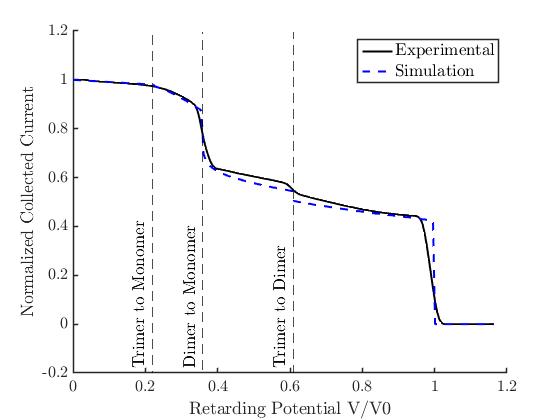}
\caption{10.1\% acceptance}
\label{fig:L2_Gaussian_meanRPA_AcceptD}
\end{minipage}
\begin{minipage}{.5\linewidth}
\centering
\captionsetup{width=.8\linewidth}
\includegraphics[width=1.1\linewidth]{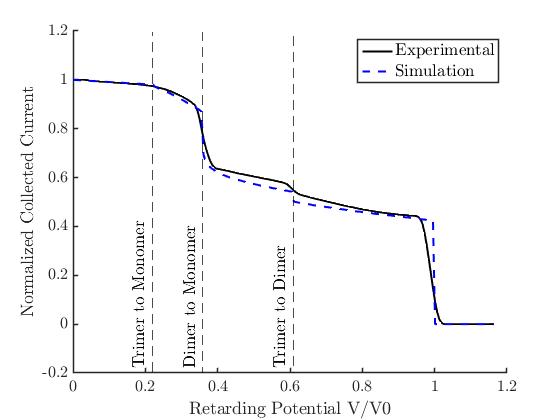}
\caption{11.9\% acceptance}
\label{fig:L2_Gaussian_meanRPA_AcceptE}
\end{minipage}\par\medskip
\end{figure}

Figures \ref{fig:L2_Gaussian_postDist_AcceptB} through \ref{fig:L2_Gaussian_postDist_AcceptE} show the posterior distributions for varying acceptance probabilities for inferring the temperatures of different cluster types. Figures \ref{fig:L2_Gaussian_meanRPA_AcceptB} through \ref{fig:L2_Gaussian_meanRPA_AcceptE} show the mean RPA curves for varying acceptance probabilities for inferring the temperatures of different cluster types.

\label{app:appb}
\renewcommand{\thechapter}{C}
\renewcommand{\chaptername}{Appendix}
\chapter{Effect of Spreading on Experimental and Simulated Experimental Data}

Electrospray ion beams spread out as they travel away from the source. The spreading angle is usually less than $45^{\circ}$ \cite{Petro2020CharacterizationEmitters}. RPA detectors determine the energy of different ions in the beam by repelling them using an electric field. If the ions reach the electric field at an angle the retarding voltage needed to stop them will not represent the full energy of the beam. Instead, the retarding voltage will represent the portion of the energy perpendicular to the detector plane and parallel to the electric field. The larger the angle between the electric field and the ion, the larger the error in the detected energy will be. Spherical RPA setups such as that used by Miller aim to reduce the spreading effect on the RPA curves by matching the curvature of the detector to the spread of the beam \cite{Miller2019CharacterizationSources}. 

\begin{figure}
    \centering
    \includegraphics[width=4.5in]{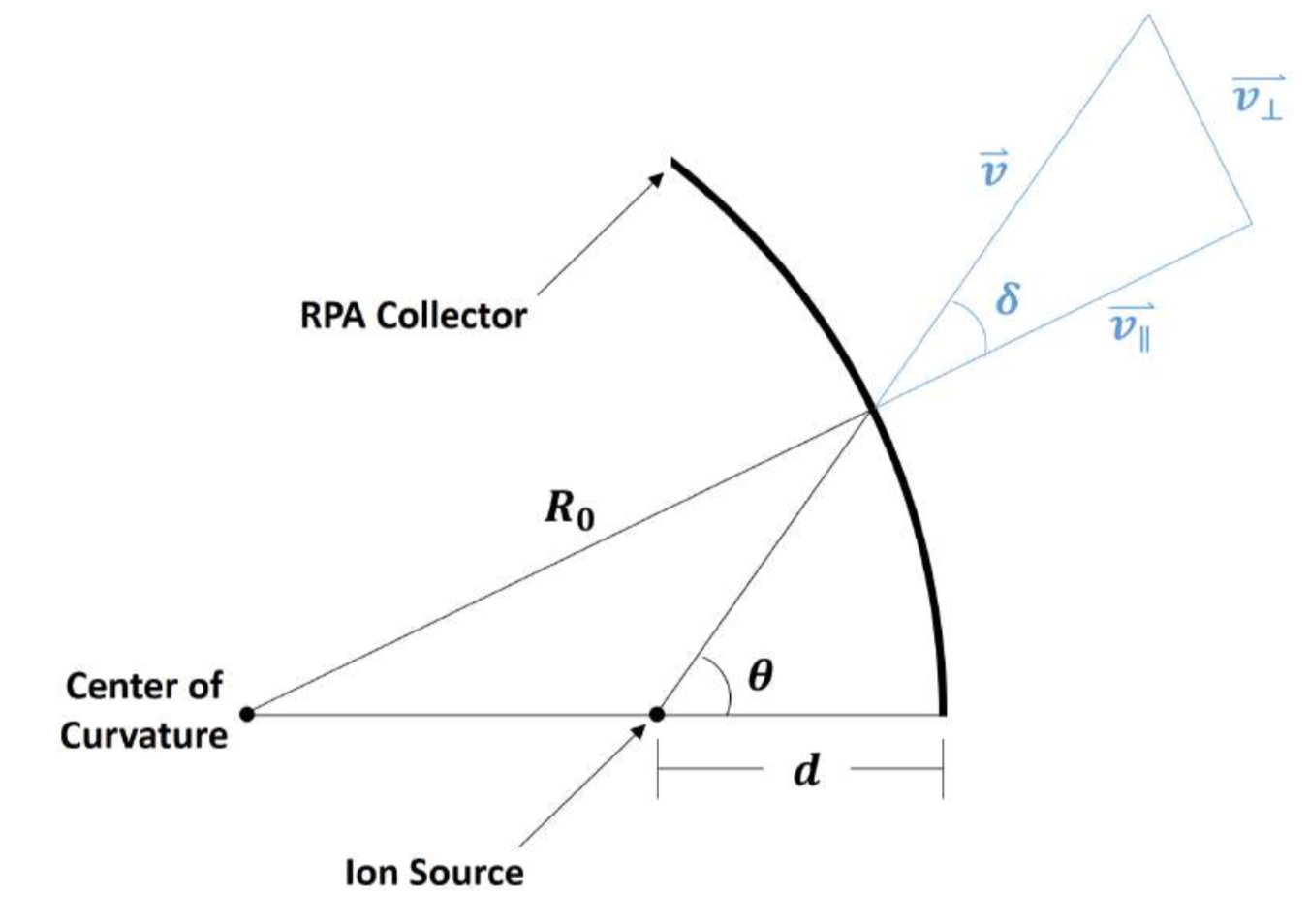}
    \caption{Spherical RPA diagram. Image from Catherine Miller \cite{Miller2019CharacterizationSources}}
    \label{fig:SphericalRPASetup}
\end{figure}

Figure \ref{fig:SphericalRPASetup} shows the geometry for a source firing at a spherical RPA. Here $R_0$ is the radius of curvature of the RPA detector, d is the distance from the source to the center of the detector, $\theta$ is the angle of a particular ion in the beam with respect to the axis of the detector. $\delta$ is the angle between the ion velocity vector and the vector parallel to the applied retarding electric field at the point where the ion reaches the detector. 

When the ion source is located exactly at the center of curvature of the RPA detector, assuming the ions travel from the point source in straight lines, there is no energy spreading because the ion velocities will be perpendicular to the detector surface upon reaching it. When the source is not located at the center of curvature of the detector some spreading will appear in the RPA curve because. The exact magnitude of this spreading effect is not known analytically for all components of the RPA curve because it depends on the total current and the distribution of the different species as a function of the angle $\theta$. However, previous work shows that it primarily results in a smoothing effect on the curve, making sharp jumps into slopes \cite{Miller2019CharacterizationSources}. This is particularly noticeable at the sharp steps which result from monoenergetic ions and ions and clusters that fragment in the field free region. The spreading effect could significantly impact our ability to determine cluster fragmentation rates, particularly in the high electric field region near the monoenergetic step. It can also negatively impact our ability to determine the solvation energy losses incurred during ion evaporation from the difference between the applied voltage and the voltage of the monoenergetic step.

\begin{figure}
    \centering
    \includegraphics[width=5in]{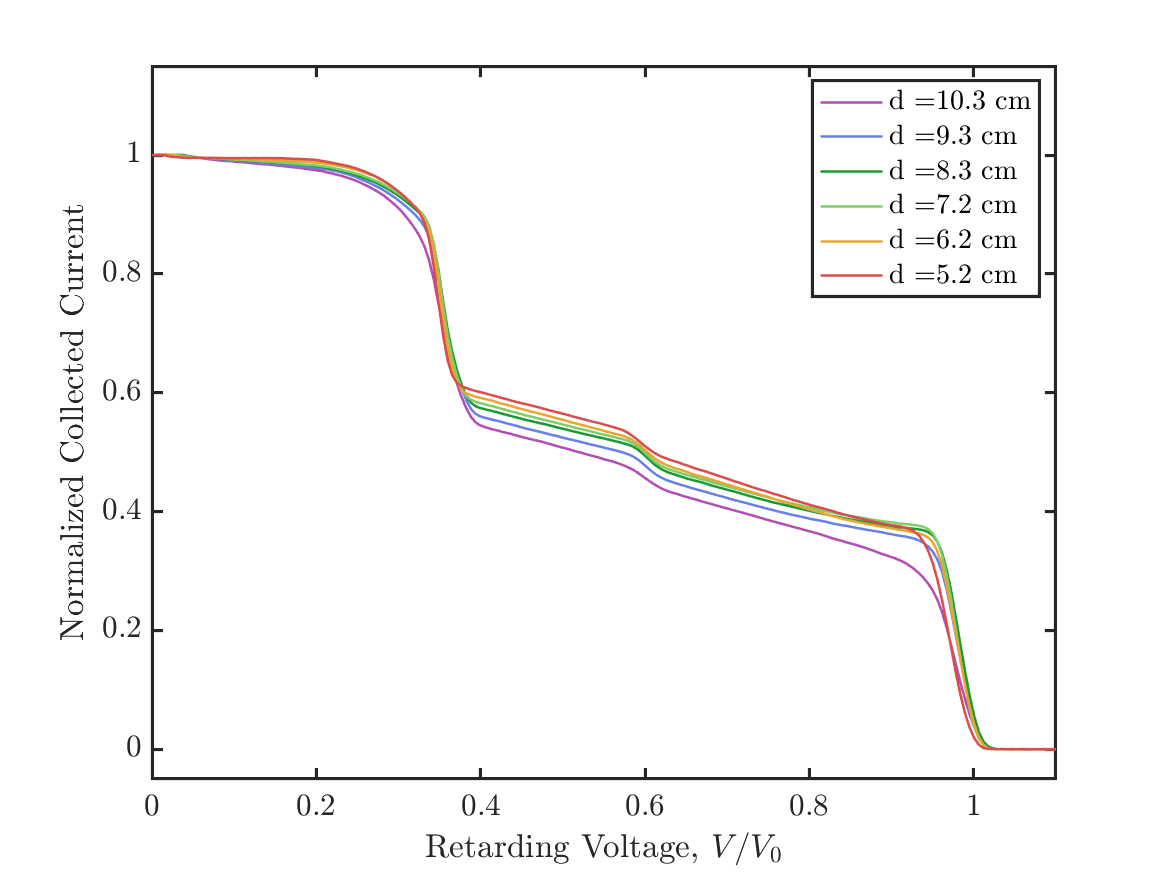}
    \caption{Experimental RPA curves for an EMI-BF$_4$ source operated at 30$^{\circ}$C and 880 V with the source at different distances from the detector. Data from Catherine Miller \cite{Miller2019CharacterizationSources}}
    \label{fig:ExperimentalRPACompareDistance}
\end{figure}

Figure \ref{fig:ExperimentalRPACompareDistance} shows the results of using the spherical RPA with radius of curvature 8.85cm for an EMI-BF$_4$ source with the source at varying distances from the detector. As seen in Figure \ref{fig:ExperimentalRPACompareDistance} the amount of fragmentation in the field free region indicated by the size of the dimer step depends on the distance between the source and the detector. The shape of the steps also depends on the distance to the center of curvature of the detector. The data from experiments with the largest distance between the center of curvature of the detector and the source resulted in significant smoothing of the monoenergetic and field free dimer fragmentation steps. This smoothing effect is worse when the source is beyond the center of curvature and is not as significant when the source is between the center of curvature and the detector. This can be seen in the comparison between the curves with distances to the detector of 10.3 cm and 5.2 cm. 

\begin{figure}
    \centering
    \includegraphics[width=5in]{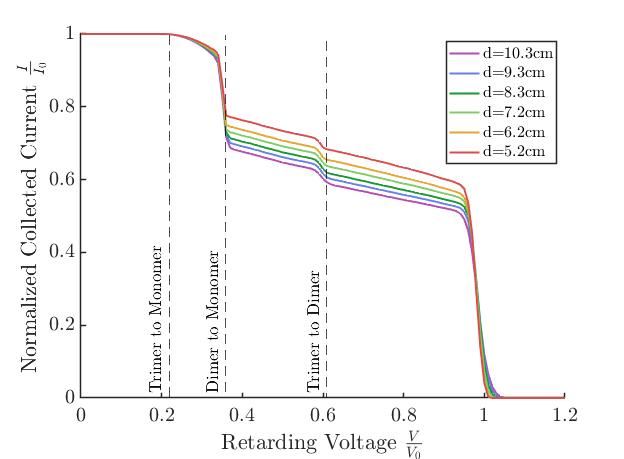}
    \caption{Simulated RPA curves for an EMI-BF$_4$ source firing at 324nA with the source at different distances from the detector.}
    \label{fig:SimulatedRPASpreading}
\end{figure}

Simulations of RPA curves performed using the results of the SOLVEiT N-body simulator were adjusted to account for the RPA detector geometry. Figure \ref{fig:SimulatedRPASpreading} shows the results of these simulations for an EMI-BF$_4$ beam firing at 324 nA with different distances between the source and the detector. The fragmentation model for the simulation was calibrated using the RPA data for the source to detector distance of 8.3cm, which is closest to the 8.85 cm radius of curvature of the detector. As in the experimental data, decreasing distance to the detector increases the size of the monoenergetic step and decreases the size of the dimer step. This is because there is less time for the emitted dimers to fragment before reaching the detector. As with the experimental data bringing the detector closer reduces the spread of the RPA curve at the base of the monoenergetic step. This is possibly due to the effect of the Poisson electric field from the other ions in the beam. The ion density closer to the emitter is larger so the Poisson field adds to the acceleration of the clusters. This results in some of the ions in the monoenergetic step having energies larger than the voltage applied to the source. As the detector is brought closer the amount of time for which the Poisson field acts on the clusters is smaller and thus the energy spreading at the base of the monoenergetic step is smaller. One significant difference between the simulated RPA curves and the experimental ones is the effect of the distance to the detector on the smoothing of the monomer and dimer steps. The smoothing effect is much larger in the experimental curves than in the simulated curves. In particular the bend in the monenergetic step when the detector is 10.3 cm from the source is much less sharp. It is possible that this is due to an interaction of secondary electrons on the beam.

\label{app:appc}
\renewcommand{\thechapter}{D}
\renewcommand{\chaptername}{Appendix}
\chapter{Effect of Aperture Size on Electric Field and Fragmentation}

Fragmentation rates in the acceleration region of electrospray sources depend primarily on the temperature of the emitted species, the types of species, the ionic liquid, and the force of the electric field on the clusters. The temperature of the emitted species and the types of species are mainly controlled by the ionic liquid type, the geometry of the emitter setup, and the voltage applied to the source. The electric field depends both on the voltage applied to the source as well as the geometry of the setup, in particular the aperture hole size on the extractor grid. Changes in the size of the aperture are likely to cause little change in the electric field at the apex of the tip as the curvature of the tip remains the same and the distance to the grid is orders of magnitude larger than the scale on which the electric field around the tip depends. In particular the electric field strength at the surface of the liquid meniscus will likely not be affected by changes in the aperture size. Thus, the voltage response of the source is also likely to not be affected by the aperture size. 

The electric field that changes fragmentation behavior, however, will change with aperture size. From experimental curves it is known that much of the fragmentation in the beam occurs throughout the acceleration region at a wide spread of potentials. Thus, if the potential field were to change due to a change in aperture size it is likely that the cluster residence time at each potential might change, affecting the fragmentation behavior that shows up in RPA curves.

\begin{figure}
    \centering
    \includegraphics[width=5in]{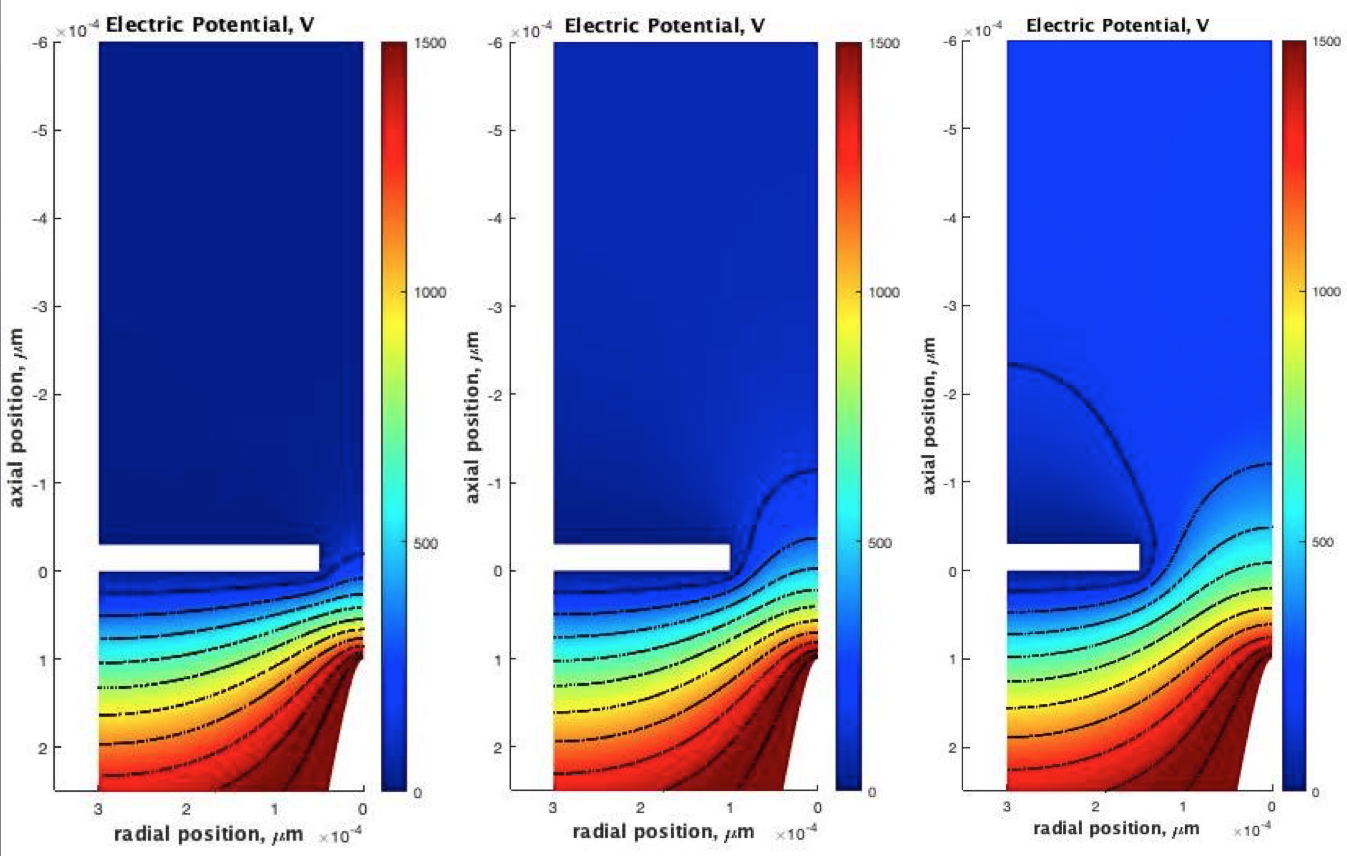}
    \caption{MATLAB Laplace solution with aperture size 50 $\mu m$, 100 $\mu m$, and 150 $\mu m$ from left to right.}
    \label{fig:PotentialFieldsApertureSize}
\end{figure}

Figure \ref{fig:PotentialFieldsApertureSize} shows the Laplacian electric field solved for the given geometry with the aperture radius of 50 $\mu m$, 100 $\mu m$, and 150 $\mu m$ respectively. Each geometry is solved with the same tip shape. The boundary condition was Neumann on each of the open spaces while the boundary condition on the tip, the base of the emitter, and the extractor grid was Dirichlet at the voltage specified for the emitter and the extractor. The voltage used was 1749 V.

\begin{figure}
    \centering
    \includegraphics[width=4.5in]{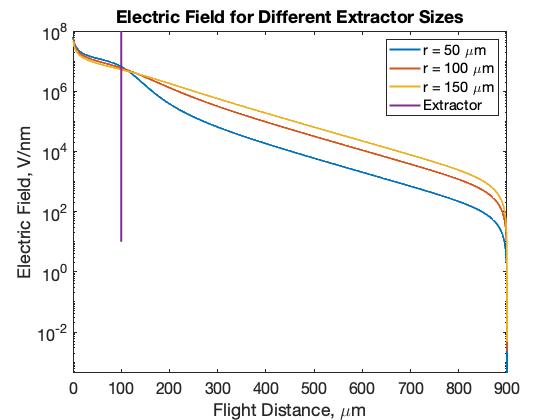}
    \caption{Electric field as a function of distance from the apex of the tip on the tip axis.}
    \label{fig:EField_HoleSizes}
\end{figure}

Figure \ref{fig:EField_HoleSizes} shows the electric field as a function of distance from the apex of the tip along the center axis of the tip. The smaller the aperture size the more the electric field is concentrated between the tip and the extractor. For the larger aperture sizes the electric field bleeds through the aperture more. The drop to zero electric field at 900 $\mu m$ is artificial as a result of the application of the Neumann boundary condition. Increasing the domain size would improve the estimation of the low electric field at this distance, however, it is likely unimportant for fragmentation as results show that fragmentation rates at electric fields this small are not significantly larger than fragmentation rates with no electric field. 

As predicted the difference between the electric fields at the apex of the tip is small. The maximum electric field on the tip axis is $6.17\times 10^7$ V/m, $5.37\times 10^7$ V/m, and $4.61\times 10^7$ V/m for aperture sizes of 50, 100, and 150 $\mu m$ respectively. This difference would result in a change of less than 5\% in the fragmentation rate at the apex. If the shape of the liquid meniscus were taken into account and a finer mesh were used the differences at the tip apex would be negligible. The differences in the electric field in the rest of the domain are larger, with half an order of magnitude difference between the electric field with an aperture size of 50 $\mu m$ or 100 $\mu m$ at 500 $\mu m$ from the apex of the tip. However, it is unclear if this would drastically affect fragmentation behavior that appears in RPA curves because the electric fields where this large difference appears between the different aperture sizes are low. It is possible that the electric fields in this region are low enough that the change in electric field between the cases will not have much of an effect. However, the changes observed closer to the extractor might be large enough to change the residence time of the clusters at each potential enough to change fragmentation behavior. This would also depend on how the aperture size changes the potential field from the tip to the extractor. To better characterize the effect of the electric field changes with different aperture sizes electrohydrodynamic solutions must be obtained for the different geometries. These results must then be used in the SOLVEiT N-body code to simulate RPA curves using the post-processing methods described in section \ref{sec:RPApostProcess}.

\label{app:appd}

\begin{singlespace}
\bibliography{main.bib}
\end{singlespace}

\end{document}